   \documentclass{article}

   \usepackage{amsmath,amssymb,amsfonts}
   \usepackage{pstcol,feynmf}

   \renewcommand{\|}{
   \setlength{\unitlength}{2pt}
   \psset{unit=2pt}
   \psset{runit=2pt}
   \psset{linewidth=0.2}
   \begin{pspicture}(0,0)(2,2)
   \psline(1,0)(1,2)
   \end{pspicture}}

   \newcommand{\Y}{
   \setlength{\unitlength}{2pt}
   \psset{unit=2pt}
   \psset{runit=2pt}
   \psset{linewidth=0.2}
   \begin{pspicture}(0,0)(2,3)
   \psline(1,0)(1,2)
   \psline(1,2)(0,3)
   \psline(1,2)(2,3)
   \end{pspicture}}

   \newcommand\deuxun{
   \setlength{\unitlength}{2pt}
   \psset{unit=2pt}
   \psset{runit=2pt}
   \psset{linewidth=0.2}
   \begin{pspicture}(0,0)(5,5)
   \psline(3,0)(3,2)
   \psline(3,2)(1,4)
   \psline(3,2)(4,3)
   \psline(2,3)(3,4)
   \end{pspicture}}

   \newcommand\deuxdeux{
   \setlength{\unitlength}{2pt}
   \psset{unit=2pt}
   \psset{runit=2pt}
   \psset{linewidth=0.2}
   \begin{pspicture}(1.5,0)(5,5)
   \psline(3,0)(3,2)
   \psline(3,2)(5,4)
   \psline(3,2)(2,3)
   \psline(4,3)(3,4)
   \end{pspicture}}

   \newcommand\troisun{
   \setlength{\unitlength}{2pt}
   \psset{unit=2pt}
   \psset{runit=2pt}
   \psset{linewidth=0.2}
   \begin{pspicture}(0,0)(5,5)
   \psline(3,0)(3,2)
   \psline(3,2)(0,5)
   \psline(3,2)(4,3)
   \psline(2,3)(3,4)
   \psline(1,4)(2,5)
   \end{pspicture}}

   \newcommand\troisdeux{
   \setlength{\unitlength}{2pt}
   \psset{unit=2pt}
   \psset{runit=2pt}
   \psset{linewidth=0.2}
   \begin{pspicture}(1,0)(5,5)
   \psline(3,0)(3,2)
   \psline(3,2)(1,4)
   \psline(3,2)(4,3)
   \psline(2,3)(4,5)
   \psline(3,4)(2,5)
   \end{pspicture}}

   \newcommand\troistrois{
   \setlength{\unitlength}{2pt}
   \psset{unit=2pt}
   \psset{runit=2pt}
   \psset{linewidth=0.2}
   \begin{pspicture}(0.7,0)(5,5)
   \psline(3,0)(3,2)
   \psline(3,2)(0.5,4.5)
   \psline(1.5,3.5)(2.5,4.5)
   \psline(3,2)(5.5,4.5)
   \psline(4.5,3.5)(3.5,4.5)
   \end{pspicture}}

   \newcommand\troisquatre{
   \setlength{\unitlength}{2pt}
   \psset{unit=2pt}
   \psset{runit=2pt}
   \psset{linewidth=0.2}
   \begin{pspicture}(1.5,0)(4,5)
   \psline(3,0)(3,2)
   \psline(3,2)(5,4)
   \psline(3,2)(2,3)
   \psline(4,3)(2,5)
   \psline(3,4)(4,5)
   \end{pspicture}}

   \newcommand\troiscinq{
   \setlength{\unitlength}{2pt}
   \psset{unit=2pt}
   \psset{runit=2pt}
   \psset{linewidth=0.2}
   \begin{pspicture}(1.5,0)(5,5)
   \psline(3,0)(3,2)
   \psline(3,2)(6,5)
   \psline(3,2)(2,3)
   \psline(4,3)(3,4)
   \psline(5,4)(4,5)
   \end{pspicture}}

   \def\C{\mathbb{C}}

   \def\Hp{{\cal{H}}^{\gamma}}
   \def\He{{\cal{H}}^{e}}
   \def\Hq{{\cal{H}}^{\mathrm{qed}}}
   \def\Hd{{\cal{H}}^{\mathrm{dif}}}
   \def\Ho{{\cal{H}}^{\mathrm{odd}}}
   \def\Gd{{\cal{G}}^{\mathrm{dif}}}
   \def\Go{{\cal{G}}^{\mathrm{odd}}}

   \def\Dp{\Delta^{\gamma}}
   \def\De{\Delta^{e}}
   \def\Dq{\Delta^{\mathrm{qed}}}
   \def\Dd{\Delta^{\mathrm{dif}}}
   \def\Do{\Delta^{\mathrm{odd}}}
   \def\DP{\Delta^{P}}
   \def\SP{S_{P}}
   \def\Sp{S^{\gamma}}

   \def\Sd{S^{\mathrm{dif}}}

   \def\Ad{{\cal{A}}^{\mathrm{dif}}}
   \def\ddd{\delta^{\mathrm{dif}}}

   \def\Fp{{\cal{F}}^{\gamma}}
   \def\Fe{{\cal{F}}^{e}}

   \def\pp{\pi^{\gamma}}
   \def\pe{\pi^{e}}
   \def\sp{\sigma^{\gamma}}
   \def\se{\sigma^{e}}
   \def\phip{\varphi^{\gamma}}
   \def\phie{\varphi^{e}}
   \def\barphip{{\bar\varphi}^{\gamma}}
   \def\barphie{{\bar\varphi}^{e}}
   \def\Lp{L^{\gamma}}
   \def\Le{L^{e}}
   \def\cp{c^{\gamma}}
   \def\ce{c^{e}}
   \def\pro{\ }
   \def\<{\langle}
   \def\>{\rangle}
   \def\dd{\mathrm{d}}
   \def\Id{\mathrm{Id}}

   \def\Qbar{\bar Q}
   
   \def\Tbar{\bar T}
   
   \def\ubf{{\mathbf{u}}}
   \def\vbf{{\mathbf{v}}}
   \def\xbf{{\mathbf{x}}}
   \def\ybf{{\mathbf{y}}}
   \def\wbf{{\mathbf{w}}}
   \def\abf{{\mathbf{a}}}
   \def\bbf{{\mathbf{b}}}
   \def\ebf{{\mathbf{e}}_0}
   \def\Dbf{{\mathbf{D}}}
   \def\Sbf{{\mathbf{S}}}
   \def\Zebf{{\mathbf{Z}_2}}
   \def\Zpbf{{\mathbf{Z}_3}}
   \def\Zobf{{\mathbf{Z}_0}}
   \def\fbf{{\mathbf{f}}}
   \def\gbf{{\mathbf{g}}}
   
   \def\1bf{{\mathbf{1}}}
   \def\Pibf{{\boldsymbol{\Pi}}}
   \def\Sigmabf{{\boldsymbol{\Sigma}}}

   \def\tr{\mathrm{tr}}
    
   \newtheorem{theorem}{Theorem}[section]
   \newtheorem{proposition}[theorem]{Proposition}
   \newtheorem{corollary}[theorem]{Corollary}
   \newtheorem{lemma}[theorem]{Lemma}
   \newtheorem{defin}[theorem]{Definition}
   \newenvironment{definition}{\begin{defin} \em}{\end{defin}}
   \newtheorem{rem}[theorem]{Remark}
   \newenvironment{remark}{\begin{rem} \em}{\end{rem}}
   \newenvironment{proof}{\noindent{\em Proof.\/}}
           {\hfill$\square$\par\vspace{.2cm}}
   \newenvironment{proof of}[1]{\noindent{\em Proof of (\ref{#1}).\/}}
           {\hfill$\square$\par\vspace{.2cm}}
    
\begin{document}

   \title{Noncommutative renormalization for massless QED}

   \author{Christian Brouder \\
   Laboratoire de Min\'eralogie-Cristallographie, CNRS UMR7590,
    Universit\'es Paris 6 et 7, \\IPGP, 4 place Jussieu,
     75252 Paris Cedex 05,
     France. {\small \tt brouder@lmcp.jussieu.fr}
   \and
   Alessandra Frabetti \\
   Institut de Math\'ematique, B\^atiment de Chimie,
   Universit\'e de Lausanne, \\
   CH-1015 Lausanne, Switzerland
   {\small \tt afrabett@magma.unil.ch}
   \date{\today}}

   \maketitle

   \begin{fmffile}{dx81}
   \setlength{\unitlength}{1mm}
   \newcommand{\setval}{\fmfset{wiggly_len}{1.5mm}\fmfset{arrow_len}{2.5mm}
           \fmfset{arrow_ang}{13}\fmfset{dash_len}{1.5mm}\fmfpen{0.25mm}
           \fmfset{dot_size}{0.8thick}}
   \newcommand{\scs}{\scriptstyle}

   %
   %
    
   \begin{abstract}
   We study the renormalization of massless QED from the point of
   view of the Hopf algebra discovered by D.~Kreimer. 
   For QED, we describe a Hopf algebra of renormalization which is 
   neither commutative nor cocommutative. 
   We obtain explicit renormalization formulas for the electron and photon
   propagators, for the vacuum polarization and the electron
   self-energy, which are equivalent to Zimmermann's forest formula 
   for the sum of all Feynman diagrams at 
   a given order of interaction. 

   Then we extend to QED the Connes-Kreimer map defined by 
   the coupling constant of the theory
   (i.e. the homomorphism between some formal diffeomorphisms 
   and the Hopf algebra of renormalization) 
   by defining a noncommutative Hopf algebra of diffeomorphisms, 
   and then showing that the renormalization of the electric charge 
   defines a homomorphism between this Hopf algebra 
   and the Hopf algebra of renormalization of QED.

   Finally we show that Dyson's formulas for the renormalization 
   of the electron and photon propagators can be given in a 
   noncommutative (e.g. matrix-valued) form. 
   \end{abstract}
    
    
   %
   %
    
   \section*{Introduction}
   \addcontentsline{toc}{section}{\bf Introduction}

   D.~Kreimer made the remarkable discovery that the combinatorics
   of renormalization was hiding a Hopf algebra \cite{Kreimer98}.
   The publication of ref.\cite{Kreimer} has spurred many developments
   in the theory of renormalization and in the application of Hopf algebras
   to quantum field theories. Among these developments, one
   of the most intriguing is the construction, by
   A.~Connes and D.~Kreimer, of a homomorphism  between 
   the Hopf algebra of renormalization of the massless $\varphi^3$ 
   quantum field theory and 
   the Connes-Moscovici Hopf algebra of
   some formal diffeomorphisms on the complex line (see \cite{CKII}).
   Such a homomorphism will be called a Connes-Kreimer map.
   In the present paper, we build a Connes-Kreimer map
   for the case of the  noncommutative (nor cocommutative) Hopf algebra
   of renormalization of massless quantum electrodynamics (QED).

   In the $\varphi^3$ quantum field theory, the field $\varphi$ is scalar.
   Thus, the amplitudes given by Feynman diagrams are scalars,
   and their product is commutative. In QED, the fields are
   either spinors (for the electrons) or vectors (for the photons),
   and the amplitudes are 4x4 complex matrices. Therefore, their
   product is not commutative. If we want the Hopf algebra
   describing the renormalization of QED to be compatible with
   the product of amplitudes, the algebraic product cannot
   be commutative. This choice was made in Ref.\cite{BF},
   where we described the renormalization of massless QED by means 
   of a coproduct on planar binary trees. 

   In the present paper, we first show that this coproduct defines
   a noncommutative nor cocommutative Hopf algebra over trees.
   Then we define the natural noncommutative extension of 
   the Connes-Moscovici algebra of formal diffeomorphisms and we prove
   that this is also a Hopf algebra. Finally, we prove that 
   the relation between the bare and renormalized electric charge
   $e_0=e/\sqrt{Z_3(e)}$ defines a homomorphism between the
   noncommutative Connes-Moscovici algebra and the Hopf algebra
   of renormalization of QED.

   By ``noncommutative renormalization'' we mean that the 
   noncommutativity of the product of amplitudes is respected 
   by the renormalization method.
   In practice, the renormalization factors $Z$ multiply 
   the various terms of the Lagrangian, which is a scalar quantity. 
   Thus, all $Z$ are scalars, therefore 
   commutative, quantities.
   However, the investigation of renormalization by 
   noncommutative multiplication factors has a practical advantage. 
   When various fields
   are present, renormalization proceeds through the
   use of various $Z_{ij}$ coupling the different fields. 
   If we group the fields together in a single field vector, 
   then $\{Z_{ij}\}$ becomes a matrix. Noncommutative renormalization
   can also be used
   to investigate the transmutation of leptons due to the
   rotation of the mass matrix \cite{Bordes}.
   Such a matrix-valued renormalization factor is used even for QED,
   in the Plymouth approach, to take care of the infrared divergences, 
   cf. \cite{BaganI,BaganII}.

   From a mathematical point of view, a morphism
   between two noncommutative Hopf algebras is
   richer than the morphism between
   their abelianizations. It is surprising that
   such a morphism between the noncommutative
   Hopf algebra of formal diffeomorphisms and the algebra of
   renormalization exists, since it does not
   seem to be required by the standard 
   renormalization methods. 
   Even more surprising is the fact that such 
   a noncommutative version of the Hopf algebra 
   of formal diffeomorphisms exists. 
   Intuitively, we would like to consider it as 
   the algebra of nonscalar functions on formal series 
   with noncommutative coefficients, 
   but this point of view fails for two main reasons: 
   on one side the composition of such series 
   is not associative, while the dual coproduct is, 
   on the other side, even if the composition were 
   associative, the convolution does not preserve 
   the noncommutative characters of the Hopf algebra, 
   and by duality we could not recover the formal series 
   from the Hopf algebra of functions. 
   If we force the composition of series to be associative, 
   by introducing a suitable normal ordering, 
   we come to another surprising fact: the composition 
   in normal ordering is not a binary operation, 
   while the coproduct on the Hopf algebra is. 
   In conclusion, the only way to regard this 
   Hopf algebra is as a deformation of the set of 
   scalar functions on formal series, a kind of infinite 
   dimensional ``quantum group'' with no commutation 
   relations at all. 

   The present paper follows the concepts developed
   in the recent works by Connes and Kreimer.
   However, since our Hopf algebras are neither
   commutative nor cocommutative, we
   cannot use the Milnor-Moore theorem
   to deal with the associated Lie algebras. 
   Our proofs are elementary, mainly based on recurrence 
   relations and generating functions.
   They provide explicit expressions for the 
   coproducts and reveal an interesting  
   connection with the resolvent
   of linear operators.

   We do not consider infrared divergences.
   \bigskip

   \noindent
   {\bf Notation.} 
   We suppose that all vector spaces and algebras are defined over the 
   field $\C$ of complex numbers. 
   We denote by $\C\langle X \rangle$ the algebra of noncommutative 
   polynomials on the set of variables $X$, and by $\C[X]$ 
   the algebra of commutative polynomials. The ideal generated 
   by $x$ is denoted by $(x)$. 

   When confusion may arise, non-commutative variables are chosen 
   with boldface characters. 

   \tableofcontents

   %
   %
    
   \section{Renormalization ``\`a la Dyson''} 
   \label{juneren}

   Many experiments proved that the density of charge 
   of an electron placed in the vacuum is in fact modified 
   by the presence of this ``vacuum'' \cite{Itzykson}. 
   Following Dirac, we can consider the vacuum of QED
   as a medium were all negative energy states are
   occupied. An electromagnetic field polarizes 
   these negative energy electrons: this is 
   called vacuum polarization, and an electron interacts
   with its electromagnetic field: this is
   called electron self-energy \cite{Itzykson}. 

   The vacuum polarization and the electron self-energy 
   are usually denoted by $\Pi$ and $\Sigma$, respectively. 
   They can be deduced directly from the Lagrangian through 
   the one-particle Green functions, or propagators, 
   of quantum electrodynamics.

   The electron and photon propagators represent the 
   amplitude of probability that the corresponding particle 
   propagates from a point $x$ to a point $y$, and they 
   are denoted by $S(x,y)$ and $D_{\lambda\mu}(x,y)$ respectively. 
   In vacuum, the propagators are translation-invariant and 
   can be written as $S(x-y)$ and $D_{\lambda\mu}(x-y)$.
   Their Fourier transforms are denoted by 
   $S(q;e_0)$ and $D(q;e_0)$, where $e_0$ is the (fictive)
   electron charge of the electrons ``before renormalization'' 
   and is called the bare charge.

   The vacuum polarization and the self-energy 
   can be deduced from the propagators through the identities 
   \begin{eqnarray}
           D^{-1}(q;e_0) &=& D_0^{-1}(q;e_0)-\Pi(q;e_0), 
           \label{vacuumpol}\\
           S^{-1}(q;e_0) &=& S_0^{-1}(q;e_0)-\Sigma(q;e_0), 
           \label{selfen} 
   \end{eqnarray}
   where $D^{-1}$ and $S^{-1}$ are the inverse of the operators 
   $D$ and $S$. 

   In fact, many interesting physical properties of the electromagnetic 
   field can be obtained directly from the propagators. 
   From the electron propagator $S(x,y)$, for instance, 
   one can describe the charge density 
   $\rho(x) = i e \tr[\gamma^0 S(x,x)]$ and the 
   electromagnetic current $J^k(x) = i e \tr[\gamma^k S(x,x)]$. 
   From the photon propagator $D_{\lambda\mu}(x,y)$, instead, 
   one can describe the electromagnetic potential $A_\lambda(x)$ 
   due to an external electromagnetic current $j^\mu(y)$
   (to first order) by $A_\lambda(x)=\int dy D_{\lambda\mu}(x,y)j^\mu(y)$.
   In other words, the photon propagator describes the
   linear response of the system to an external electromagnetic
   current.

   Most methods available to calculate the propagators
   are perturbative.
   In perturbative quantum electrodynamics, the electron 
   self-energy, the vacuum polarization and the propagators 
   are written as power series over the coupling constant $e_0^2$, 
   \begin{eqnarray} 
           \Sigma(q;e_0) &=& \sum_{n=1}^{\infty} e_0^{2n} \Sigma_n(q), 
           \label{Sigma(q)} \\ 
           \Pi(q;e_0) &=& \sum_{n=1}^{\infty} e_0^{2n} \Pi_n(q), \label{Pi(q)} \\ 
           S(q;e_0) &=& \sum_{n=0}^{\infty} e_0^{2n} S_n(q), \label{S(q)} \\ 
           D(q;e_0) &=& \sum_{n=0}^{\infty} e_0^{2n} D_n(q). \label{D(q)} 
   \end{eqnarray}
   The terms $\Sigma_n(q)$, $\Pi_n(q)$, $S_n(q)$ and $D_n(q)$ 
   are finite sums of integrals. 
   Each of these integrals is classically described by 
   a suitable Feynman diagram $\Gamma$ with $n$ loops.  
   The graphs $\Gamma$ should be chosen according 
   to the propagator they describe, and they provide 
   a very figurative and elegant way to remember 
   all possible interactions which should be considered 
   in the expansion at a fixed order. 
   For instance, for $n=0$ we consider the free propagators, 
   i.e. the propagators for QED without electromagnetic interactions 
   ($e_0=0$), which are usually denoted by 
   simple lines 
   \begin{eqnarray*}
           S_0(q) = 
   \parbox{12mm}{\begin{center}
           \begin{fmfgraph}(10,5)
           \setval
           \fmfforce{0w,0.5h}{v1}
           \fmfforce{1w,0.5h}{v2}
           \fmf{plain}{v2,v1}
           \end{fmfgraph}
           \end{center}}, &\qquad&  
           D_0(q) = 
   \parbox{12mm}{\begin{center}
           \begin{fmfgraph}(10,5)
           \setval
           \fmfforce{0w,0.5h}{v2}
           \fmfforce{1w,0.5h}{v3}
           \fmf{boson}{v2,v3}
           \end{fmfgraph}
           \end{center}};  
   \end{eqnarray*} 
   and for $n=1$ we consider the graphs with one loop, 
   \begin{eqnarray*}
           S_1(q) = 
   \parbox{22mm}{\begin{center}
     \begin{fmfgraph}(20,5)
     \setval
     \fmfforce{0w,0.5h}{v1}
     \fmfforce{1/3w,0.5h}{v2}
     \fmfforce{2/3w,0.5h}{v3}
     \fmfforce{1w,0.5h}{v4}
     \fmf{plain}{v4,v3,v2,v1}
     \fmf{boson,left}{v2,v3}
     \end{fmfgraph}
     \end{center}}, &\qquad&  
           D_1(q) =  
   \parbox{22mm}{\begin{center}
           \begin{fmfgraph}(20,5)
           \setval
     \fmfforce{0w,0.5h}{v1}
     \fmfforce{1/3w,0.5h}{v2}
     \fmfforce{2/3w,0.5h}{v3}
     \fmfforce{1w,0.5h}{v4}
           \fmf{boson}{v1,v2}
           \fmf{fermion,right}{v2,v3,v2}
           \fmf{boson}{v3,v4}
           \end{fmfgraph}
           \end{center}}. 
   \end{eqnarray*} 

   More precisely, denote by $\Fe$ and $\Fp$ the sets of graphs which 
   describe the electron and the photon propagators respectively. 
   The rules to assign an integral to each graph $\Gamma$ 
   were given by Feynman \cite{Feynman}. 
   Denote by $U$ the ``Feynman rules'' map, which assign 
   to each graph $\Gamma$ the suitable integral $U(\Gamma)$. 
   (An explicit description of the sets $\Fe$, $\Fp$ and of 
   the Feynman rules can be found in any textbook of quantum field theory, 
   for instance \cite{Itzykson}, Appendix 4.)
   Then, the coefficients $S_n(q)$ and $D_n(q)$ can be described as 
   \begin{eqnarray*} 
           S_n(q) &=& \sum_{{\Gamma \in \Fe},\ {|\Gamma|=n}} 
           U(\Gamma;q),  \label{Sn(q)} \\ 
           D_n(q) &=& \sum_{{\Gamma \in \Fp},\ {|\Gamma|=n}} 
           U(\Gamma;q),  \label{Dn(q)} 
   \end{eqnarray*}
   where $|\Gamma|$ is half the number of vertices 
   in the diagram $\Gamma$. 

   The graphs which contribute to the self-energy and to 
   the vacuum polarization are the so-called 
   1-particule irreducible (1PI) Feynman diagrams, 
   i.e. those diagrams which can not be seen as 
   junction of two or more meaningful adjacent diagrams. 
   If we denote by $\Fe_{1PI} \subset \Fe$ and 
   $\Fp_{1PI} \subset \Fp$ the subsets of 
   1PI Feynman diagrams, the coefficients $\Sigma_n(q)$ 
   and $\Pi_n(q)$ can be described as 
   \begin{eqnarray*} 
           \Sigma_n(q) &=& \sum_{{\Gamma \in \Fe_{1PI}},\ {|\Gamma|=n}} 
           U(\Gamma;q),  \label{Sigman(q)} \\ 
           \Pi_n(q) &=& \sum_{{\Gamma \in \Fp_{1PI}},\ {|\Gamma|=n}} 
           U(\Gamma;q).  \label{Pin(q)} 
   \end{eqnarray*}

   The problem is that the most of the integrals $U(\Gamma;q)$ 
   involved in these expansions are divergent.

   Renormalization theory is a prescription to remove 
   the divergences from the integrals $U(\Gamma;q)$, 
   yielding finite numbers in accordance with experimental data. 
   In practice, renormalization consists in two main steps. 

   The first step is a regularization of  
   the divergent integrals to avoid manipulating infinities. 
   Many regularizations are possible, the integrals can be limited 
   to an upper value $q^2 < \Lambda^2$ (cutoff), they can be 
   carried out in a non integer dimension $d=4-\epsilon$ 
   (dimensional regularization), etc. 
   Various regularizations are described in 
   Refs.\cite{Itzykson,Collins}. Thus, the terms $U(\Gamma;q)$,
   and consequently the coefficients $S_n(q)$ and $D_n(q)$ 
   of the perturbative expansions, usually depend on the regularization
   variables (e.g. $\Lambda$, $\epsilon$, etc.) which are left 
   implicit for notational convenience. 
   The original, divergent, integrals are recovered when 
   the regularization parameters tend to the limit which 
   gives the divergency ($\epsilon\rightarrow 0$,
   $\Lambda\rightarrow \infty$, etc.).

   The second step consists in extracting, from the regularized 
   integrals, a term which contains the divergent part, 
   once the regularization parameter is sent to his limit. 
   We take the example of dimensional regularization, 
   for which the regularization parameter is $\epsilon$.
   To find the finite value corresponding to $U_\epsilon(\Gamma)$, 
   we first remove all
   subdivergences from $U_\epsilon(\Gamma)$ to obtain a
   quantity denoted by $\bar R_\epsilon(\Gamma)$. In a renormalizable
   theory, such as QED, the term $\bar R_\epsilon(\Gamma)$ 
   has a given order of divergence. 
   For instance, the self-energy is
   linearly divergent, so that
   \begin{eqnarray*} 
   \bar R_\epsilon(\Gamma;q) &=& -C_0(\Gamma,\epsilon)- 
   C_1(\Gamma,\epsilon) q + R_\epsilon(\Gamma;q),
   \end{eqnarray*}
   where $R_\epsilon(\Gamma;q)$ has now a finite value
   for $\epsilon\rightarrow 0$, this value being
   the renormalized value of $\Gamma$ denoted
   by $R(\Gamma;q)$.
   The divergent counterterms $C_0(\Gamma,\epsilon)$
   and $C_1(\Gamma,\epsilon)$ are determined by the
   so-called renormalization conditions, which are
   boundary conditions for the finite part
   $R_\epsilon(\Gamma;q)$. For instance,
   we may require that $R_\epsilon(\Gamma;q)$
   and its derivative with respect to $q$ are
   0 for $q=0$. Many other renormalization
   conditions are possible  (cf. \cite{Collins}).
   The purpose of the renormalization conditions
   is determine the finite and the infinite
   parts of $\bar R_\epsilon(\Gamma)$. 

   For comparison with experiment,
   the renormalization conditions use experimental
   results, such as the electron mass or the
   electron charge \cite{Itzykson}.

   The renormalization of 1-loop diagrams is usually 
   an accessible computation. 
   It becomes a hard task when 
   divergent integrals contain divergent subintegrals, 
   soon as the order of interaction is greater than $1$. 
   In such cases, the presence of loops with subloops in 
   Feynman diagrams translates into a dependency 
   of the counterterm of the diagram from those of the subdiagrams.  

   Renormalization of Feynman diagrams was throughly 
   studied, culminating first in Zimmermann's forest
   formula \cite{Zimmermann}, which describes the intricate 
   dependency of the counterterms of several-loops graphs, 
   and then in Kreimer's discovery of a Hopf algebra structure 
   hidden in the procedure, cf. \cite{Kreimer98}. 
   Since the renormalization of Feynman diagrams is not our subject, 
   for more details we refer the reader to standard textbooks 
   \cite{Itzykson,Collins} and to the recent publications 
   \cite{Connes,CKI,CKII,KreimerBook}.

   After renormalization, by summing up all the finite integrals 
   we obtain a finite value of the propagators at each order 
   of interaction, 
   \begin{eqnarray*} 
           \bar{S}_n(q) &=& \sum_{{\Gamma \in \Fe},\ {|\Gamma|=n}} 
           R(\Gamma;q) , \\ 
           \bar{D}_n(q) &=& \sum_{{\Gamma \in \Fp},\ {|\Gamma|=n}} 
           R(\Gamma;q) , 
   \end{eqnarray*}
   and similarly for the self-energy and the vacuum polarization. 
   The renormalized propagators 
   $\bar{S}(q;e)$ and $\bar{D}(q;e)$ can then be described as
   power series over the square of the (finite) effective charge $e$, 
   \begin{eqnarray} 
           \bar{S}(q;e) &=& \sum_{n \geq 0} e^{2n} \bar{S}_n(q), \\ 
           \bar{D}(q;e) &=& \sum_{n \geq 0} e^{2n} \bar{D}_n(q).
   \end{eqnarray}
   In the simplest case, $e$ is the charge of the electron
   experimentally measured.

   In a strike of genius, Freeman Dyson \cite{Dyson} noticed 
   that the relation between the renormalized propagators 
   $\bar S(q;e)$, $\bar D(q;e)$ and the bare propagators 
   $S(q;e_0)$, $D(q;e_0)$ can be given by 
   introducing two (scalar) renormalization factors 
   $Z_2(e)$ and $Z_3(e)$ which withdraw the divergent part 
   of $S(q;e_0)$ and $D(q;e_0)$ according to the equations 
   \begin{eqnarray}
           Z_2(e) \bar S(q,e) &=& S(q,e_0), \label{DysonS}\\ 
           Z_3(e) \bar D(q,e) &=& D(q,e_0). \label{DysonD}
   \end{eqnarray}
   Equations (\ref{DysonS}) and (\ref{DysonD}) are known
   as Dyson's formulas. 
   J.C.~Ward then showed \cite{Ward} that the renormalized charge $e$ 
   is related to the bare charge $e_0$, which is an infinite value, 
   by means of the photon renormalization factor, 
   \begin{eqnarray} 
           Z_3(e)^{-1/2} e &=& e_0. \label{Z3e0}
   \end{eqnarray}

   In modern textbooks, cf. \cite{WeinbergBook,PeskinSchroeder}, 
   the renormalization factors $Z_2$ and $Z_3$ are introduced from the ratio
   between renormalized and bare fields, from the counterterms
   of the Lagrangian, from the residue of the pole of the
   propagators, sometimes also from the field commutation
   relations. We will show that Dyson's formulas
   are very effective for an extension to nonscalar renormalization
   factors.
   In fact, in this paper we discuss how they determine 
   some relations between the coefficients of the perturbative
   expansions, and finally how these relations determine a new 
   noncommutative form of the Dyson's formulas (\ref{DysonS},\ref{DysonD}) 
   and of the charge renormalization (\ref{Z3e0}). 

   Since the renormalization conditions are given via 
   the self-energy and the vacuum polarization, which are 
   expanded over 1PI Feynman diagrams, it is clear that 
   the renormalization factors $Z_2$ and $Z_3$ 
   are related to the counterterms $C(\Gamma)$ of 
   1PI diagrams \cite{BF}.

   In particular, to fix a notation, let us expand 
   the renormalization factors over $e^2$ as  
   \begin{eqnarray} 
           Z_2(e) &=& 1+ \sum_{n \geq 1} e^{2n} Z_{2,n}, \label{Z2}\\ 
           Z_3(e) &=& 1- \sum_{n \geq 1} e^{2n} Z_{3,n}, \label{Z3}
   \end{eqnarray}
   where $Z_{2,n}$ and $Z_{3,n}$ are the renormalization 
   terms which appear at the $n$-th order of interaction. 
   The minus sign in the definition of $Z_3(e)$ is chosen
   for later convenience (it is similar to the minus sign
   used in the Connes-Kreimer homomorphism \cite{CKII}). 
   Then, the coefficients $Z_{2,n}$ and $Z_{3,n}$ are 
   related to the counterterms as follows, cf. \cite{BF}, 
   \begin{eqnarray*} 
           Z_{2,n} &=& -\sum_{{\Gamma \in \Fe_{1PI}},\ {|\Gamma|=n}} 
           C_2(\Gamma;q),  \label{Z2n} \\ 
           Z_{3,n} &=& \sum_{{\Gamma \in \Fp_{1PI}},\ {|\Gamma|=n}} 
           C_3(\Gamma;q).  \label{Z3n} 
   \end{eqnarray*}

   One of the main features of Feynman diagrams is the fact that 
   the amplitude of two graphs joined by a single line 
   is computed as the product of the amplitudes of the disjoint graphs. 
   In other words, the junction of adjacent Feynman diagrams 
   defines an internal associative product on the sets 
   $\Fe$ and $\Fp$ compatible with the map $U$, 
   with generators given by the subspaces $\Fe_{1PI}$ 
   and $\Fp_{1PI}$. 

   In fact, this assertion is not entirely correct. 
   The propagators $S(q;e_0)$ and $D(q;e_0)$ of 
   Eqs.~(\ref{S(q)},\ref{D(q)}) are described by graphs 
   which start and end with a free propagator (a simple line). 
   When we joint two such graphs, one of the two central lines 
   of the new graph disappears, and this produces a new factor 
   in the amplitudes, proportional to the inverse of the 
   amplitude of the free propagator. 
   For instance, 
   \begin{eqnarray*}
   U \big( \parbox{32mm}{\begin{center}
   \begin{fmfgraph}(30,5)
   \setval
   \fmfforce{0w,0.5h}{v1}
   \fmfforce{1/5w,0.5h}{v2}
   \fmfforce{2/5w,0.5h}{v3}
   \fmfforce{3/5w,0.5h}{v4}
   \fmfforce{4/5w,0.5h}{v5}
   \fmfforce{1w,0.5h}{v6}
   \fmf{boson}{v1,v2}
   \fmf{fermion,right}{v2,v3,v2}
   \fmf{boson}{v3,v4}
   \fmf{fermion,right}{v4,v5,v4}
   \fmf{boson}{v5,v6}
   \end{fmfgraph}
   \end{center}} \big) 
   &=& 
   U \big( \parbox{17mm}{\begin{center}
   \begin{fmfgraph}(15,5)
   \setval
   \fmfforce{0w,0.5h}{v1}
   \fmfforce{1/3w,0.5h}{v2}
   \fmfforce{2/3w,0.5h}{v3}
   \fmfforce{1w,0.5h}{v4}
   \fmf{boson}{v1,v2}
   \fmf{fermion,right}{v2,v3,v2}
   \fmf{boson}{v3,v4}
   \end{fmfgraph}
   \end{center}} \big) 
   \frac{1}
   {U \big( \parbox{12mm}{\begin{center}
   \begin{fmfgraph}(10,5)
   \setval
   \fmfforce{0w,0.5h}{v2}
   \fmfforce{1w,0.5h}{v3}
   \fmf{boson}{v2,v3}
   \end{fmfgraph}
   \end{center}} \big)} 
   U \big(\parbox{17mm}{\begin{center}
   \begin{fmfgraph}(15,5)
   \setval
   \fmfforce{0w,0.5h}{v1}
   \fmfforce{1/3w,0.5h}{v2}
   \fmfforce{2/3w,0.5h}{v3}
   \fmfforce{1w,0.5h}{v4}
   \fmf{boson}{v1,v2}
   \fmf{fermion,right}{v2,v3,v2}
   \fmf{boson}{v3,v4}
   \end{fmfgraph}
   \end{center}} \big). 
   \end{eqnarray*}

   To avoid this factor we choose a notation of Feynman diagrams 
   which differs slightly from the usual one: we drop the first 
   simple line from all diagrams. 
   This modification of the diagrams corresponds to divide 
   all propagators by the free propagator on the left, 
   and to multiply the vacuum polarization and the self-energy 
   by the free propagators on the right. 
   In particular, the free propagators $D^0$ and $S^0$ become 
   unit matrices and both correspond to no graph. 
   We denote them by $1$.  
   The sets of Feynman diagrams that we consider are then the 
   following: 
   \begin{eqnarray*}
   \Fe &=& \big\{ 1, 
   \parbox{12mm}{\begin{center}
           \begin{fmfgraph}(10,5)
           \setval
           \fmfforce{0w,0.5h}{v1}
           \fmfforce{1/2w,0.5h}{v2}
           \fmfforce{1w,0.5h}{v3}
           \fmf{plain}{v3,v2,v1}
           \fmf{boson,left}{v1,v2}
           \fmfdot{v1,v2}
           \end{fmfgraph}
           \end{center}}, 
   \parbox{17mm}{\begin{center}
     \begin{fmfgraph}(15,5)
     \setval
     \fmfforce{0w,0.5h}{v1}
     \fmfforce{1/4w,0.5h}{v2}
     \fmfforce{2/4w,0.5h}{v3}
     \fmfforce{3/4w,0.5h}{v4}
     \fmfforce{1w,0.5h}{v5}
     \fmf{plain}{v5,v4,v3,v2,v1}
     \fmf{boson,left}{v1,v3}
     \fmf{boson,right}{v2,v4}
     \fmfdot{v1,v2,v3,v4}
     \end{fmfgraph}
     \end{center}} , 
   \parbox{17mm}{\begin{center}
     \begin{fmfgraph}(15,5)
     \setval
     \fmfforce{0w,0.5h}{v1}
     \fmfforce{1/4w,0.5h}{v2}
     \fmfforce{2/4w,0.5h}{v3}
     \fmfforce{3/4w,0.5h}{v4}
     \fmfforce{1w,0.5h}{v5}
     \fmf{plain}{v5,v4,v3,v2,v1}
     \fmf{boson,left}{v1,v4}
     \fmf{boson,left}{v2,v3}
     \fmfdot{v1,v2,v3,v4}
     \end{fmfgraph}
     \end{center}}, 
   \parbox{17mm}{\begin{center}
     \begin{fmfgraph}(15,5)
     \setval
     \fmfforce{0w,0.5h}{v1}
     \fmfforce{1/4w,0.5h}{v2}
     \fmfforce{2/4w,0.5h}{v3}
     \fmfforce{3/4w,0.5h}{v4}
     \fmfforce{1w,0.5h}{v5}
     \fmf{plain}{v5,v4,v3,v2,v1}
     \fmf{boson,left}{v1,v2}
     \fmf{boson,left}{v3,v4}
     \fmfdot{v1,v2,v3,v4}
     \end{fmfgraph}
     \end{center}}, 
   \parbox{17mm}{\begin{center}
     \begin{fmfgraph}(15,8)
     \setval
     \fmfforce{0w,0.5h}{v1}
     \fmfforce{1/4w,1h}{v2}
     \fmfforce{2/4w,1h}{v3}
     \fmfforce{3/4w,0.5h}{v4}
     \fmfforce{1w,0.5h}{v5}
     \fmf{plain}{v5,v4,v1}
     \fmf{boson,left=1/3}{v1,v2}
     \fmf{boson,left=1/3}{v3,v4}
     \fmf{fermion,right}{v2,v3,v2}
     \fmfdot{v1,v2,v3,v4}
     \end{fmfgraph}
     \end{center}},
           ... \big\} \\ 
   \Fp &=& \big\{ 1, 
   \parbox{12mm}{\begin{center}
           \begin{fmfgraph}(10,5)
           \setval
           \fmfforce{0w,0.5h}{v1}
           \fmfforce{1/2w,0.5h}{v2}
           \fmfforce{1w,0.5h}{v3}
           \fmf{fermion,right}{v1,v2,v1}
           \fmf{boson}{v2,v3}
           \fmfdot{v1,v2}
           \end{fmfgraph}
           \end{center}}, 
   \parbox{17mm}{\begin{center}
   \begin{fmfgraph}(15,5)
   \setval
   \fmfforce{1/3w,0.5h}{v2}
   \fmfforce{1/2w,1h}{v2b}
   \fmfforce{2/3w,0.5h}{v3}
   \fmfforce{1/2w,0h}{v3b}
   \fmfforce{1w,0.5h}{v4}
   \fmf{fermion,right=0.45}{v2,v3b,v3,v2b,v2}
   \fmf{boson}{v3,v4}
   \fmf{boson}{v2b,v3b}
   \fmfdot{v2,v2b,v3,v3b}
   \end{fmfgraph}
           \end{center}}, 
   \parbox{17mm}{\begin{center}
           \begin{fmfgraph}(15,5)
           \setval
           \fmfforce{1/3w,0.5h}{v2}
           \fmfforce{0.4008w,0.933h}{v2b}
           \fmfforce{0.5992w,0.933h}{v2c}
           \fmfforce{2/3w,0.5h}{v3}
           \fmfforce{1w,0.5h}{v4}
           \fmf{fermion,right=0.33}{v3,v2c,v2b,v2}
           \fmf{fermion,right=1}{v2,v3}
           \fmf{boson,right=0.8}{v2b,v2c}
           \fmf{boson}{v3,v4}
           \fmfdot{v2,v2b,v2c,v3}
           \end{fmfgraph}
           \end{center}}, 
   \parbox{17mm}{\begin{center}
           \begin{fmfgraph}(15,5)
   \setval
   \fmfforce{1/3w,0.5h}{v2}
   \fmfforce{2/3w,0.5h}{v3}
   \fmfforce{0.4008w,0.067h}{v3b}
   \fmfforce{0.5992w,0.067h}{v3c}
   \fmfforce{1w,0.5h}{v4}
   \fmf{fermion,right=1}{v3,v2}
   \fmf{fermion,right=0.33}{v2,v3b,v3c,v3}
   \fmf{boson,left=0.8}{v3b,v3c}
   \fmf{boson}{v3,v4}
   \fmfdot{v2,v3b,v3c,v3}
   \end{fmfgraph}
   \end{center}}, 
   \parbox{17mm}{\begin{center}
           \begin{fmfgraph}(15,5)
           \setval
           \fmfforce{0w,0.5h}{v2}
           \fmfforce{1/4w,0.5h}{v3}
           \fmfforce{2/4w,0.5h}{v4}
           \fmfforce{3/4w,0.5h}{v5}
           \fmfforce{1w,0.5h}{v6}
           \fmf{fermion,right}{v2,v3,v2}
           \fmf{boson}{v3,v4}
           \fmf{fermion,right}{v4,v5,v4}
           \fmf{boson}{v5,v6}
           \fmfdot{v2,v3,v4,v5}
           \end{fmfgraph}
           \end{center}},
           ... \big\} 
   \end{eqnarray*} 

   With these notations, the map $U$ becomes multiplicative, 
   that is, $U(\Gamma_1) U(\Gamma_2) = U(\Gamma_1 \Gamma_2)$. 
   For instance, 
   \begin{eqnarray*}
   U \big(
   \parbox{22mm}{\begin{center}
   \begin{fmfgraph}(20,5)
   \setval
   \fmfforce{0w,0.5h}{v2}
   \fmfforce{1/4w,0.5h}{v3}
   \fmfforce{2/4w,0.5h}{v4}
   \fmfforce{3/4w,0.5h}{v5}
   \fmfforce{1w,0.5h}{v6}
   \fmf{fermion,right}{v2,v3,v2}
   \fmf{boson}{v3,v4}
   \fmf{fermion,right}{v4,v5,v4}
   \fmf{boson}{v5,v6}
   \fmfdot{v2,v3,v4,v5}
   \end{fmfgraph}
   \end{center}} \big)
   &=& 
   U \big(
   \parbox{12mm}{\begin{center}
   \begin{fmfgraph}(10,5)
   \setval
   \fmfforce{0w,0.5h}{v1}
   \fmfforce{1/2w,0.5h}{v2}
   \fmfforce{1w,0.5h}{v3}
   \fmf{fermion,right}{v1,v2,v1}
   \fmf{boson}{v2,v3}
   \fmfdot{v1,v2}
   \end{fmfgraph}
   \end{center}} \big) 
   U \big(
   \parbox{12mm}{\begin{center}
   \begin{fmfgraph}(10,5)
   \setval
   \fmfforce{0w,0.5h}{v1}
   \fmfforce{1/2w,0.5h}{v2}
   \fmfforce{1w,0.5h}{v3}
   \fmf{fermion,right}{v1,v2,v1}
   \fmf{boson}{v2,v3}
   \fmfdot{v1,v2}
   \end{fmfgraph}
   \end{center}} \big)
   \end{eqnarray*}
   and 
   \begin{eqnarray*}
   U\big(
     \parbox{22mm}{\begin{center}
     \begin{fmfgraph}(20,5)
     \setval
     \fmfforce{0w,0.5h}{v1}
     \fmfforce{1/4w,0.5h}{v2}
     \fmfforce{2/4w,0.5h}{v3}
     \fmfforce{3/4w,0.5h}{v4}
     \fmfforce{1w,0.5h}{v5}
     \fmf{plain}{v5,v4,v3,v2,v1}
     \fmf{boson,left}{v1,v2}
     \fmf{boson,left}{v3,v4}
     \fmfdot{v1,v2,v3,v4}
     \end{fmfgraph}
     \end{center}} \big)
   &=& 
   U \big(
     \parbox{12mm}{\begin{center}
     \begin{fmfgraph}(10,5)
     \setval
     \fmfforce{0w,0.5h}{v1}
     \fmfforce{1/2w,0.5h}{v2}
     \fmfforce{1w,0.5h}{v3}
     \fmf{plain}{v3,v2,v1}
     \fmf{boson,left}{v1,v2}
     \fmfdot{v1,v2}
     \end{fmfgraph}
     \end{center}} \big) 
   U \big(
     \parbox{12mm}{\begin{center}
     \begin{fmfgraph}(10,5)
     \setval
     \fmfforce{0w,0.5h}{v1}
     \fmfforce{1/2w,0.5h}{v2}
     \fmfforce{1w,0.5h}{v3}
     \fmf{plain}{v3,v2,v1}
     \fmf{boson,left}{v1,v2}
     \fmfdot{v1,v2}
     \end{fmfgraph}
     \end{center}} \big) . 
   \end{eqnarray*} 

   The 1PI Feynman diagrams are then the generators 
   of the algebra of all diagrams with the ``junction'' 
   product. 
   Since no other relationship between the diagrams is  
   forced by the amplitude map $U$, the algebra of 
   Feynman diagrams can also be seen as the free 
   algebra on the 1PI diagrams, that is, 
   \begin{eqnarray*}
           (\Fp, \mbox{junction}) &\cong& 
           (T(\Fp_{1PI}), \otimes) , \\ 
           (\Fe, \mbox{junction}) &\cong& 
           (T(\Fe_{1PI}), \otimes) . 
   \end{eqnarray*}

    
   %
   %

   \section{From Feynman graphs to planar binary trees}
   \label{junetree}

   In \cite{BrouderEPJC2}, \cite{BF} we proposed another way to describe 
   the coefficients $S_n(q)$ and $D_n(q)$ of Eqs. (\ref{S(q)}, \ref{D(q)}).  
   It is an intermediate step between the calculation of single Feynman diagrams
   and the value of the sums $S_n(q)$ and $D_n(q)$. 
   The idea is to represent the sum of ``similar'' Feynman diagrams 
   with a single symbol, namely a planar binary tree.
   The purpose is twofold: on one side we can make use of the rich 
   algebraic structure of the set of trees to have new informations 
   on the perturbative terms; on the other side 
   we can still confortably deal with the restricted sums 
   of Feynman diagrams. 
   For instance, when the fermions are heavy, it is convenient 
   to limit the calculation to a small number of fermion loops. 
   The main example is quenched QED, cf. \cite{BroadhurstQI}, 
   where the photon propagator is described by the Feynman diagrams 
   without vacuum polarization insertions. 

   The choice of planar binary trees is motivated by the fact, showed in
   \cite{BrouderEPJC1}, that the functional Schwinger equations 
   satisfied by the QED propagators have a very natural 
   perturbative solution in terms of such trees. 
   As we said, the expansion of the propagators 
   over trees gives more detailed information than the
   sums $S_n(q)$ and $D_n(q)$. This situation is analogue to 
   Jean-Louis Loday's arithmetree \cite{LodayATree}, 
   where each natural number $n$ is written as the sum 
   of all planar binary trees with $n$ internal vertices, 
   and operations are defined on trees, that are compatible 
   with the addition and the product of integers.

   Passing from Feynman graphs to trees has the disadvantage 
   of losing the intuition of each `physical interaction', 
   but not that of `order of interaction', since each leaf 
   of the tree detects the presence of a loop in the 
   corresponding graphs. 
   The main advantage of the tree representation of interactions is that
   overlapping divergencies do not appear isolated, and their
   renormalization is encoded by the same procedure used
   for simple divergencies. Moreover, the Ward identities
   are built into the tree representation, so their consequences
   are automatically taken into account. 

   We remark, once more, that the present work is intended 
   to deal only with the algebraic and combinatorical structure 
   which encodes renormalization. In no way it simplifies 
   the analytical computations of the amplitudes. 
   In particular, in \cite{BrouderEPJC1} and \cite{BF} 
   we showed that the tree-expansion method 
   has a peculiarity, if compared to the graph-expansion one. 
   Namely, the coefficients of the expansions satisfy an explicit 
   recursive formula on the order of the trees, in both cases before 
   and after the renormalization. However, as kindly remarked 
   to the authors by D.~Kreimer and D.~Broadhurst, the difficulty 
   arising in the analytical computation of each integral 
   described by Feynman graphs is only postponed to the 
   computation of the recursive terms described by trees. 
   Moreover, it should be added that the renormalization 
   in the tree representation
   does not produce a finite result for each tree,
   as explained in Ref.\cite{BF}. The tree representation
   is used only as a powerful method to generate 
   recursive equations which, when summed over all trees
   of a given order, will give a finite result.
   These recursive equations are required to 
   define the Hopf algebra of renormalization.

    
   \subsection{Correspondence between trees and graphs}

   Planar binary trees are connected planar graphs 
   with trivalent internal vertices and no cycle. 
   The external edges are called the leaves. 
   We draw them with one preferred leaf on the bottom, 
   that we call the root. 
   Let $Y$ denote the set of planar binary trees, 
   simply called ``trees''. 
   Trees are naturally graded by the number of internal vertices. 
   We denote by $|t|$ the number of internal vertices of the tree $t$, 
   and by $Y_n$ the set of trees $t$ with $|t|=n$. 
   Then, $Y = \bigcup_{n \geq 0} Y_n$, where $Y_0 = \{ \| \}$,
   $Y_1 = \{ \Y \}$, $Y_2 = \{ \deuxun, \deuxdeux \}$, etc.

   The correspondence between trees and Feynman's diagrams can 
   be chosen as two maps 
   \begin{eqnarray*} 
           \se : Y \longrightarrow \Fe , \\
           \sp : Y \longrightarrow \Fp ,  
   \end{eqnarray*}
   where $\Fe$ and $\Fp$ now denote the complex vector 
   spaces spanned by the sets of graphs, such that 
   their image exhausts all the 
   2-legs Feynman diagrams of QED at a fixed order 
   of interaction, that is 
   \begin{eqnarray*}
           \sum_{{t \in Y},\ {|t|=n}} \se(t) 
           = \sum_{{\Gamma \in \Fe},\ {|\Gamma|=n}} \Gamma, \\ 
           \sum_{{t \in Y},\ {|t|=n}} \sp(t) 
           = \sum_{{\Gamma \in \Fp},\ {|\Gamma|=n}} \Gamma. 
   \end{eqnarray*}
   Moreover, each graph appears once 
   and only once as an addend in $\se(t)$ or $\sp(t)$ 
   for some tree $t$. 
   In this way, if we extend linearly the map $U$ to 
   formal sums of graphs, the Feynman rules can be read 
   directly on trees, through the compositions 
   \begin{eqnarray}
   \label{phi=Usigma} 
           \varphi^i &=& U \circ \sigma^i, 
           \qquad\mbox{for $i = e, \gamma$}, 
   \end{eqnarray}  
   and the QED propagators can be developed as formal 
   series over the trees, with amplitudes $\varphi^i$.  

   To define the maps $\sigma^i$, for $i=e,\gamma$, 
   we follow some basic rules:
   \begin{itemize}
   \item 
   The orientation ``left to right'' in a Feynman diagram 
   corresponds to the orientation ``root to leaves'' in a tree. 
   \item 
   Electron propagators correspond to $/$-branches; 
   photon propagators correspond to 
   $\backslash$-branches. 
   \item 
   Electron loops correspond to $/$-leaves; 
   photon loops correspond to $\backslash$-leaves. 
   \item
   Subloops are loops of a subdiagram. 
   \end{itemize}  

   To make the next algorithms clear, we also 
   fix a precise definition of the electron and photon 
   loops we shall encounter:  
   \begin{itemize}
   \item 
   A photon loop in a 2-legs electron graph is a 2-legs 
   photon graph such that if we remove it we still have 
   a 2-legs electron graph. 
   For instance, in the electron graph 
   \begin{eqnarray*}
   && \parbox{42mm}
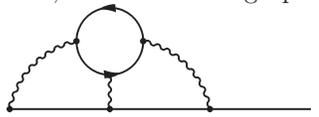
{\begin{center}
           \begin{fmfgraph}(40,15)
           \setval
           \fmfforce{0w,0.5h}{v1}
           \fmfforce{1/3w,0.5h}{v2}
           \fmfforce{2/3w,0.5h}{v3}
           \fmfforce{1w,0.5h}{v4}
           \fmfforce{2/9w,1.1h}{v5}
           \fmfforce{4/9w,1.1h}{v6}
           \fmfforce{1/3w,0.8h}{v7}
           \fmf{plain}{v4,v3,v2,v1}
           \fmf{boson}{v2,v7}
           \fmf{boson,left=0.3}{v1,v5}
           \fmf{boson,right=0.3}{v3,v6}
           \fmf{fermion,right}{v5,v6,v5}
           \fmfdot{v1,v2,v3,v5,v6,v7}
           \end{fmfgraph}
           \end{center}}, 
   \end{eqnarray*}
   if we remove 
   $\parbox{22mm}{\begin{center}
           \begin{fmfgraph}(20,8)
           \setval
           \fmfforce{0w,0.5h}{v1}
           \fmfforce{1w,0.5h}{v3}
           \fmfforce{1/3w,1.1h}{v5}
           \fmfforce{2/3w,1.1h}{v6}
           \fmf{boson,left=0.3}{v1,v5}
           \fmf{boson,right=0.3}{v3,v6}
           \fmf{fermion,right}{v5,v6,v5}
           \fmfdot{v1,v3,v5,v6}
           \end{fmfgraph}
           \end{center}}$, 
   we remain with 
   $\parbox{15mm}{\begin{center}
           \begin{fmfgraph}(12,8)
           \setval
           \fmfforce{0w,0.5h}{v1}
           \fmfforce{1/2w,0.5h}{v2}
           \fmfforce{1w,0.5h}{v3}
           \fmfforce{1/2w,1h}{v4}
           \fmf{plain}{v3,v2,v1}
           \fmf{boson}{v2,v4}
           \fmfdot{v1,v2}
           \end{fmfgraph}
           \end{center}}$, 
   which is not a 2-legs diagram. 
   Therefore, the only photon loop we can consider is the 
   free propagator 
   $\parbox{12mm}{\begin{center}
           \begin{fmfgraph}(10,5)
           \setval
           \fmfforce{0w,0.5h}{v2}
           \fmfforce{1w,0.5h}{v3}
           \fmf{boson}{v2,v3}
           \end{fmfgraph}
           \end{center}}$. 

   \item 
   An electron loop in a 2-legs photon graph is (the closure of) 
   a 2-legs electron graph such that if we remove it 
   we still have a 2-legs photon graph. 
   For instance, in the photon graph 
   \begin{eqnarray*}
   && \parbox{42mm}
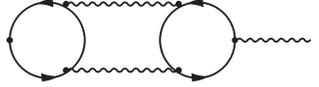
{\begin{center}
           \begin{fmfgraph}(40,8)
           \setval
           \fmfforce{0w,0.5h}{v1}
           \fmfforce{1/4w,0.5h}{v2}
           \fmfforce{2/4w,0.5h}{v3}
           \fmfforce{3/4w,0.5h}{v4}
           \fmfforce{1w,0.5h}{v5}
           \fmfforce{3/16w,1.1h}{v6}
           \fmfforce{9/16w,1.1h}{v7}
           \fmfforce{3/16w,0h}{v8}
           \fmfforce{9/16w,0h}{v9}
           \fmf{boson}{v4,v5}
           \fmf{fermion,right}{v2,v1,v2}
           \fmf{fermion,right}{v4,v3,v4}
           \fmf{boson}{v6,v7}
           \fmf{boson}{v8,v9}
           \fmfdot{v1,v4,v6,v7,v8,v9}
           \end{fmfgraph}
           \end{center}}, 
   \end{eqnarray*} 
   we can not remove 
   $\parbox{20mm}{\begin{center}
           \begin{fmfgraph}(18,7)
           \setval
           \fmfforce{0w,0.5h}{v1}
           \fmfforce{1/2w,0.5h}{v2}
           \fmfforce{3/8w,1.1h}{v6}
           \fmfforce{1w,1.1h}{v7}
           \fmf{fermion,right}{v2,v1,v2}
           \fmf{boson}{v6,v7}
           \fmfdot{v1,v6}
           \end{fmfgraph}
           \end{center}}$, 
   because it is the closure of a 3-legs diagram 
   $\parbox{15mm}{\begin{center}
           \begin{fmfgraph}(12,7)
           \setval
           \fmfforce{0w,0.5h}{v1}
           \fmfforce{1/2w,0.5h}{v2}
           \fmfforce{1w,0.5h}{v3}
           \fmfforce{1/2w,1h}{v4}
           \fmf{plain}{v3,v2,v1}
           \fmf{boson}{v2,v4}
           \fmfdot{v1,v2}
           \end{fmfgraph}
           \end{center}}$. 
   Therefore, the only electron loop we can consider is the 
   diagram 
   \begin{eqnarray*}
   && \parbox{32mm}
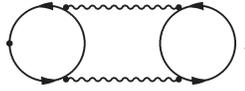
{\begin{center}
           \begin{fmfgraph}(30,12)
           \setval
           \fmfforce{0w,0.5h}{v1}
           \fmfforce{1/3w,0.5h}{v2}
           \fmfforce{2/3w,0.5h}{v3}
           \fmfforce{1w,0.5h}{v4}
           \fmfforce{3/12w,0.9h}{v6}
           \fmfforce{9/12w,0.9h}{v7}
           \fmfforce{3/12w,0.1h}{v8}
           \fmfforce{9/12w,0.1h}{v9}
           \fmf{fermion,right}{v2,v1,v2}
           \fmf{fermion,right}{v4,v3,v4}
           \fmf{boson}{v6,v7}
           \fmf{boson}{v8,v9}
           \fmfdot{v1,v6,v7,v8,v9}
           \end{fmfgraph}
           \end{center}}, 
   \end{eqnarray*} 
   which is the closure of the 2-legs diagram 
   \begin{eqnarray*}
   && \parbox{42mm}
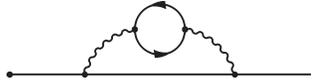
{\begin{center}
           \begin{fmfgraph}(40,12) 
           \setval
           \fmfforce{0w,0.5h}{v1}
           \fmfforce{1/4w,0.5h}{v2}
           \fmfforce{3/4w,0.5h}{v3}
           \fmfforce{1w,0.5h}{v4}
           \fmfforce{5/12w,1h}{v5}
           \fmfforce{7/12w,1h}{v6}
           \fmf{plain}{v4,v3,v2,v1}
           \fmf{boson,left=0.3}{v2,v5}
           \fmf{boson,right=0.3}{v3,v6}
           \fmf{fermion,right}{v6,v5,v6}
           \fmfdot{v1,v2,v3,v5,v6}
           \end{fmfgraph}
           \end{center}} . 
   \end{eqnarray*} 
   \end{itemize}  

   Starting from the Schwinger-Dyson equations for the 
   propagators $S(q;e_0)$ and $D(q;e_0)$, translated into 
   a recursive formula of perturbative amplitudes 
   in \cite{BrouderEPJC2}, 
   we are forced to define the maps $\se$ and $\sp$ 
   through a recursive algorithm, which is better 
   understood as the inverse of the following one. 

   \begin{definition}
   \label{defpi} 
   Define the maps $\pe:\Fe \longrightarrow Y$ and 
   $\pp:\Fp \longrightarrow Y$ recursively, starting 
   from the free propagators $\pe(1),\pp(1) = \|$. 
   For any 2-legs Feynman diagram $\Gamma$,  
   follow this algorithm. 
   If $\Gamma$ is an electron graph:  
   \begin{itemize}
   \item 
   Draw a $/$-leaf and read the graph from left to right. 
   \item
   Graft a $\backslash$-leaf at the bottom of the $/$-leaf for the first 
   photon loop encountered. 
   \item
   If the photon loop has subloops, draw a photon subtree 
   following the algorithm below for photon graphs. 
   \item 
   Return to the first vertex of the photon loop and 
   start the algorithm from the top of the rightmost $/$-leaf. 
   \item 
   Never read the second vertex of the photon loops. 
   The algorithm ends when all the first vertices of the photon 
   loops are met. At the end, draw the root. 
   \end{itemize}
   If $\Gamma$ is a photon graph:  
   \begin{itemize}
   \item 
   Draw a $\backslash$-leaf and read the graph 
   from left to right. 
   \item
   Graft a $/$-leaf at the bottom of the $\backslash$-leaf 
   for the first electron loop encoutered. 
   \item
   If the electron loop has subloops, draw an electron subtree 
   following the algorithm above for electron graphs. 
   \item 
   Return to the second vertex of the electron loop and 
   start the algorithm from the top of the leftmost 
   $\backslash$-leaf. 
   \item 
   The algorithm ends when all the electron loops are met. 
   At the end, draw the root. 
   \end{itemize}
   \end{definition}

   As an example, step by step, let us compute 
   $\pe(\parbox{12mm}{\begin{center}
           \begin{fmfgraph}(10,5)
           \setval
           \fmfforce{0w,0.5h}{v1}
           \fmfforce{1/2w,0.5h}{v2}
           \fmfforce{1w,0.5h}{v3}
           \fmf{plain}{v3,v2,v1}
           \fmf{boson,left}{v1,v2}
           \fmfdot{v1,v2}
           \end{fmfgraph}
           \end{center}})$, 
   knowing that 
   $\pp(\parbox{6mm}{\begin{center}
   \begin{fmfgraph}(5,5)
   \setval
   \fmfforce{0w,0.5h}{v2}
   \fmfforce{1w,0.5h}{v3}
   \fmf{boson}{v2,v3}
   \end{fmfgraph}
   \end{center}}) = \pp(1) = \|$: 
   \begin{eqnarray*}
   && \pe(\parbox{12mm}{\begin{center}
           \begin{fmfgraph}(10,5)
           \setval
           \fmfforce{0w,0.5h}{v1}
           \fmfforce{1/2w,0.5h}{v2}
           \fmfforce{1w,0.5h}{v3}
           \fmf{plain}{v3,v2,v1}
           \fmf{boson,left}{v1,v2}
           \fmfdot{v1,v2}
           \end{fmfgraph}
           \end{center}}) : 
   \qquad 
   {\setlength{\unitlength}{6pt}
   \psset{unit=6pt}
   \psset{runit=6pt}
   \psset{linewidth=0.1}
   \begin{pspicture}(0,0)(3,1)
   \psline(1,0)(2,1)
   \end{pspicture}, 
   \qquad 
   \begin{pspicture}(0,0)(3,1)
   \psline(1,0)(2,1)
   \psline(1,0)(0,1)
   \end{pspicture}, 
   \qquad 
   \begin{pspicture}(0,0)(3,1)
   \psline(1,-1)(1,0)
   \psline(1,0)(0,1)
   \psline(1,0)(2,1)
   \end{pspicture}}. 
   \end{eqnarray*}
   Similarly, let us compute 
   $\pp(\parbox{12mm}{\begin{center}
           \begin{fmfgraph}(10,5)
           \setval
           \fmfforce{0w,0.5h}{v1}
           \fmfforce{1/2w,0.5h}{v2}
           \fmfforce{1w,0.5h}{v3}
           \fmf{fermion,right}{v1,v2,v1}
           \fmf{boson}{v2,v3}
           \fmfdot{v1,v2}
           \end{fmfgraph}
           \end{center}})$, 
   knowing that 
   $\pe(\parbox{6mm}{\begin{center}
   \begin{fmfgraph}(5,5)
   \setval
   \fmfforce{0w,0.5h}{v2}
   \fmfforce{1w,0.5h}{v3}
   \fmf{plain}{v2,v3}
   \end{fmfgraph}
   \end{center}}) = \pe(1) = \|$: 
   \begin{eqnarray*}
   &&\pp(\parbox{12mm}{\begin{center}
           \begin{fmfgraph}(10,5)
           \setval
           \fmfforce{0w,0.5h}{v1}
           \fmfforce{1/2w,0.5h}{v2}
           \fmfforce{1w,0.5h}{v3}
           \fmf{fermion,right}{v1,v2,v1}
           \fmf{boson}{v2,v3}
           \fmfdot{v1,v2}
           \end{fmfgraph}
           \end{center}}): 
   \qquad 
   {\setlength{\unitlength}{6pt}
   \psset{unit=6pt}
   \psset{runit=6pt}
   \psset{linewidth=0.1}
   \begin{pspicture}(0,0)(3,1)
   \psline(1,0)(0,1)
   \end{pspicture}, 
   \qquad 
   \begin{pspicture}(0,0)(3,1)
   \psline(1,0)(0,1)
   \psline(1,0)(2,1)
   \end{pspicture} , 
   \qquad
   \begin{pspicture}(0,0)(3,1)
   \psline(1,-1)(1,0)
   \psline(1,0)(0,1)
   \psline(1,0)(2,1)
   \end{pspicture}}. 
   \end{eqnarray*}
   More complicated examples: 
   \begin{eqnarray*}
   &&\pe(\parbox{17mm}{\begin{center}
     \begin{fmfgraph}(15,5)
     \setval
     \fmfforce{0w,0.5h}{v1}
     \fmfforce{1/4w,0.5h}{v2}
     \fmfforce{2/4w,0.5h}{v3}
     \fmfforce{3/4w,0.5h}{v4}
     \fmfforce{1w,0.5h}{v5}
     \fmf{plain}{v5,v4,v3,v2,v1}
     \fmf{boson,left}{v1,v3}
     \fmf{boson,right}{v2,v4}
     \fmfdot{v1,v2,v3,v4}
     \end{fmfgraph}
     \end{center}}): 
   \qquad 
   {\setlength{\unitlength}{6pt}
   \psset{unit=6pt}
   \psset{runit=6pt}
   \psset{linewidth=0.1}
   \begin{pspicture}(1.5,0)(6,3)
   \psline(3,0)(4,1)
   \end{pspicture} , 
   \qquad
   \begin{pspicture}(1.5,0)(6,3)
   \psline(3,0)(4,1)
   \psline(3,0)(2,1)
   \end{pspicture} , 
   \qquad
   \begin{pspicture}(1.5,0)(6,3)
   \psline(3,0)(5,2)
   \psline(3,0)(2,1)
   \psline(4,1)(3,2)
   \end{pspicture} , 
   \qquad
   \begin{pspicture}(1.5,0)(6,3)
   \psline(3,-1)(3,0)
   \psline(3,0)(5,2)
   \psline(3,0)(2,1)
   \psline(4,1)(3,2)
   \end{pspicture}} ; 
   \\ 
   &&\pe(\parbox{17mm}{\begin{center}
     \begin{fmfgraph}(15,8)

     \setval
     \fmfforce{0w,0.5h}{v1}
     \fmfforce{1/4w,1h}{v2}
     \fmfforce{2/4w,1h}{v3}
     \fmfforce{3/4w,0.5h}{v4}
     \fmfforce{1w,0.5h}{v5}
     \fmf{plain}{v5,v4,v1}
     \fmf{boson,left=1/3}{v1,v2}
     \fmf{boson,left=1/3}{v3,v4}
     \fmf{fermion,right}{v2,v3,v2}
     \fmfdot{v1,v2,v3,v4}
     \end{fmfgraph}
     \end{center}}): 
   \qquad 
   {\setlength{\unitlength}{6pt}
   \psset{unit=6pt}
   \psset{runit=6pt}
   \psset{linewidth=0.1}
   \begin{pspicture}(0,0)(6,3)
   \psline(3,0)(4,1)
   \end{pspicture} , 
   \qquad 
   \begin{pspicture}(0,0)(6,3)
   \psline(3,0)(4,1)
   \psline(3,0)(1,2)
   \psline(2,1)(3,2)
   \end{pspicture} , 
   \qquad 
   \begin{pspicture}(0,0)(6,3)
   \psline(3,-1)(3,0)
   \psline(3,0)(1,2)
   \psline(3,0)(4,1)
   \psline(2,1)(3,2)
   \end{pspicture}}; 
   \\ 
   &&\pp(\parbox{17mm}{\begin{center}
   \begin{fmfgraph}(15,5)
   \setval
   \fmfforce{1/3w,0.5h}{v2}
   \fmfforce{1/2w,1h}{v2b}
   \fmfforce{2/3w,0.5h}{v3}
   \fmfforce{1/2w,0h}{v3b}
   \fmfforce{1w,0.5h}{v4}
   \fmf{fermion,right=0.45}{v2,v3b,v3,v2b,v2}
   \fmf{boson}{v3,v4}
   \fmf{boson}{v2b,v3b}
   \fmfdot{v2,v2b,v3,v3b}
   \end{fmfgraph}
           \end{center}}): 
   \qquad 
   {\setlength{\unitlength}{6pt}
   \psset{unit=6pt}
   \psset{runit=6pt}
   \psset{linewidth=0.1}
   \begin{pspicture}(1.5,0)(6,3)
   \psline(3,0)(2,1)
   \end{pspicture}  , 
   \qquad 
   \begin{pspicture}(1.5,0)(6,3)
   \psline(3,0)(5,2)
   \psline(3,0)(2,1)
   \psline(4,1)(3,2)
   \end{pspicture} , 
   \qquad 
   \begin{pspicture}(1.5,0)(6,3)
   \psline(3,-1)(3,0)
   \psline(3,0)(5,2)
   \psline(3,0)(2,1)
   \psline(4,1)(3,2)
   \end{pspicture}}; 
   \\ 
   &&\pp(\parbox{17mm}{\begin{center}
           \begin{fmfgraph}(15,5)
           \setval
           \fmfforce{0w,0.5h}{v2}
           \fmfforce{1/4w,0.5h}{v3}
           \fmfforce{2/4w,0.5h}{v4}
           \fmfforce{3/4w,0.5h}{v5}
           \fmfforce{1w,0.5h}{v6}
           \fmf{fermion,right}{v2,v3,v2}
           \fmf{boson}{v3,v4}
           \fmf{fermion,right}{v4,v5,v4}
           \fmf{boson}{v5,v6}
           \fmfdot{v2,v3,v4,v5}
           \end{fmfgraph}
           \end{center}}): 
   \qquad 
   {\setlength{\unitlength}{6pt}
   \psset{unit=6pt}
   \psset{runit=6pt}
   \psset{linewidth=0.1}
   \begin{pspicture}(0,0)(6,3)
   \psline(3,0)(2,1)
   \end{pspicture} , 
   \qquad 
   \begin{pspicture}(0,0)(6,3)
   \psline(3,0)(2,1)
   \psline(3,0)(4,1)
   \end{pspicture} , 
   \qquad 
   \begin{pspicture}(0,0)(6,3)
   \psline(3,0)(1,2)
   \psline(3,0)(4,1)
   \psline(2,1)(3,2)
   \end{pspicture} , 
   \qquad 
   \begin{pspicture}(0,0)(6,3)
   \psline(3,-1)(3,0)
   \psline(3,0)(1,2)
   \psline(3,0)(4,1)
   \psline(2,1)(3,2)
   \end{pspicture}}. 
   \end{eqnarray*}

   The maps $\pe$ and $\pp$ are not injective, 
   but they are surjective and ``finite-to-one''. 
   Therefore, they admit two linear sections $\se$ 
   and $\sp$ (left inverse maps) which can be described, 
   up to some scalar factors, as follows. 

   Let $\vee : Y_n \times Y_m \longrightarrow Y_{n+m+1}$ 
   be the map which grafts two trees on a new root. 
   Then, each tree $t$ except the root tree $\|$ is 
   the grafting of two trees with strictly smaller order, 
   \begin{eqnarray*} 
           t &=& t^l \vee t^r = \vee(t^l,t^r), \\
           |t| &=& |t^l|+|t^r|+1, 
   \end{eqnarray*} 
   the first one on the left and the second one on the 
   right-hand side of the root. 
   For instance, 
   \begin{eqnarray*}
           \troisdeux = \deuxdeux \vee \| , &\qquad& 
           \troistrois = \Y \vee \Y . 
   \end{eqnarray*}

   \begin{definition} 
   \label{defsigma} 
   Define the maps $\se:Y \longrightarrow \Fe$ and 
   $\sp:Y \longrightarrow \Fp$ recursively, starting 
   from the free propagators $\se(\|),\sp(\|) = 1$. 
   For any tree $t \neq \|$, proceed in two different 
   ways to build $\se(t)$ and $\sp(t)$. 
   For electron trees: 
   \begin{itemize}
   \item 
   add a new vertex and a free electron propagator on the 
   left-hand side of the electron graphs in the sum $\se(t_r)$, 
   graft a new free photon propagator on the first vertex; 
   \item 
   add a new vertex in the middle of each free electron 
   component, call the new sum $\Gamma$, 
   \item 
   graft the free photon propagator 
   of all the electron graphs of $\Gamma$ onto the first vertex 
   of all the photon graphs of $\sp(t_l)$, 
   and graft the last free photon propagator of $\sp(t_l)$ 
   to each free vertex of $\Gamma$, 
   the sum of the graphs is now $\se(t)$. 
   \end{itemize} 
   For photon trees: 
   \begin{itemize}
   \item 
   add a vertex and a free electron propagator on the 
   left-hand side of the electron graphs in the sum $\se(t_r)$; 
   \item 
   add a new vertex and a new free photon propagator 
   in the middle of each free electron component, 
   call the new sum $\Gamma$, 
   \item 
   close the graphs of $\Gamma$ in a loop, by grafting 
   the last free electron propagator to the first vertex, 
   and graft the new free photon propagator of the loop 
   onto the first vertex of the photon graphs of $\sp(t_l)$, 
   the sum of the graphs is now $\sp(t)$. 
   \end{itemize} 
   \end{definition} 

   For example, let us compute $\se(\Y)$, 
   knowing that $\se(\|)=1$ and $\sp(\|)=1$: 
   \begin{eqnarray*}
   && \Gamma = \parbox{22mm}{\begin{center}
           \begin{fmfgraph}(20,10)
           \setval
           \fmfforce{0w,1/2h}{v1}
           \fmfforce{1/2w,1/2h}{v2}
           \fmfforce{1w,1/2h}{v3}
           \fmfforce{0w,1h}{v4}
           \fmf{plain}{v3,v2,v1}
           \fmf{boson}{v1,v4}
           \fmfdot{v1,v2}
           \end{fmfgraph}
           \end{center}}, \qquad 
   \se(\Y) = \parbox{22mm}{\begin{center}
           \begin{fmfgraph}(20,5)
           \setval
           \fmfforce{0w,0.5h}{v1}
           \fmfforce{1/2w,0.5h}{v2}
           \fmfforce{1w,0.5h}{v3}
           \fmf{plain}{v3,v2,v1}
           \fmf{boson,left}{v1,v2}
           \fmfdot{v1,v2}
           \end{fmfgraph}
           \end{center}}. 
   \end{eqnarray*}  
   In a similar way compute $\sp(\Y)$:
   \begin{eqnarray*}
   && \Gamma = \parbox{22mm}{\begin{center}
           \begin{fmfgraph}(20,10)
           \setval
           \fmfforce{0w,1/2h}{v1}
           \fmfforce{1/2w,1/2h}{v2}
           \fmfforce{1w,1/2h}{v3}
           \fmfforce{1/2w,1h}{v4}
           \fmf{plain}{v3,v2,v1}
           \fmf{boson}{v2,v4}
           \fmfdot{v1,v2}
           \end{fmfgraph}
           \end{center}}, \qquad  
   \sp(\Y) = \parbox{25mm}{\begin{center}
           \begin{fmfgraph}(20,5)
           \setval
           \fmfforce{0w,0.5h}{v1}
           \fmfforce{1/2w,0.5h}{v2}
           \fmfforce{1w,0.5h}{v3}
           \fmf{fermion,right}{v1,v2,v1}
           \fmf{boson}{v2,v3}
           \fmfdot{v1,v2}
           \end{fmfgraph}
           \end{center}} . 
   \end{eqnarray*}  

   A more complicated example is the computation of 
   $\se(\troistrois)$: 
   \begin{eqnarray*}
   && \Gamma = \parbox{28mm}{\begin{center}
           \begin{fmfgraph}(25,5)
           \setval
           \fmfforce{0w,0.5h}{v1}
           \fmfforce{1/4w,0.5h}{v2}
           \fmfforce{2/4w,0.5h}{v3}
           \fmfforce{3/4w,0.5h}{v4}
           \fmfforce{1w,0.5h}{v5}
           \fmfforce{0w,1.5h}{v6}
           \fmf{plain}{v5,v4,v3,v2,v1}
           \fmf{boson,left}{v3,v4}
           \fmf{boson}{v1,v6}
           \fmfdot{v1,v2,v3,v4}
           \end{fmfgraph}
           \end{center}} 
   + \parbox{28mm}{\begin{center}
           \begin{fmfgraph}(25,5)
           \setval
           \fmfforce{0w,0.5h}{v1}
           \fmfforce{1/4w,0.5h}{v2}
           \fmfforce{2/4w,0.5h}{v3}
           \fmfforce{3/4w,0.5h}{v4}
           \fmfforce{1w,0.5h}{v5}
           \fmfforce{0w,1.5h}{v6}
           \fmf{plain}{v5,v4,v3,v2,v1}
           \fmf{boson,left}{v2,v4}
           \fmf{boson}{v1,v6}
           \fmfdot{v1,v2,v3,v4}
           \end{fmfgraph}
           \end{center}} 
   + \parbox{28mm}{\begin{center}
           \begin{fmfgraph}(25,5)
           \setval
           \fmfforce{0w,0.5h}{v1}
           \fmfforce{1/4w,0.5h}{v2}
           \fmfforce{2/4w,0.5h}{v3}
           \fmfforce{3/4w,0.5h}{v4}
           \fmfforce{1w,0.5h}{v5}
           \fmfforce{0w,1.5h}{v6}
           \fmf{plain}{v5,v4,v3,v2,v1}
           \fmf{boson,left}{v2,v3}
           \fmf{boson}{v1,v6}
           \fmfdot{v1,v2,v3,v4}
           \end{fmfgraph}
           \end{center}} , \\ 
   && \se(\troistrois) = \parbox{42mm}{\begin{center}
           \begin{fmfgraph}(40,8)
           \setval
           \fmfforce{0w,0.5h}{v1}
           \fmfforce{1/4w,0.5h}{v2}
           \fmfforce{2/4w,0.5h}{v3}
           \fmfforce{3/4w,0.5h}{v4}
           \fmfforce{1w,0.5h}{v5}
           \fmfforce{1/16w,1h}{v6}
           \fmfforce{3/16w,1h}{v7}
           \fmf{plain}{v5,v4,v3,v2,v1}
           \fmf{boson,left}{v3,v4}
           \fmf{boson,left=1/3}{v1,v6}
           \fmf{boson,right=1/3}{v2,v7}
           \fmf{fermion,right}{v6,v7,v6}
           \fmfdot{v1,v2,v3,v4,v6,v7}
           \end{fmfgraph}
           \end{center}} 
   + \parbox{42mm}{\begin{center}
           \begin{fmfgraph}(40,5)
           \setval
           \fmfforce{0w,0.5h}{v1}
           \fmfforce{1/4w,0.5h}{v2}
           \fmfforce{2/4w,0.5h}{v3}
           \fmfforce{3/4w,0.5h}{v4}
           \fmfforce{1w,0.5h}{v5}
           \fmfforce{1/6w,-1h}{v6}
           \fmfforce{2/6w,-1h}{v7}
           \fmf{plain}{v5,v4,v3,v2,v1}
           \fmf{boson,left}{v2,v4}
           \fmf{boson,right=0.3}{v1,v6}
           \fmf{boson,left=0.3}{v3,v7}
           \fmf{fermion,right}{v6,v7,v6}
           \fmfdot{v1,v2,v3,v4,v6,v7}
           \end{fmfgraph}
           \end{center}} 
   + \parbox{42mm}{\begin{center}
           \begin{fmfgraph}(40,8)
           \setval
           \fmfforce{0w,0.5h}{v1}
           \fmfforce{1/4w,0.5h}{v2}
           \fmfforce{2/4w,0.5h}{v3}
           \fmfforce{3/4w,0.5h}{v4}
           \fmfforce{1w,0.5h}{v5}
           \fmfforce{1/4w,-0.6h}{v6}
           \fmfforce{2/4w,-0.6h}{v7}
           \fmf{plain}{v5,v4,v3,v2,v1}
           \fmf{boson,left}{v2,v3}
           \fmf{boson,right=0.3}{v1,v6}
           \fmf{boson,left=0.3}{v4,v7}
           \fmf{fermion,right}{v6,v7,v6}
           \fmfdot{v1,v2,v3,v4,v6,v7}
           \end{fmfgraph}
           \end{center}}. 
   \end{eqnarray*}
   Similarly, here is the computation of $\sp(\troistrois)$:
   \begin{eqnarray*}
   && \Gamma = \parbox{28mm}{\begin{center}
           \begin{fmfgraph}(25,5)
           \setval
           \fmfforce{0w,0.5h}{v1}
           \fmfforce{1/4w,0.5h}{v2}
           \fmfforce{2/4w,0.5h}{v3}
           \fmfforce{3/4w,0.5h}{v4}
           \fmfforce{1w,0.5h}{v5}
           \fmfforce{1/4w,1.5h}{v6}
           \fmf{plain}{v5,v4,v3,v2,v1}
           \fmf{boson,left}{v3,v4}
           \fmf{boson}{v2,v6}
           \fmfdot{v1,v2,v3,v4}
           \end{fmfgraph}
           \end{center}} 
   + \parbox{28mm}{\begin{center}
           \begin{fmfgraph}(25,5)
           \setval
           \fmfforce{0w,0.5h}{v1}
           \fmfforce{1/4w,0.5h}{v2}
           \fmfforce{2/4w,0.5h}{v3}
           \fmfforce{3/4w,0.5h}{v4}
           \fmfforce{1w,0.5h}{v5}
           \fmfforce{1/2w,-0.5h}{v6}
           \fmf{plain}{v5,v4,v3,v2,v1}
           \fmf{boson,left}{v2,v4}
           \fmf{boson}{v3,v6}
           \fmfdot{v1,v2,v3,v4}
           \end{fmfgraph}
           \end{center}} 
   + \parbox{28mm}{\begin{center}
           \begin{fmfgraph}(25,5)
           \setval
           \fmfforce{0w,0.5h}{v1}
           \fmfforce{1/4w,0.5h}{v2}
           \fmfforce{2/4w,0.5h}{v3}
           \fmfforce{3/4w,0.5h}{v4}
           \fmfforce{1w,0.5h}{v5}
           \fmfforce{3/4w,1.5h}{v6}
           \fmf{plain}{v5,v4,v3,v2,v1}
           \fmf{boson,left}{v2,v3}
           \fmf{boson}{v4,v6}
           \fmfdot{v1,v2,v3,v4}
           \end{fmfgraph}
           \end{center}} , \\ 
   && \sp \big( \troistrois \big) = 
           \parbox{32mm}{\begin{center}
           \begin{fmfgraph}(30,5) 
           \setval
           \fmfforce{0w,1/2h}{v1}
           \fmfforce{1/4w,1/2h}{v2}
           \fmfforce{1/16w,-0.1h}{v3}
           \fmfforce{3/16w,-0.1h}{v4}
           \fmfforce{2/4w,1/2h}{v5}
           \fmfforce{3/4w,1/2h}{v6}
           \fmfforce{1w,1/2h}{v7}
           \fmf{fermion,left}{v1,v2,v1}
           \fmf{boson,left=0.8}{v3,v4}
           \fmf{boson}{v2,v5}
           \fmf{fermion,right}{v5,v6,v5}
           \fmf{boson}{v6,v7}
           \fmfdot{v1,v2,v3,v4,v5,v6}
           \end{fmfgraph}
           \end{center}} 
   + \parbox{32mm}{\begin{center}
           \begin{fmfgraph}(30,5) 
           \setval
           \fmfforce{0w,1/2h}{v1}
           \fmfforce{1/4w,1/2h}{v2}
           \fmfforce{1/8w,1.2h}{v3}
           \fmfforce{1/8w,-0.2h}{v4}
           \fmfforce{2/4w,1/2h}{v5}
           \fmfforce{3/4w,1/2h}{v6}
           \fmfforce{1w,1/2h}{v7}
           \fmf{fermion,right=0.4}{v1,v4,v2,v3,v1}
           \fmf{boson}{v3,v4}
           \fmf{boson}{v2,v5}
           \fmf{fermion,right}{v5,v6,v5}
           \fmf{boson}{v6,v7}
           \fmfdot{v1,v2,v3,v4,v5,v6}
           \end{fmfgraph}
           \end{center}}
   + \parbox{32mm}{\begin{center}
           \begin{fmfgraph}(30,5) 
           \setval
           \fmfforce{0w,1/2h}{v1}
           \fmfforce{1/4w,1/2h}{v2}
           \fmfforce{1/16w,1.1h}{v3}
           \fmfforce{3/16w,1.1h}{v4}
           \fmfforce{2/4w,1/2h}{v5}
           \fmfforce{3/4w,1/2h}{v6}
           \fmfforce{1w,1/2h}{v7}
           \fmf{fermion,left}{v1,v2,v1}
           \fmf{boson,right=0.8}{v3,v4}
           \fmf{boson}{v2,v5}
           \fmf{fermion,right}{v5,v6,v5}
           \fmf{boson}{v6,v7}
           \fmfdot{v1,v2,v3,v4,v5,v6}
           \end{fmfgraph}
           \end{center}}. 
   \end{eqnarray*}
   Other examples of the maps $\se$ and $\sp$ 
   are given in appendix 3 of Ref.\cite{BF}. 


   In conclusion, using the amplitudes $\phie,\phip$ 
   defined by Eq.~(\ref{phi=Usigma}), 
   we can consider the coefficients of the 
   perturbative expansion of the Green functions as finite sums 
   labelled by planar binary trees $t$, 
   \begin{eqnarray} 
           S_n(q) &=& \sum_{|t|=n} \phie(t;q)  \label{S(t;q)} \\ 
           D_n(q) &=& \sum_{|t|=n} \phip(t;q) ,  \label{D(t;q)} 
   \end{eqnarray}
   where the order of interaction is now given by the number $|t|$ 
   of internal vertices of the tree $t$. 

   After the renormalization of each finite sum of integrals contained in 
   $\phie(t;q)$ and $\phip(t;q)$, we denote the finite parts by 
   $\barphie(t;q)$ and $\barphip(t;q)$. The coefficients
   $\bar{S}_n(q)$ and $\bar{D}_n(q)$ of the renormalized propagators 
   are then given by 
   \begin{eqnarray} 
           \bar{S}_n(q) &=& \sum_{|t|=n} \barphie(t;q),  \label{barS(t;q)} \\ 
           \bar{D}_n(q) &=& \sum_{|t|=n} \barphip(t;q).  \label{barD(t;q)}   
   \end{eqnarray}

   The relationship between the coefficients of the tree-expansion 
   before and after the massless renormalization was found in \cite{BF} 
   by means of a Hopf algebra structure on the set of trees. 
   We recall it at the end of section \ref{junehopf}, where we introduce
   the necessary notations. 

   In the next two sections we define a multiplicative structure 
   on the set of trees which is compatible with the (noncommutative) 
   junction of adjacent 2-legs Feynman graphs of QED. 


   \subsection{Algebra of photon trees and vacuum polarization}

   For photon trees, the Schwinger-Dyson equation for $D(q)$ provides
   the relation $\varphi^\gamma(t^l\vee t^r)=\varphi^\gamma(\|\vee t^r)
   \varphi^\gamma(t^l)$ for the amplitudes (see Refs.\cite{BrouderEPJC2,BF}).
   For example, applying the definition (\ref{defpi}) of the map $\sp$, we find 
   \begin{eqnarray*} 
   \phip(\Y;q) &=& U \big( 
   \parbox{12mm}{\begin{center}
   \begin{fmfgraph}(10,5)
   \setval
   \fmfforce{0w,0.5h}{v2}
   \fmfforce{1/2w,0.5h}{v3}
   \fmfforce{1w,0.5h}{v4}
   \fmf{fermion,right}{v2,v3,v2}
   \fmf{boson}{v3,v4}
   \fmfdot{v2,v3}
   \end{fmfgraph}
   \end{center}} \big) 
   \\ 
   \phip(\deuxdeux;q) &=& U \big(
   \parbox{18mm}{\begin{center}
   \begin{fmfgraph}(15,5)
   \setval
   \fmfforce{1/3w,0.5h}{v2}
   \fmfforce{1/2w,1h}{v2b}
   \fmfforce{2/3w,0.5h}{v3}
   \fmfforce{1/2w,0h}{v3b}
   \fmfforce{1w,0.5h}{v4}
   \fmf{fermion,right=0.45}{v2,v3b,v3,v2b,v2}
   \fmf{boson}{v3,v4}
   \fmf{boson}{v2b,v3b}
   \fmfdot{v2,v2b,v3,v3b}
   \end{fmfgraph}
   \end{center}}+
   \parbox{18mm}{\begin{center}
   \begin{fmfgraph}(15,5)
   \setval
   \fmfforce{1/3w,0.5h}{v2}
   \fmfforce{0.4008w,0.933h}{v2b}
   \fmfforce{0.5992w,0.933h}{v2c}
   \fmfforce{2/3w,0.5h}{v3}
   \fmfforce{1w,0.5h}{v4}
   \fmf{fermion,right=0.33}{v3,v2c,v2b,v2}
   \fmf{fermion,right=1}{v2,v3}
   \fmf{boson,right=0.8}{v2b,v2c}
   \fmf{boson}{v3,v4}
   \fmfdot{v2,v2b,v2c,v3}
   \end{fmfgraph}
   \end{center}}
   +
   \parbox{18mm}{\begin{center}
   \begin{fmfgraph}(15,5)
   \setval
   \fmfforce{1/3w,0.5h}{v2}
   \fmfforce{2/3w,0.5h}{v3}
   \fmfforce{0.4008w,0.067h}{v3b}
   \fmfforce{0.5992w,0.067h}{v3c}
   \fmfforce{1w,0.5h}{v4}
   \fmf{fermion,right=1}{v3,v2}
   \fmf{fermion,right=0.33}{v2,v3b,v3c,v3}
   \fmf{boson,left=0.8}{v3b,v3c}
   \fmf{boson}{v3,v4}
   \fmfdot{v2,v3b,v3c,v3}
   \end{fmfgraph}
   \end{center}} \big) 
   \\ 
   \phip(\troisdeux;q) &=& U \big(
   \parbox{33mm}{\begin{center}
   \begin{fmfgraph}(28,8)
   \setval
   \fmfforce{5/28w,0.5h}{v2}
   \fmfforce{10/28w,0.5h}{v3}
   \fmfforce{15/28w,0.5h}{v4}
   \fmfforce{19/28w,1h}{v5}
   \fmfforce{19/28w,0h}{v6}
   \fmfforce{23/28w,0.5h}{v7}
   \fmfforce{1w,0.5h}{v8}
   \fmf{fermion,right}{v2,v3,v2}
   \fmf{boson}{v3,v4}
   \fmf{fermion,right}{v4,v7,v4}
   \fmf{boson}{v5,v6}
   \fmf{boson}{v7,v8}
   \fmfdot{v2,v3,v4,v5,v6,v7}
   \end{fmfgraph}
   \end{center}}
   +
   \parbox{33mm}{\begin{center}
   \begin{fmfgraph}(28,8)
   \setval
   \fmfforce{5/28w,0.5h}{v2}
   \fmfforce{10/28w,0.5h}{v3}
   \fmfforce{15/28w,0.5h}{v4}
   \fmfforce{.55485w,0.75h}{v5}
   \fmfforce{.80229w,0.75h}{v6}
   \fmfforce{23/28w,0.5h}{v7}
   \fmfforce{1w,0.5h}{v8}
   \fmf{fermion,right}{v2,v3,v2}
   \fmf{boson}{v3,v4}
   \fmf{fermion,right}{v4,v7,v4}
   \fmf{boson}{v5,v6}
   \fmf{boson}{v7,v8}
   \fmfdot{v2,v3,v4,v5,v6,v7}
   \end{fmfgraph}
   \end{center}}
   +
   \parbox{33mm}{\begin{center}
   \begin{fmfgraph}(28,8)
   \setval
   \fmfforce{5/28w,0.5h}{v2}
   \fmfforce{10/28w,0.5h}{v3}
   \fmfforce{15/28w,0.5h}{v4}
   \fmfforce{.55485w,0.25h}{v5}
   \fmfforce{.80229w,0.25h}{v6}
   \fmfforce{23/28w,0.5h}{v7}
   \fmfforce{1w,0.5h}{v8}
   \fmf{fermion,right}{v2,v3,v2}
   \fmf{boson}{v3,v4}
   \fmf{fermion,right}{v4,v7,v4}
   \fmf{boson}{v5,v6}
   \fmf{boson}{v7,v8}
   \fmfdot{v2,v3,v4,v5,v6,v7}
   \end{fmfgraph}
   \end{center}} \big)
   \end{eqnarray*}
   so we see that 
   $\phip(\troisdeux;q) = \phip(\Y;q)\phip(\deuxdeux;q)$. 
   We transform this relation between amplitudes
   $\phip(t;q)$ into a relation between trees.

   \begin{definition} 
   \label{algebraHp}
   Denote by $\Hp=Y$ the graded vector space generated by $Y$  
   (with $\Hp_n = Y_n$), equipped with the graded product
   $\pro:\Hp_n \otimes \Hp_m \longrightarrow \Hp_{n+m}$\footnote{
   The product taken in $\Hp$ is the opposite of the operation {\sl over} 
   defined by Loday and Ronco in \cite{LodayRonco2}.}, 
   defined by the recurrence relation
   \begin{eqnarray*}
           s \pro t &:=& (s^l \pro t) \vee s^r 
           \quad\mathrm{for}\quad s=s^l\vee s^r, \\
           \| \pro t &:=& t. 
   \end{eqnarray*}
   This product consists in grafting the tree $t$ 
   onto the leftmost leaf of tree $s$. Therefore, it is associative and 
   has a unit given by the root tree $\|$.  
   In particular, $(1\vee s) \pro t = t\vee s$, which is the relation
   we wanted to implement.
   Examples of products are 
   $$
   \begin{array}{c}
           \Y \pro \Y = \deuxun , \\ 
           \Y \pro \deuxun = \troisun = \deuxun \pro \Y = \Y\pro\Y\pro\Y , \\ 
           \Y \pro \deuxdeux = \troisdeux ,\quad 
           \deuxdeux \pro \Y = \troistrois. 
   \end{array}
   $$
   \end{definition}

   The $\pro$ product is defined in such a way that the following 
   property automatically holds: 

   \begin{lemma}
   The map $\sp: \Hp \longrightarrow \Fp$ is an algebra morphism, 
   that is, the $\pro$ product between photon trees is transformed 
   into the junction of adjacent Feynman graphs. 
   Therefore, the amplitude map $\phip$ (and $\barphip$) is also 
   an algebra morphism on trees. 
   \end{lemma} 

   Denote by $V: Y_n \longrightarrow Y_{n+1}$ the left grafting by the root, 
   \begin{eqnarray}
   \label{defV(t)}
           V(t) = \| \vee t. 
   \end{eqnarray} 
   Then, any tree $t$ can be decomposed as 
   \begin{eqnarray*}
           t = t^l\vee t^r = V(t^r) \pro t^l , 
   \end{eqnarray*}
   where the left tree $t^l$ can be further decomposed until we reach a 
   root tree $\|$ on the left.  
   Therefore, the elements $V(t)$, for all $t \in Y$, form a system 
   of generators for the algebra $\Hp$\footnote{ 
   In fact, no other relation than associativity holds 
   between any product of generators, hence $\Hp$ is isomorphic to 
   the free associative unital algebra $\C\langle Y\rangle$ generated by the set 
   of trees, where the unit $1 \in \C$ 
   corresponds to the root tree $\|$.}.   

   \begin{proposition}
   Under the map $\sp$, the generators $V(t)$ of the algebra $\Hp$
   correspond to sums of 1PI 2-legs Feynman diagrams for the photon propagator. 

   Therefore, the bare and renormalized vacuum polarizations are 
   expanded only on the generator trees as
   \begin{eqnarray}
   \Pi(q;e_0) &=& \sum_{t} e_0^{2|t|+2} \phip(V(t);q), 
           \label{defPibare}\\ 
   \bar\Pi(q;e) &=& \sum_{t} e^{2|t|+2} \barphip(V(t);q). 
           \label{defPiren}
   \end{eqnarray}
   \end{proposition}

   This fact is evident from the algorithm defining 
   $\sp$ in (\ref{defsigma}). 

   \begin{lemma}
   \label{Inversephoton}
   Any formal series in $\Hp[[x]]$ starting with $\|$ 
   can be inverted, w.r.t. the $\pro$ product given in (\ref{algebraHp}). 
   Moreover the inverse series is developped only on generator trees, 
   \begin{eqnarray} 
   \big(\|+\sum_{|t|>0} x^{|t|} t\big)^{-1} &=& \|-\sum_{t} x^{|t|+1}
   V(t).
   \label{inversephoton}
   \end{eqnarray} 
   \end{lemma}

   \begin{proof}
   Since $\phip$ is an algebra homomorphism, 
   formula (\ref{inversephoton}) could be deduced 
   from the equation (\ref{vacuumpol}). 
   But we give here a direct proof, based 
   on the typical property of trees, cf. \cite{LodayRonco,Frabetti}, 
   \begin{eqnarray}
           \sum_{|t|=n} t = 
           \sum_{k=0}^{n-1}\sum_{|t^l|=k}\sum_{|t^r|=n-k-1} t^l\vee t^r,
   \label{property}
   \end{eqnarray}
   valid for $n>0$. We obtain
   \begin{eqnarray*}
   \|+\sum_{|t|>0} x^{|t|} t &=& \| + \sum_{t^l}\sum_{t^r} 
     x^{|t^l|+|t^r|+1} t^l\vee t^r = \| + \sum_{t^l}\sum_{t^r} 
     x^{|t^l|+|t^r|+1} V(t^r)\pro t^l\\
   &=& \| + \big(\sum_{t} x^{|t|+1} V(t)\big)\pro
       \big(\|+\sum_{|t|>0} x^{|t|} t\big).
   \end{eqnarray*}
   Hence,
   \begin{eqnarray*}
   \big(\|-\sum_{t} x^{|t|+1} V(t)\big)\pro
   \big(\|+\sum_{|t|>0} x^{|t|} t\big)=\|. 
   \end{eqnarray*} 
   \end{proof}

    
   \subsection{Algebra of electron trees and self-energy}

   The algebra of trees for electrons is a little more subtle,
   because no particular subset of trees corresponds to 
   1PI graphs under the map $\se$, and no internal product 
   analogue to $\pro$ is forced from the Schwinger-Dyson equations. 
   For instance, if $\phie(\Y)$ is a contribution
   to the electron propagator, $\phie(\Y)\phie(\Y)$ cannot
   be represented as $\phie(t)$ for a tree $t$.
   Therefore, the product is not given by an internal low, 
   and we have to consider the free (tensor) product. 

   If we want the map $\se$ to be an algebra morphism, 
   however, the set of generators for the electron algebra
   must correspond to 1PI electron graphs. 
   Therefore, we start by describing the 1PI 2-legs 
   Feynman diagrams for the electron propagator, using the 
   pruning operator $P$ introduced in \cite{BrouderEPJC2}. 
   We recall its definition and main properties. 

   \begin{definition} 
   Define an operator $\DP: Y \longrightarrow Y \otimes Y$ 
   recursively as 
   \begin{eqnarray}
   \DP \| &=& \| \otimes \|, \nonumber \\
   \DP (s\vee t) &=& \| \otimes (s\vee t) + 
   \big( \vee \otimes \Id \big) (s \otimes \DP t), 
   \end{eqnarray}
   where $\vee(s,t)=s\vee t$.
   For example: 
   \begin{eqnarray*} 
           \DP \Y &=& \| \otimes \Y + \Y \otimes \|, \\ 
           \DP \deuxdeux &=& \| \otimes \deuxdeux + \Y \otimes \Y + 
           \deuxdeux \otimes \|, \\ 
           \DP \deuxun &=& \| \otimes \deuxun + \deuxun \otimes \|.
   \end{eqnarray*}  
   The pruning operator of \cite{BrouderEPJC2} is then 
   $P(t) = \DP(t) - t \otimes \| - \| \otimes t$. 
   In ref.~\cite{BF}, it was shown that $\DP$ is coassociative. 
   Moreover, the coproduct $\DP$ preserves the order of the trees, 
   in fact $\DP(Y_n) \subset \bigoplus_{p+q=n} Y_p \otimes Y_q$. 
   Therefore, if we extend it multiplicatively on tensor product 
   of trees, it is easy to see that the operator $\DP$ defines 
   a graded Hopf algebra over $\C\langle Y \rangle/(\|-1)$, 
   with respect to the grading given by the total order of products 
   of trees, that is 
   \begin{eqnarray*}
           |t_1 \otimes\cdots\otimes t_k| = |t_1|+\cdots+|t_k|. 
   \end{eqnarray*}
   The counit is the algebra morphism defined on generators 
   as $\epsilon(t)=0$ for all $t \neq \|$ and $\epsilon(\|)=1$, 
   and the pruning antipode $\SP$ is the (graded) algebra anti-morphism 
   automatically defined on generators by the recursive formula $\SP(\|)=\|$ and
   \begin{eqnarray*}
           \SP(t) &=& -t -\sum_{P(t)}  \SP(t_{(1)}) t_{(2)} 
           = -t -\sum_{P(t)}  t_{(1)} \SP(t_{(2)}),  
   \end{eqnarray*}
   where we intentionally omit the tensor product symbol between trees. 
   We give a few examples: 
   \begin{eqnarray*}
           \SP(\Y) &=& -\Y , \\ 
           \SP(\deuxdeux) &=& -\deuxdeux+\Y^2, \\ 
           \SP(\deuxun) &=& -\deuxun, \\ 
           \SP(\troiscinq) &=& -\troiscinq + \deuxdeux \Y
           + \Y \deuxdeux - \Y^3, \\ 
           \SP(\troisquatre) &=& - \troisquatre + \Y \deuxun . 
   \end{eqnarray*}
   Notice that the coproduct $\DP$ is neither commutative 
   nor cocommutative, and $\SP^2\not= \Id$.
   \end{definition}

   \begin{remark}
   An alternative way to define $\DP$ was pointed out to
   us by J.-L.~Loday. In Ref.\cite{LodayRonco2}, the
   operation {\sl under} was defined recursively by
   $\| \backslash t = t$ and
   $s \backslash t = s^l \vee (s^r\backslash t)$ for
   $s=s^l\vee s^r$. Then the coproduct $\DP$ is 
   the dual operation to the product {\sl under}, 
   in the sense that 
   \begin{eqnarray*}
           \DP t &=& t \otimes \| + \| \otimes t + 
           \sum_{s_1\backslash s_2=t} s_1\otimes s_2,
   \end{eqnarray*}
   with the convention that the sum is over all
   $s_1\not= \|$, $s_2\not=\|$ such that $s_1\backslash s_2=t$. 
   If $t$ cannot be written as $s_1\backslash s_2$ for $s_1\not= \|$ 
   and $s_2\not=\|$, then $\DP t = t \otimes \| + \| \otimes t$.
   \end{remark}

   \begin{lemma}
   \label{Inverseelectron}
   Any formal series in $\C\langle Y \rangle/(\|-1)[[x]]$ 
   starting with $\|$ can be inverted, w.r.t. the tensor product. 
   Moreover the inverse series is developped only on 
   trees of the form $\SP(t)$, with $t \in Y$, that is, 
   \begin{eqnarray}
   \label{inverseelectron}
           {\big( \| + \sum_{|t|>0} x^{|t|} t \big)}^{-1} &=&
           \| + \sum_{|t|>0} x^{|t|} \SP(t).
   \end{eqnarray} 
   \end{lemma}

   \begin{proof}
   The main property of $\DP$ was noticed in 
   Ref.\cite{BrouderEPJC2}:
   \begin{eqnarray}
           \big( \| + \sum_{|t|>0} x^{|t|} t \big) \otimes
           \big( \| + \sum_{|t|>0} x^{|t|} t \big) 
           &=& \|\otimes\| +\sum_{|t|>0} x^{|t|} 
           \big( t \otimes \| + \| \otimes t +
           \sum_{P(t)}  t_{(1)} \otimes t_{(2)} \big) \nonumber \\ 
           &=& \|\otimes\| + \sum_{|t|>0} x^{|t|} \DP t .
           \label{mainproDP}
   \end{eqnarray}
   From Eq.~(\ref{mainproDP}) and the definition of 
   the antipode we obtain
   \begin{eqnarray*}
           \big( \| + \sum_{|t|>0} x^{|t|} \SP(t) \big) 
           \big( \| + \sum_{|t|>0} x^{|t|} t \big) &=&
           \|+ \sum_{|t|>0} x^{|t|} \big( \SP(t) + t +
           \sum_{s_1\backslash s_2=t} \SP(s_1)  s_2\big) = \|.
   \end{eqnarray*}
   \end{proof} 

   By definition (\ref{selfen}) of the self-energy, 
   and knowing that it should be expanded on 1PI Feynman graphs, 
   we obtain: 

   \begin{proposition}
   The bare and renormalized self-energy are expanded as 
   \begin{eqnarray*}
           \Sigma(q;e_0) &=& 
           - \sum_{|t|>0} e_0^{2|t|} \phie(\SP(t);q) , 
           \label{defSigmabare}\\
           \bar\Sigma(q;e) &=& 
           - \sum_{|t|>0} e^{2|t|} \barphie(\SP(t);q) . 
           \label{defSigmaren}\\
   \end{eqnarray*}

   Therefore, if we extend multiplicatively the map 
   $\se:Y \longrightarrow \Fe$ to tensor powers of $Y$, 
   the elements $\SP(t) \in \C\langle Y \rangle/(\|-1)$ 
   correspond to sums of 1PI 2-legs Feynman diagrams 
   for the electron propagator. 
   \end{proposition}

   Now we know which must be the generators for the 
   electron algebra. 

   \begin{definition}
   \label{algebraHe}
   We define $\He=\C\langle \SP(Y) \rangle /(1- \| )$ as the free 
   associative algebra on the set of trees transformed by $\SP$, 
   where we identify the formal unit $1$ with the root tree $\|$.
   For notational convenience, we omit the tensor product 
   symbols in the algebra $\He$. 

   The vector space of planar binary trees $Y$ can be seen as 
   a subspace of $\He$, because by definition 
   of the antipode $\SP$ each tree $t \in Y$ can be seen 
   as a (noncommutative) polynomial 
   \begin{eqnarray*}
           t &=& -\SP(t) - \sum_{P(t)} \SP(t_{(1)}) t_{(2)},  
   \end{eqnarray*}
   where $t_{(2)}$ can be further developped as a 
   polynomial in variables $\SP(t')$. 
   Therefore the two algebras $\C\langle Y\rangle/(\|-1)$ 
   and $\C\langle \SP(Y)\rangle/(\|-1) = \He$
   are isomorphic and the map $\se:Y \longrightarrow \Fe$ 
   can be extended as an algebra morphism to the space $\He$. 
   \end{definition}

   \begin{remark}
   \label{gradingHe}
   Beside the natural grading coming from the tensor powers, 
   on a product of electron trees we can define 
   a total order as the sum of the orders of the trees, 
   \begin{eqnarray*}
           |t_1 \cdots t_k| &=& |t_1| +\cdots+ |t_k|. 
   \end{eqnarray*}
   Then the algebra $\He$ of electron trees can be 
   decomposed into homogeneous components 
   \begin{eqnarray*}
           \He_n = \bigoplus_{n_1+\cdots+n_k=n} 
           \SP(Y_{n_1}) \otimes\dots\otimes \SP(Y_{n_k}), 
   \end{eqnarray*}
   and since the pruning antipode is a homogeneous operator, 
   these components can be written as 
   \begin{eqnarray*}
           \He_n = \bigoplus_{n_1+\cdots+n_k=n} 
           Y_{n_1} \otimes\dots\otimes Y_{n_k}. 
   \end{eqnarray*}
   \end{remark}

   To end this section, we give an example in terms of 
   Feynman diagrams. Let us draw the perturative expansion 
   of $S(q;e_0)$ and $\Sigma(q;e_0)$ up to order $2$. 
   The propagator is 
   \begin{eqnarray*}
   S(q;e_0)=1+e_0^2\phie(\Y;q) + e_0^4 \phie(\deuxdeux;q) 
   +e_0^4 \phie(\deuxun;q)+ {\cal{O}}(e_0^6),
   \end{eqnarray*}
   where
   \begin{eqnarray*}
   \phie(\Y) &=& U \big( 
     \parbox{27mm}{\begin{center}
     \begin{fmfgraph}(22,5)
     \setval
     \fmfforce{0w,0.5h}{v1}
     \fmfforce{1/3w,0.5h}{v2}
     \fmfforce{2/3w,0.5h}{v3}
     \fmfforce{1w,0.5h}{v4}
     \fmf{plain}{v4,v3,v2}
     \fmf{boson,left}{v2,v3}
     \fmfdot{v2,v3}
     \end{fmfgraph}
     \end{center}} \big) \\
   \phie(\deuxdeux) &=& U \big( 
   \parbox{28mm}{\begin{center}
   \begin{fmfgraph}(25,5)
   \setval
   \fmfforce{0w,0.5h}{v1}
   \fmfforce{1/5w,0.5h}{v2}
   \fmfforce{2/5w,0.5h}{v3}
   \fmfforce{3/5w,0.5h}{v4}
   \fmfforce{4/5w,0.5h}{v5}
   \fmfforce{1w,0.5h}{v6}
   \fmf{plain}{v6,v5,v4,v3,v2}
   \fmf{boson,left=0.6}{v2,v4}
   \fmf{boson,right=0.6}{v3,v5}
   \fmfdot{v2,v3,v4,v5}
   \end{fmfgraph}
   \end{center}}
   +
   \parbox{28mm}{\begin{center}
   \begin{fmfgraph}(25,5)
   \setval
   \fmfforce{0w,0.5h}{v1}
   \fmfforce{1/5w,0.5h}{v2}
   \fmfforce{2/5w,0.5h}{v3}
   \fmfforce{3/5w,0.5h}{v4}
   \fmfforce{4/5w,0.5h}{v5}
   \fmfforce{1w,0.5h}{v6}
   \fmf{plain}{v6,v5,v4,v3,v2}
   \fmf{boson,left=0.6}{v2,v5}
   \fmf{boson,left=0.6}{v3,v4}
   \fmfdot{v2,v3,v4,v5}
   \end{fmfgraph}
   \end{center}}
   +
   \parbox{28mm}{\begin{center}
   \begin{fmfgraph}(25,5)
   \setval
   \fmfforce{0w,0.5h}{v1}
   \fmfforce{1/5w,0.5h}{v2}
   \fmfforce{2/5w,0.5h}{v3}
   \fmfforce{3/5w,0.5h}{v4}
   \fmfforce{4/5w,0.5h}{v5}
   \fmfforce{1w,0.5h}{v6}
   \fmf{plain}{v6,v5,v4,v3,v2}
   \fmf{boson,left}{v2,v3}
   \fmf{boson,left}{v4,v5}
   \fmfdot{v2,v3,v4,v5}
   \end{fmfgraph}
   \end{center}} \big) , 
   \\
   \phie(\deuxun) &=& U \big( 
   \parbox{28mm}{\begin{center}
   \begin{fmfgraph}(25,8)
   \setval
   \fmfforce{0w,0h}{v1}
   \fmfforce{0.2w,0h}{v2}
   \fmfforce{0.4w,11/16h}{v2b}
   \fmfforce{0.8w,0h}{v3}
   \fmfforce{0.6w,11/16h}{v3b}
   \fmfforce{1w,0h}{v4}
   \fmf{plain}{v4,v3,v2}
   \fmf{boson,left=0.33}{v2,v2b}
   \fmf{boson,right=0.33}{v3,v3b}
   \fmf{plain,right}{v2b,v3b,v2b}
   \fmfdot{v2,v3,v3b,v2b}
   \end{fmfgraph}
   \end{center}} \big) . 
   \end{eqnarray*}
   Therefore, the self-energy is now written as
   \begin{eqnarray*}
   \Sigma(q;e_0)=-e_0^2\phie(\SP(\Y);q) - e_0^4 \phie(\SP(\deuxdeux);q)
   -e_0^4 \phie(\SP(\deuxun);q)-{\cal{O}}(e_0^6),
   \end{eqnarray*}
   where
   \begin{eqnarray*}
   \phie(\SP(\Y)) &=& -\phie(\Y) = -U \big( 
   \parbox{27mm}{\begin{center}
     \begin{fmfgraph}(22,5)
     \setval
     \fmfforce{0w,0.5h}{v1}
     \fmfforce{1/3w,0.5h}{v2}
     \fmfforce{2/3w,0.5h}{v3}
     \fmfforce{1w,0.5h}{v4}
     \fmf{plain}{v4,v3,v2}
     \fmf{boson,left}{v2,v3}
     \fmfdot{v2,v3}
     \end{fmfgraph}
     \end{center}} \big) \\
   \phie(\SP(\deuxdeux)) &=& -\phie(\deuxdeux)
   +\phie(\Y)\phie(\Y) = - U \big( 
   \parbox{28mm}{\begin{center}
   \begin{fmfgraph}(25,5)
   \setval
   \fmfforce{0w,0.5h}{v1}
   \fmfforce{1/5w,0.5h}{v2}
   \fmfforce{2/5w,0.5h}{v3}
   \fmfforce{3/5w,0.5h}{v4}
   \fmfforce{4/5w,0.5h}{v5}
   \fmfforce{1w,0.5h}{v6}
   \fmf{plain}{v6,v5,v4,v3,v2}
   \fmf{boson,left=0.6}{v2,v4}
   \fmf{boson,right=0.6}{v3,v5}
   \fmfdot{v2,v3,v4,v5}
   \end{fmfgraph}
   \end{center}} \big) 
   - U \big( 
   \parbox{28mm}{\begin{center}
   \begin{fmfgraph}(25,5)
   \setval
   \fmfforce{0w,0.5h}{v1}
   \fmfforce{1/5w,0.5h}{v2}
   \fmfforce{2/5w,0.5h}{v3}
   \fmfforce{3/5w,0.5h}{v4}
   \fmfforce{4/5w,0.5h}{v5}
   \fmfforce{1w,0.5h}{v6}
   \fmf{plain}{v6,v5,v4,v3,v2}
   \fmf{boson,left=0.6}{v2,v5}
   \fmf{boson,left=0.6}{v3,v4}
   \fmfdot{v2,v3,v4,v5}
   \end{fmfgraph}
   \end{center}} \big) , 
   \\
   \phie(\SP(\deuxun)) &=& -\phie(\deuxun) = - U \big( 
   \parbox{28mm}{\begin{center}
   \begin{fmfgraph}(25,8)
   \setval
   \fmfforce{0w,0h}{v1}
   \fmfforce{0.2w,0h}{v2}
   \fmfforce{0.4w,11/16h}{v2b}
   \fmfforce{0.8w,0h}{v3}
   \fmfforce{0.6w,11/16h}{v3b}
   \fmfforce{1w,0h}{v4}
   \fmf{plain}{v4,v3,v2}
   \fmf{boson,left=0.33}{v2,v2b}
   \fmf{boson,right=0.33}{v3,v3b}
   \fmf{plain,right}{v2b,v3b,v2b}
   \fmfdot{v2,v3,v3b,v2b}
   \end{fmfgraph}
   \end{center}} \big) . 
   \end{eqnarray*}


   \subsection{Gradings and loops}
   \label{gradings}

   By construction (\ref{defpi}), in any planar binary 
   tree each leaf (except one!) corresponds to a loop 
   in the associated Feynman diagrams. 
   Photon and electron loops are distinguished through 
   the orientation of the corresponding leaf: 
   leaves growing to the right (i.e. $/$-leaves) correspond 
   to electron loops, leaves growing to the left 
   (i.e. $\backslash$-leaves) correspond to photon loops. 
   The remaining leaf is just a prolongation of the root, 
   and we can identify it with the leftmost one for 
   photon trees, and with the rightmost one for electron trees. 

   For any tree $t$, we denote by $\Le(t)$ and $\Lp(t)$ the number 
   of electron and photon loops respectively. 
   Since in planar binary trees each leaf (except one) corresponds 
   to an internal vertex, we have exactly 
   \begin{eqnarray*} 
           \Le(t) + \Lp(t) = |t|.
   \end{eqnarray*}  
   For instance, in quenched QED, the photon trees $t$
   have $\Lp(t)=|t|-1$, $\Le(t)=1$, and the electron trees $s$
   have $\Lp(s)=|s|$, $\Le(s)=0$.

   To identify the space of photon and electron trees with 
   a fixed number of photon or electron loops, we 
   recall the double decomposition of the set of trees 
   first introduced in \cite{Frabetti}. 

   For any $p,q \geq 0$, let $Y_{p,q}$ be the set of 
   trees with $p$ leaves oriented like $\backslash$ 
   (except the leftmost one) and $q$ leaves oriented 
   like $/$ (except the rightmost one). 
   Hence, the order of a $(p,q)$-tree $t$ is 
   $|t|=p+q+1$, and the set of trees with order $n>0$ 
   can be decomposed as 
   \begin{eqnarray}
           Y_n &=& \bigoplus_{p+q+1=n} Y_{p,q} , 
   \end{eqnarray}
   while $Y_0$ can not be decomposed. 

   \begin{definition} 
   The set of photon trees with no photon loops and 
   no electron loops is, of course, $Y_0$. 
   Since a photon tree can not have any photon loop 
   if there is no electron loop which supports it, 
   we can single out the set of photon trees $t$ with 
   $\Lp(t)=p$ and $\Le(t)=q$ for any $p \geq 0$ 
   and any $q>0$ as the set $Y_{p,q-1}$. 
   Therefore we define a loops bigrading 
   on the vector space $\Hp$ as 
   \begin{eqnarray*}
           \Hp_{0,0} &=& Y_{0}, \\ 
           \Hp_{p,0} &=& \emptyset, \qquad\mbox{for any $p>0$}, \\ 
           \Hp_{p,q} &=& Y_{p,q-1}, 
           \qquad\mbox{for $p\geq 0$ and $q>0$}. 
   \end{eqnarray*}
   It is easy to see that this bigrading yields a complete 
   decomposition $\Hp_n = \oplus_{p+q=n} \Hp_{p,q}$, 
   which is compatible with the $\pro$ product, that is 
   \begin{eqnarray*}
           \Hp_{p_1,q_1} \pro \Hp_{p_2,q_2} 
           \subset \Hp_{p_1+p_2,q_1+q_2}. 
   \end{eqnarray*}

   Similarly, the set of electron trees with 
   no loops at all is $Y_0$. 
   Since on electron trees any electron loop needs 
   at least one photon loop as support, this time 
   we can identify the set of electron trees $s$ 
   with $\Lp(s)=p$ and $\Le(s)=q$ for any $p>0$ 
   and $q \geq 0$ as the set $Y_{p-1,q}$. 
   In order to have a consistent extension of the 
   loops bigrading with the free product of 
   electron trees we define it as 
   \begin{eqnarray*}
           \He_{0,0} &=& Y_0 , \\ 
           \He_{0,q} &=& \emptyset, \qquad\mbox{for any $q>0$}, \\ 
           \He_{p,q} &=& \bigoplus_{k=1}^p 
           \bigoplus_{{p_1+...+p_k=p}\atop{q_1+...+q_k=q}}
           Y_{p_1-1,q_1} \otimes\cdots\otimes Y_{p_k-1,q_k}, 
           \qquad\mbox{for $p>0$ and $q \geq 0$}. 
   \end{eqnarray*} 
   Since the pruning antipode preserve the number of loops, 
   to have an algebra decomposition the above vector space 
   can be written as 
   \begin{eqnarray*}
           \He_{p,q} &=& \bigoplus_{k=1}^p 
           \bigoplus_{{p_1+...+p_k=p}\atop{q_1+...+q_k=q}}
           \SP(Y_{p_1-1,q_1}) \otimes\cdots\otimes \SP(Y_{p_k-1,q_k}). 
   \end{eqnarray*} 
   Then at each order we have a complete decomposition 
   $\He_n = \oplus_{p+q=n} \He_{p,q}$. 
   \end{definition}

   \begin{remark}
   In \cite{LodayDias}, J.-L.~Loday introduced a new type of 
   associative algebra, called dialgebras, and their 
   Koszul duals, simply called dual dialgebras. 
   He then showed that the set of planar binary trees 
   can be interpretated as the free dual dialgebra generated 
   by one element. Therefore, in virtue of the operadic homology 
   introduced by V.~Ginzburg and M.~Kapranov in \cite{GinzburgKapranov}, 
   the set $Y$ appears naturally in the dialgebra homology $HY_*$ 
   (moreover the order $n$ of trees determines the degree 
   of the homology, i.e. $HY_n$).
   Loday showed that the boundary operator of this homology 
   is very elegantly expressed in terms of trees:  
   it acts by deleting the leaves (see \cite{L-P}). 

   More precisely, Loday showed that the dialgebra boundary 
   operator is in fact simplicial, and each $i$-th face 
   acts by deleting the $i$-th leaf of the trees. 
   In \cite{Frabetti}, the simplicial complex on the 
   set $Y$ was recognized as the total complex of 
   a bisimplicial structure, with horizontal and vertical 
   face maps acting by deleting $/$ and $\backslash$-leaves, 
   respectively. 
   The resulting bicomplex was shown to be acyclic. 

   Now, translating these results in terms of QED propagators, 
   we obtain an acyclic bicomplex on the set of 
   2-legs Feynman diagrams, whose horizontal and vertical 
   boundary operators act by deleting respectively 
   electron and photon loops. 
   How this can be interpretated in quantum field theory 
   is an open question. 
   \end{remark}

   \begin{remark}
   It is known that the number of trees with order $n$ is 
   the Catalan number $c_n= (2n)!/n!(n+1)!$. 
   The number of trees with order $n$ and a fixed quantity of 
   photon loops $p$ and of electron loops $q$ has been computed 
   in \cite{Frabetti}, as
   \begin{eqnarray*}
           \cp_{p,q} &=& \frac{(p+q+1)!(p+q+2)!}{p!(p+1)!(q-1)!q!} 
           \qquad\mbox{for photon trees}, \\ 
           \ce_{p,q} &=& \frac{(p+q+1)!(p+q+2)!}{(p-1)!p!q!(q+1)!} 
           \qquad\mbox{for electron trees}. 
   \end{eqnarray*}
   There is an obvious symmetry $\cp_{p,q} = \ce_{q,p}$. 
   \end{remark}


   %
   %
    
   \section{Hopf algebras on planar binary trees}
   \label{junehopf}

   It is nowadays established that the relationship between 
   the amplitudes of Feynman diagrams before and after 
   the renormalization has a combinatorical description 
   in terms of a Hopf algebra structure defined on the set of 
   diagrams, cf. \cite{Kreimer98}. 

   In \cite{BF}, we showed that on the sets of trees there 
   exists a structure of a Hopf algebra which encodes 
   the renormalization of QED propagators. 
   Looking at trees as sums of Feynman diagrams, 
   the coproduct on trees is much simpler then the sum 
   of the coproducts on each Feynman diagram. 
   We pointed out that the collections of amplitudes 
   described by each tree do not yield a finite value, 
   but the sum of amplitudes at each order of interaction, 
   renormalized through the tree coproduct, indeed give 
   the usual value obtained by summing up the renormalized 
   amplitudes of all Feynman diagrams, with a procedure 
   which automatically encodes the Ward identities. 

   In this section we give the precise definition 
   of the Hopf algebras defined on photon and electron trees, 
   as quoted in \cite{BF}. 
   Using these structures, in section \ref{juneorde} 
   we will find the explicit 
   formulas for the renormalization of QED propagators 
   at each order of interaction.

    
   \subsection{Hopf algebra for the photon trees}
   \label{hopfphotons}

   \begin{definition} 
   Define a coproduct $\Dp: \Hp \longrightarrow \Hp \otimes \Hp$ 
   by the recurrence relations 
   \begin{eqnarray}
           \Dp \| &=& \| \otimes \|, \nonumber\\ 
           \Dp V(t) &=& V(t) \otimes \| 
           + \sum_{\Dp t} t_{(1)} \otimes V(t_{(2)})  
           - \sum_{\Dp t^l} (t^l_{(1)} \vee t^r) \otimes V(t^l_{(2)})  
           \nonumber \\ 
           &=& V(t) \otimes \| 
           + (\Id \otimes V) \Dp t - (\Id \otimes V) 
           \big[(V(t^r) \otimes \|) \pro \Dp t^l \big], 
           \label{photoncoproduct} \\ 
           \Dp (t \vee s) &=& \Dp V(s) \pro \Dp t, \nonumber
   \end{eqnarray}
   where we adopt Sweedler's convention 
   $\Dp t = \sum_{\Dp t} t_{(1)} \otimes t_{(2)}$
   and where $t=t^l\vee t^r$.
    
   Let $\epsilon : \Hp \longrightarrow \C$ be the map 
   which sends all the trees to $0$ except the root 
   $\|$ which is sent to $1$. 
   \end{definition} 

   A few examples might be useful:
   \begin{eqnarray*}
   \Dp \Y &=& \Y\otimes \| + \| \otimes \Y , \\ 
   \Dp \deuxun &=& \deuxun\otimes \| + \| \otimes \deuxun +
   2\Y \otimes \Y , \\ 
   \Dp \deuxdeux &=& \deuxdeux\otimes \| + \| \otimes \deuxdeux , \\ 
   \Dp\troisun &=& \troisun \otimes \| + \| \otimes \troisun
   +3 \Y\otimes \deuxun+ 3 \deuxun \otimes \Y,\\
   \Dp\troisdeux&=&\troisdeux \otimes \| + \| \otimes \troisdeux
   +\Y\otimes \deuxdeux + \deuxdeux\otimes \Y,\\
   \Dp\troistrois &=& \troistrois\otimes \| + \| \otimes \troistrois
   +\Y \otimes \deuxdeux + \deuxdeux\otimes \Y,\\
   \Dp\troisquatre &=& \troisquatre \otimes \| + \| \otimes \troisquatre
   +\Y \otimes \deuxdeux,\\
   \Dp\troiscinq &=& \troiscinq \otimes \| + \| \otimes \troiscinq.
   \end{eqnarray*}
    
   \begin{proposition}
   \label{hopfHp}
   The space $\Hp$ is a non-commutative nor cocommutative bigraded 
   Hopf algebra. 
   \end{proposition}
    
   \begin{proof} 
   We first observe that the coproduct preserves 
   the bigrading of $\Hp$, that is, 
   \begin{eqnarray*}
           \Dp(\Hp_{p,q}) \subset \bigoplus_{\atop{p_1+p_2=p},{q_1+q_2=q}} 
           \Hp_{p_1,q_1} \otimes \Hp_{p_2,q_2},  
   \end{eqnarray*}
   and therefore it also preserves the total order of trees 
   defined in (\ref{algebraHp}). 

   By recursion arguments, it is easy to see that the only terms 
   of total order $(n,0)$ and $(0,n)$, in the image of $\Dp$, 
   are given by the primitive part $t \otimes \|$ and $\| \otimes t$ 
   of any tree $t$.
   Then, the map $\epsilon$ is a counit for $\Dp$. 
   Moreover (see ref.~\cite{Sweedler} for details), the antipode 
   $\Sp: \Hp \longrightarrow \Hp$ 
   is automatically defined by the recursive formula 
   \begin{eqnarray} 
   \label{photonantipode}
           \Sp(t) = -t - (\Sp \bar{\star} \Id)(t) 
   \end{eqnarray}
   which involves the convolution $\bar{\star}$ of the 
   reduced coproduct 
   $\bar{\Dp}(t)=\Dp(t) - t \otimes \| - \| \otimes t$. 

   To prove that $\Dp$ is coassociative, it is useful to introduce 
   another operator $F: \Hp \longrightarrow \Hp \otimes \Hp$, 
   such that 
   \begin{eqnarray}
   \label{DeltaF}
           \Dp V(t) &=& V(t) \otimes \| + F(V(t)).   
   \end{eqnarray}
    
   We achieve this by defining $F$ again with some recurrence conditions 
   \begin{eqnarray*}
   \label{defF}
           F(\|) &=& \| \otimes \|, \\  
           F(V(t)) &=& (\Id \otimes V) F(t) 
           = \sum_{F(t)} t_{(1)} \otimes V(t_{(2)}), 
   \end{eqnarray*}
   and supposing that it is a multiplicative map, in the sense that 
   \begin{eqnarray}
   \label{blackcoaction}
           F(t \vee s) &=& F(V(s)) \pro \Dp t 
           = \sum_{\Dp(t),F(s)} (s_{(1)}\pro t_{(1)}) \otimes t_{(2)} \vee s_{(2)}.  
   \end{eqnarray}
   In fact, if we want $\Dp$ to be given by Eq. (\ref{DeltaF}), 
   the operator $F$ must be defined in such a way that 
   \begin{eqnarray*}
           F(V(t)) &=& 
           \sum_{\Dp t} t_{(1)} \otimes V(t_{(2)})  
           - \sum_{\Dp t^l} (t^l_{(1)} \vee t^r) \otimes V(t^l_{(2)}) \\
           &=& (\Id \otimes V) \left[ \Dp t - (V(t^r) \otimes \|) 
           \pro \Dp t^l \right]. 
   \end{eqnarray*}
   Therefore, if on a generator $V(t)$ we assume that 
   $F(V(t)) = (\Id \otimes V) F(t)$, we have 
   \begin{eqnarray*}
           F(t) &=& \Dp t - (V(t^r) \otimes \|) \pro \Dp t^l  \\ 
           &=& \left[ \Dp V(t^r) - (V(t^r)) \otimes \|) \right] \pro \Dp t^l \\ 
           &=& F(V(t^r)) \pro \Dp t^l. 
   \end{eqnarray*} 

   First we prove by induction that the operator $F$ defines a left
   $\Dp$-coaction of $\Hp$ on itself, that is, it satisfies 
   $(\Id \otimes F) F = (\Dp \otimes \Id) F$. 
   Since $F$ is multiplicative, we only need to show it on the 
   generators $V(t)$. It is true on $t=\|$. Suppose that it is true 
   for all the trees up to order $n$, and let $V(t)$ has order $n+1$. 
   On one side we have
   \begin{eqnarray*} 
           (\Id \otimes F) F(V(t)) &=& (\Id \otimes F \circ V) F(t) 
           = (\Id \otimes \Id \otimes V) (\Id \otimes F) F(t), 
   \end{eqnarray*}
   on the other side we have 
   \begin{eqnarray*} 
           (\Dp \otimes \Id) F(V(t)) &=& (\Dp \otimes V) F(t) 
           = (\Id \otimes \Id \otimes V) (\Dp \otimes \Id) F(t). 
   \end{eqnarray*}
   So the equality holds by inductive hypothesis. 

   Now we prove by induction that the operator $\Dp$ is coassociative, 
   using the fact that $F$ is a coaction. 
   Since $\Dp$ is multiplicative, we only need to prove it 
   on the generators $V(t)$. It is true on $t=\|$. 
   Suppose that $\Dp$ is coassociative on all the trees with order 
   up to $n$, and let $V(t)$ be a generator with order $n+1$. 
   Then by the expression (\ref{DeltaF}) of $\Dp$ we have on one side 
   \begin{eqnarray*}
           (\Dp \otimes \Id) \Dp V(t) &=& 
           \Dp V(t) \otimes \| + (\Dp \otimes \Id) F(V(t)) \\ 
           &=& V(t) \otimes \| \otimes \| + F(V(t)) \otimes \| 
           + (\Dp \otimes \Id) F(V(t)), 
   \end{eqnarray*}
   and on the other side 
   \begin{eqnarray*}
           (\Id \otimes \Dp) \Dp V(t) &=& 
           V(t) \otimes \Dp(\|) +  (\Id \otimes \Dp) F(V(t)) \\  
           &=& V(t) \otimes \| \otimes \| 
           + (\Id \otimes \Dp \circ V) F(t) \\
           &=& V(t) \otimes \| \otimes \| 
           + (\Id \otimes V \otimes \Id) F(t) 
           + (\Id \otimes F \circ V) F(t) \\ 
           &=& V(t) \otimes \| \otimes \| 
           + (\Id \otimes V) F(t) \otimes \|  
           + (\Id \otimes F) (\Id \otimes V) F(t) \\ 
           &=& V(t) \otimes \| \otimes \| 
           + F(V(t)) \otimes \|  
           + (\Id \otimes F) F(V(t)).   
   \end{eqnarray*}
   Then, the two sides are equal if and only if 
   $(\Dp \otimes \Id) F(t) = (\Id \otimes F) F(t)$, 
   which holds by inductive hypothesis. 
   \end{proof}

    
   \subsection{Hopf structure for the electron trees}
   \label{junehopfelectrons}

   On the electron algebra $\He$, the coproduct operator 
   analogue to the one on $\Hp$ is not an internal operation. 
   To describe it, we have to extend it to a bigger space 
   which describes at the same time the electron and the photon 
   propagators of QED. 
    
   Consider the tensor algebra $\Hq = \Hp \otimes \He$ with 
   product defined componentwise as the product given in (\ref{photoncoproduct}) 
   on the photon trees and the tensor product on the electron trees. 
   The unit is $\| \otimes \|$, and the two identifications 
   $\Hp = \Hp \otimes \|$ and $\He = \| \otimes \He$ 
   allow us to see $\Hp$ and $\He$ as subalgebras of $\Hq$.  
   For the simplicity of notation, we denote by couples 
   the elements of the tensor product between $\Hp$ and $\He$. 
   Hence we write $1 = (\|,\|)$ for the unit $\| \otimes \| \in \Hq$ 
   and $(t,s_1...s_n)$ for $t \otimes (s_1 \dots s_n) 
   \in \Hp \otimes \He$. 

   On the algebra $\Hq$ we chose a grading given by the sum of 
   the orders of all the trees appearing in a monomial. 
   Hence, $\Hq$ can be written as a direct sum 
   $\Hq = \bigoplus_{n\geq 0} \Hq_n$ of homogeneous components 
   $\Hq_n = \bigoplus_{p+q=n} \Hp_p \otimes \He_q$, 
   where $\Hp_p$ was defined in (\ref{algebraHp}) 
   and $\He_q$ in (\ref{gradingHe}). 

   \begin{definition} 
   We define a coproduct $\Dq: \Hq \longrightarrow \Hq \otimes \Hq$ 
   by its action on the photon and electron trees, and then we extend it on 
   products as a morphism of algebras, that is 
   \begin{eqnarray*}
           \Dq 1 &=& 1 \otimes 1, \\ 
           \Dq(t,s_1 \ldots s_n) &=& 
           \Dp(t) \De(s_1) \ldots \De(s_n).   
   \end{eqnarray*}
   On the photon trees, $\Dq$ agrees with the coproduct $\Dp$ defined above. 
   On the electron trees, we define the coproduct 
   $\De:\He \longrightarrow \Hq \otimes \He$ (notice the photon 
   component on the lefthand side!) by the recurrence relations 
   \begin{eqnarray}
           \De \| &=& 1 \otimes \|, \nonumber \\ 
           \De (t \vee s) &=& (\|,t \vee s)\otimes \|  
           + \sum_{\Dp(t),\De(s)} (t_{(1)}\pro s_{(1)}^{\gamma},s_{(1)}^{e}) 
           \otimes t_{(2)} \vee s_{(2)} \\ 
           &=& (\|,t \vee s) \otimes \|  
           + \big(\Id \otimes \vee\big) \Dp(t) \De(s), 
   \nonumber
           \label{electroncoproduct} 
   \end{eqnarray}
   where we adopt a modified version of Sweedler's notation 
   \begin{eqnarray*}
           \De s &=& \sum_{\De(s)} (s_{(1)}^{\gamma},s_{(1)}^{e}) 
           \otimes s_{(2)},  
   \end{eqnarray*} 
   which takes into account the photon component produced by $\De$. 
   For instance, 
   \begin{eqnarray*}
   \De \Y &=& 1 \otimes \Y + (\|, \Y) \otimes \|\\
   \De \deuxun &=& 1 \otimes \deuxun 
           + (\Y , \|) \otimes \Y + (\|,\deuxun) \otimes \|\\
   \De \deuxdeux &=& 1 \otimes \deuxdeux
           + (\|, \Y) \otimes \Y + (\|,\deuxdeux) \otimes \|\\
   \De\troisun &=& 1 \otimes \troisun + 2(\Y,\|) \otimes \deuxun 
           + (\deuxun,\|)\otimes \Y + (\|,\troisun)\otimes\|,\\
   \De\troisdeux &=& 1 \otimes \troisdeux + (\deuxdeux, \|)\otimes \Y 
           +(\|,\troisdeux) \otimes \|\\
   \De\troistrois &=& 1 \otimes \troistrois + (\Y,\|)\otimes \deuxdeux 
           + (\|,\deuxun) \otimes \Y + (\|,\troistrois)\otimes \|\\
   \De\troisquatre &=& 1 \otimes \troisquatre + (\Y,\|\otimes) \deuxdeux 
           + (\|,\Y) \otimes \deuxun + (\Y,\Y) \otimes\Y 
           + (\|,\troisquatre) \otimes \|,\\
   \De\troiscinq &=& 1 \otimes \troiscinq + (\|,\Y) \otimes \deuxdeux 
           + (\|,\deuxdeux) \otimes \Y + (\|,\troiscinq) \otimes \|.
   \end{eqnarray*}

   Let $\epsilon: \Hq \longrightarrow \C$ be the map 
   which sends everything to $0$ except the unit,  
   which is sent to $1$.
   \end{definition} 
    
   \begin{proposition}
   The space $\Hq$ is a bigraded Hopf algebra, 
   not commutative nor cocommutative,  
   of which $\Hp$ is a sub-Hopf algebra. 
   \end{proposition} 

   \begin{proof}
   Since all the operations involved in the definition of 
   $\Dq$ are graded, it is easy to check that the coproduct is  
   a graded operation, that is, it preserves the total degree 
   of each addend. 
   Then, the extended map $\epsilon$ is a counit for $\Dq$, 
   because each $\Dq (t,s)$ contains exactly the two terms 
   $(t,s) \otimes \|$ and $\| \otimes (t,s)$ in degree 
   $(n,0)$ and $(0,n)$ respectively, where $n=|t|+|s|$. 
   The proof is by induction. Since by proposition (\ref{hopfHp}) 
   it is true on all photon trees, we only need to check it on electron trees, 
   where $\Dq$ coincides with $\De$. 
   It is true for $\De(\|)$, so assume that it is
   true for trees up to order $n-1$, and consider a tree $t \vee s$ with
   order $n$. Then, using (\ref{DeltaF}) and (\ref{blackcoaction}) we write 
   $$
           \Dp(t) = (V(t^r) \otimes \|) \pro \Dp(t^l) + F(t), 
   $$
   and therefore by (\ref{electroncoproduct}) we have
   \begin{eqnarray*}
           \De(t \vee s) &=& (\|,t \vee s) \otimes \| + 
           (\Id \otimes \vee) [(V(t^r) \otimes \|) \pro \Dp(t^l) + F(t)] \De(s) \\ 
           &=& (\|,t \vee s) \otimes \| + 
           (\Id \otimes \vee) [(V(t^r) \otimes \|) \pro \Dp(t^l) \De (s) 
           + (\Id \otimes \vee) F(t) \De (s). 
   \end{eqnarray*} 
   Because of the presence of the operator $\Id \otimes \vee$, a pair of
   trees with bidegree $(n,0)$ can not be contained in the second and
   third addends, therefore the only such pair is the first addend 
   $t \vee s \otimes \|$. 
   Similarly, a pair of trees with bidegree $(0,n)$ can not be contained 
   in the second addend because of the term $(V(t^r) \otimes \Id)$. 
   Therefore it must be contained in the third addend where, by
   inductive hypothesis, it can only appear as the product of 
   the corresponding terms of bidegrees $(0,|s|)$ and $(0,|t|)$. 

   Again, the antipode $Sq: \Hq \longrightarrow \Hq$ is automatically 
   defined by the recursive formula (\ref{photonantipode}). 
   Since, by definition, $\Dq$ is a multiplicative map, 
   it only remains to show that it is coassociative. 
   Again, we prove it by induction on the order of trees. 
   Of course, we only need to consider electron trees. 
    
   The coassociativity is easy to check for $t= \|$.
   Suppose that it holds for all trees with order 
   up to $n$, and let $t \vee s \in Y$ be 
   a tree with order $n+1$. Then $t$ and $s$ have at most order $n$, 
   hence, the coassociativity of $\Dq$ holds for the couple  
   $(t,s) \in Y \otimes Y \subset \Hq$. 
    
   Then, by definition of $\Dq$, on one side we have 
   \begin{eqnarray*}
           (\Dq \otimes \Id) \Dq (\|, t \vee s) 
           &=& (\Dq \otimes \Id) \De (t \vee s) \\ 
           &=& \De (t \vee s) \otimes \| + \sum_{\Dp(t),\De(s)} 
           \Dq (t_{(1)} \pro s_{(1)}^{\gamma} , s_{(1)}^e) 
           \otimes t_{(2)} \vee s_{(2)} \\ 
           &=& (\|,t \vee s) \otimes 1 \otimes \| + \sum_{\Dp(t),\De(s)} 
           (t_{(1)} \pro s_{(1)}^{\gamma} , s_{(1)}^e)
           \otimes (\|,t_{(2)} \vee s_{(2)}) \otimes \| \\ 
           && + \sum_{\Dp(t),\De(s)} 
           \Dq (t_{(1)} \pro s_{(1)}^{\gamma} , s_{(1)}^e) 
           \otimes t_{(2)} \vee s_{(2)}, 
   \end{eqnarray*}
   and on the other side 
   \begin{eqnarray*}
           (\Id \otimes \Dq) \Dq (\|, t \vee s) 
           &=& (\Id \otimes \Dq) \De (t \vee s) \\ 
           &=& (\|,t \vee s) \otimes 1 \otimes \| + 
           \sum_{\Dp(t),\De(s)} (t_{(1)},\|) (s_{(1)}^{\gamma} , s_{(1)}^e) 
           \otimes \De (t_{(2)} \vee s_{(2)}) \\ 
           &=& (\|,t \vee s) \otimes 1 \otimes \| +  
           \sum_{\Dp(t),\De(s)} (t_{(1)} \pro s_{(1)}^{\gamma} , s_{(1)}^e)  
           \otimes (\|,t_{(2)} \vee s_{(2)}) \otimes \|  \\ 
           && + \sum_{{\Dp(t),\De(s)}\atop{\Dp(t_{(2)}),\De(s_{(2)})}} 
           (t_{(1)} \pro s_{(1)}^{\gamma} , s_{(1)}^e) \otimes 
           (t_{(21)} \pro s_{(21)}^{\gamma} , s_{(21)}^e)  
           \otimes t_{(22)} \vee s_{(22)} .    
   \end{eqnarray*}
   Then, the two sides are equal if and only if 
   \begin{eqnarray*}
           \sum_{\Dp(t),\De(s)} 
           \Dq (t_{(1)} \pro s_{(1)}^{\gamma} , s_{(1)}^e) 
           \otimes t_{(2)} \vee s_{(2)} 
           &=& \sum_{{\Dp(t),\De(s)}\atop{\Dp(t_{(2)}),\De(s_{(2)})}} 
           (t_{(1)} \pro s_{(1)}^{\gamma} , s_{(1)}^e) \otimes 
           (t_{(21)} \pro s_{(21)}^{\gamma} , s_{(21)}^e)  
           \otimes t_{(22)} \vee s_{(22)}.   
   \end{eqnarray*}
   Since $t \vee s = t' \vee s'$ if and only if 
   $(t,s) = (t',s')$, this equality holds in $\Hq\otimes\Hq\otimes\He$ 
   if and only if in $\Hq\otimes\Hq\otimes\Hq$ we have  
   \begin{eqnarray*}
           \sum_{\Dp(t),\De(s)} 
           \Dq (t_{(1)} \pro s_{(1)}^{\gamma} , s_{(1)}^e) 
           \otimes (t_{(2)},s_{(2)}) 
           &=& \sum_{{\Dp(t),\De(s)}\atop{\Dp(t_{(2)}),(s_{(2)})}} 
           (t_{(1)} \pro s_{(1)}^{\gamma} , s_{(1)}^e) \otimes 
           (t_{(21)} \pro s_{(21)}^{\gamma} , s_{(21)}^e) \otimes 
           (t_{(22)},s_{(22)}) . 
   \end{eqnarray*}
   The last equality means that 
   $$
           (\Dq \otimes \Id) \Dq(t,s) = 
           (\Id \otimes \Dq) \Dq(t,s), 
   $$  
   and holds by inductive hypothesis. 
   \end{proof}

   \begin{remark}
   On the generators $\SP(t)$ of the algebra $\He$, the electron 
   coproduct $\De$ takes the form 
   \begin{eqnarray*}
           \De(\SP(t)) &=& \sum_{\De(t)} 
           (t_{(1)}^{\gamma},\SP(t_{(1)}^e)) \otimes \SP(t_{(2)}). 
   \end{eqnarray*}
   \end{remark}


   \subsection{Tree expansion for renormalized propagators} 
   \label{ren-tree}

   Following Ref.~\cite{CKII}, and using the fact that 
   the renormalization factors depend only on the vacuum 
   polarization and the self-energy, we know that the photon 
   renormalization factor $Z_3$ is defined only on the generators
   $\|\vee t$, and that the electron renormalization factor $Z_2$ 
   is defined only on elements of the form $\Sp(t)$. 
   Therefore, the coefficients $Z_{3,n}$ and $Z_{2,n}$ 
   of the perturbative expansion defined by Eqs.~(\ref{Z2}) and 
   (\ref{Z3}), are defined as 
   \begin{eqnarray}
           Z_{3,n} &=& \sum_{|t|=n} \zeta_3(V(t)),  
           \label{zeta3t} \\ 
           Z_{2,n} &=& \sum_{|t|=n} \zeta_2(\SP(t)) ,  
           \label{zeta2t}. 
   \end{eqnarray}

   The maps $\zeta_3: \Hp \longrightarrow \C$ and 
   $\zeta_2: \He \longrightarrow \C$ are determined 
   recursively on the generators, through a renormalization 
   prescription (cf. \cite{BF}),
   \begin{eqnarray*}
           \zeta_3(V(t)) &=& \sum_{\pp(\Gamma)=V(t)} C_3(\Gamma), \\
           \zeta_2(\SP(t)) &=& -\sum_{\pe(\Gamma)=\SP(t)} C_2(\Gamma),
   \end{eqnarray*}
   and then extended multiplicatively. 

   Denote by $\zeta: \Hq \longrightarrow \C$ the product map 
   of $\zeta_3$ and $\zeta_2$, that is,  
   \begin{eqnarray}
           \zeta (t,s) &:=& 
           \zeta_3(V(t^1))\dots\zeta_3(V(t^l)) 
           \zeta_2(\SP(s_1))\dots\zeta_2(\SP(s_k)),
   \end{eqnarray}
   where $(t,s) = (V(t^1)\pro\dots\pro V(t^l),\SP(s_1)\dots\SP((s_k))$ 
   is a generic element of $\Hq$. 

   In Ref.~\cite{BF}, the renormalized amplitudes corresponding to 
   each tree were found by a recurrence procedure based on 
   the Schwinger equations for the renormalized propagators. 
   In the case of massless QED, it was shown that the 
   relationship between the amplitudes before and after 
   renormalization was encoded by the coproduct on trees. 
   We recall the main results of \cite{BF}. 

   \begin{theorem}
   The relation between the coefficients of the expansion 
   (\ref{D(t;q)}, \ref{barD(t;q)}) of the 
   renormalized and bare photon propagators is
   \begin{eqnarray}
           \barphip(t;q) &=& 
           \sum_{\Dp(t)} \zeta(t_{(1)}, \|) \phip(t_{(2)};q).  
           \label{renphotont}
   \end{eqnarray}
   The relation between the coefficients of the expansion 
   (\ref{S(t;q)}, \ref{barS(t;q)}) of the renormalized 
   and bare electron propagators is 
   \begin{eqnarray}
           \barphie(t;q) &=& 
           \sum_{\De(t)} \zeta(t_{(1)}^{\gamma} , t_{(1)}^e) 
           \phie(t_{(2)};q) . \label{renelectront}
   \end{eqnarray}
   \end{theorem}

   In particular, choosing only the generator trees, 
   the same result holds for the renormalization 
   of the vacuum polarization $\Pi(q;e_0)$ and of 
   the self-energy $\Sigma(q;e_0)$. 

   \begin{theorem}
   The relation between the coefficients of the expansion 
   (\ref{defPibare},\ref{defPiren}) of the 
   renormalized and bare vacuum polarizations is
   \begin{eqnarray}
           \barphip(V(t);q) &=& 
           \sum_{\Dp(V(t))} \zeta(V(t)_{(1)}, \|) 
           \phip(V(t)_{(2)};q) . \label{renvacuumt}
   \end{eqnarray}
   The relation between the coefficients of the expansion 
   (\ref{defSigmabare},\ref{defSigmaren}) of the 
   renormalized and bare self-energies is
   \begin{eqnarray}
           \barphie(\SP(t);q) &=& 
           \sum_{\De(t)} \zeta(t_{(1)}^{\gamma} , \SP(t_{(1)}^e)) 
           \phie(\SP(t_{(2)});q).\label{renselft}
   \end{eqnarray}
   \end{theorem}

   %
   %
    
   \section{Renormalization ``at the order of interaction''}
   \label{juneorde} 

   In section \ref{juneren}, the terms $S_n(q)$ and
   $D_n(q)$ at each order of interaction are defined using the 
   Feynman's rules. 
   In fact, they can be regarded as the evaluation of the map $U$ 
   on some formal sums of Feynman's graphs, that is 
   \begin{eqnarray*} 
           S_n(q) &=& 
           U \Big(\sum_{{\Gamma \in \Fe},\ {|\Gamma|=n}} \Gamma \Big), \\ 
           D_n(q) &=& 
           U \Big(\sum_{{\Gamma \in \Fp},\ {|\Gamma|=n}} \Gamma \Big).   
   \end{eqnarray*}
   Now, the same terms can be regarded as the evaluation 
   of the map $\varphi = \phip \otimes \phie$ on some formal sums of trees 
   \begin{eqnarray} 
           S_n(q) &=& \phie \Big( \sum_{|t|=n} t \Big)
           = \langle \varphi, (\|,\xbf_n) \rangle , \label{evalS} \\ 
           D_n(q) &=& \phip \Big( \sum_{|t|=n} t \Big) 
           = \langle \varphi, (\ubf_n,\|) \rangle , \label{evalD} 
   \end{eqnarray}
   where we define the following elements in the Hopf algebra $\Hq$: 
   \begin{eqnarray*} 
           \xbf_n &=& \sum_{|t|=n} t \in \He \\ 
           \ubf_n &=& \sum_{|t|=n} t \in \Hp . 
   \end{eqnarray*}

   To determine the relationship between the terms of the Green functions 
   at each order of interaction before and after the renormalization, 
   we need to compute the photon and electron coproducts of the 
   elements $\xbf_n$ and $\ubf_n$. 


   \subsection{Hopf algebra for photons at order $n$}

   Let $\ubf_n$ and $\vbf_n$ be the elements of $\Hp$ defined by 
   \begin{eqnarray}
           \ubf_n &=& \sum_{|t|=n} t , \qquad n \geq 0, \label{defubf}\\ 
           \vbf_n &=& \sum_{|t|=n-1} V(t) , \qquad n \geq 1. \label{defvbf}
   \end{eqnarray} 
   Of course, $\vbf_n$ is then related to $\ubf_n$ by the 
   relation $\vbf_n = V(\ubf_{n-1})$. 
   Conversely, it is not difficult to show that 
   \begin{eqnarray}
           \ubf_n &=& \sum_{k=1}^n P_n^{(k)}(\vbf_*),   
   \end{eqnarray} 
   where for any $m \geq 1$ and any $k \leq m$, we let  
   \begin{eqnarray}
           P_m^{(k)}(\vbf_*) &=& 
           \sum_{ {j_1 > 0,\dots, j_k > 0}\atop{j_1+\cdots+j_k=m}} 
           \vbf_{j_1} \pro\cdots\pro \vbf_{j_k} 
   \end{eqnarray}
   be the symmetric (noncommutative) polynomial
   of degree $m$ in $k$ (noncommutative) variables $\vbf_*$. 

   \begin{proposition} 
   \label{Deltavbf} 
   The photon coproduct on trees induces the following coproduct
   on $\vbf_n$:
   \begin{eqnarray}
           \Dp \vbf_{1} &=& \vbf_{1} \otimes \| + \| \otimes \vbf_{1}, 
           \nonumber \\
           \Dp \vbf_{2} &=& \vbf_{2} \otimes \| + \| \otimes \vbf_{2}, 
           \nonumber \\
           \Dp \vbf_n &=& \vbf_n \otimes \| + \| \otimes \vbf_n
            + \sum_{m=1}^{n-2} R^{n-2}_m(\vbf_*) \otimes \vbf_{n-m},  
           \qquad n \geq 3  \label{deltavbf}
   \end{eqnarray}
   where we set
   \begin{eqnarray}
   \label{R(v)}
           R^l_m(\vbf_*)&=& 
           \sum_{k=1}^m \binom{l-m+k}{k} P^{(k)}_m(\vbf_*), 
           \qquad m, l \geq 1. 
   \end{eqnarray}
   \end{proposition}

   A few examples of the values of the polynomials $R^n_m(\vbf_*)$ 
   might be useful: 
   \begin{eqnarray*}
           R^l_1(\vbf_*) &=& l\vbf_1,\\
           R^l_2(\vbf_*) &=& (l-1)\vbf_2+\frac{l(l-1)}{2}\vbf_1^2,\\
           R^l_l(\vbf_*) &=& \sum_{k=1}^l P^{(k)}_l(\vbf_*) = \ubf_l . 
   \end{eqnarray*}
   Conversely, we can determine each $\vbf_n$ as a combination of 
   $R^l_m$'s: 
   \begin{eqnarray*}
           \vbf_n &=& -\sum_{k=1}^{n} (-1)^k 
           \binom{n+1}{k+1} R^{n+k-1}_n(\vbf_*).
   \end{eqnarray*}

   The computation of $\Dp(\vbf_n)$ for the 
   cases $n=1$ and $n=2$ can be done explicitly starting 
   from the coproduct over trees, and using
   $\Dp \Y = \Y \otimes \| + \| \otimes \Y$ and
   $\Dp \deuxdeux = \deuxdeux \otimes \| + \| \otimes \deuxdeux$.
   The next terms are: 
   \begin{eqnarray*}
           \Dp \vbf_3 &=& \vbf_3 \otimes \| + \| \otimes \vbf_3 
           + \vbf_1 \otimes \vbf_2 ,\\
           \Dp \vbf_4 &=& \vbf_4 \otimes \| + \| \otimes \vbf_4 
           + 2\vbf_1 \otimes \vbf_3 +(\vbf_2+\vbf_1^2) \otimes \vbf_2.
   \end{eqnarray*}

   \begin{proposition} 
   \label{Deltaubf} 
   The photon coproduct evaluated on $\ubf_n$ gives:
   \begin{eqnarray}
           \Dp \ubf_{1} &=& \ubf_{1} \otimes \| + \| \otimes \ubf_{1},
           \nonumber \\
           \Dp \ubf_{n} &=& \ubf_{n} \otimes \| + \| \otimes \ubf_{n} 
           + \sum_{m=1}^{n-1} R^{n}_m(\vbf_*) \otimes \ubf_{n-m}, 
           \quad n \geq 2 . \label{deltaubfold} 
   \end{eqnarray}
   \end{proposition}

   The simplest examples are: 
   \begin{eqnarray*}
           \Dp \ubf_2 &=& \ubf_1 \otimes \| + \| \otimes \ubf_1 
           + 2\vbf_1 \otimes \ubf_1 \\
           \Dp \ubf_3 &=& \ubf_3 \otimes \| + \| \otimes \ubf_3 
           + 3\vbf_1 \otimes \ubf_2 + (2\vbf_2+\vbf_1^2) \otimes \ubf_1 .
   \end{eqnarray*}

   \begin{proof of}{Deltavbf}
   The statements in propositions~(\ref{Deltavbf}) and  
   (\ref{Deltaubf}) are closely related. 
   We prove them by induction simultaneously. 

   Assume that Eq.~(\ref{deltaubfold}) is true up to order $n-1$ 
   and that Eq.~(\ref{deltavbf}) is true up order $n$. 
   Then, using Eq.~(\ref{deltavbf}) we show that relation 
   (\ref{deltaubfold}) is true at order $n$.
   Finally, we show that Eq.~(\ref{deltavbf}) is true at order $n+1$. 

   The first part is quite simple. We calculate
   \begin{eqnarray*}
           \Dp \ubf_n- \sum_{k=0}^{n-1} 
           \big(\vbf_{k+1}\otimes 1\big) \pro \Dp \ubf_{n-k-1} &=&
           \Dp \ubf_n-
           \sum_{k=1}^{n-1}\big(\vbf_{k}\otimes 1\big) 
           \pro \Dp \ubf_{n-k} -\vbf_n\otimes \| \\ 
           &&\hspace*{-30mm}=
           \|\otimes \ubf_n +\sum_{m=1}^{n-1}
           \big( R^n_m - \vbf_m -\sum_{k=1}^{m-1}\vbf_k \pro R^{n-k}_{m-k}\big)
           \otimes \ubf_{n-m} \\ 
           &&\hspace*{-30mm}=
           \|\otimes \ubf_n +\sum_{m=1}^{n-1} R^{n-1}_m\otimes \ubf_{n-m}.
   \end{eqnarray*} 
   Introducing this into Eq.(\ref{deltavbf}) yields
   \begin{eqnarray*}
           \Dp \vbf_{n+1} &=& \vbf_{n+1}\otimes \|+(\Id\otimes V)
           \big(\|\otimes \ubf_n +\sum_{m=1}^{n-1} R^{n-1}_m\otimes 
           \ubf_{n-m}\big) \\ 
           &=& \vbf_{n+1}\otimes \|+
           \|\otimes \vbf_{n+1} +\sum_{m=1}^{n-1} R^{n-1}_m\otimes 
           \vbf_{n-m+1}.
   \end{eqnarray*}
   To derive the last line, we used the fact that 
   $V(\ubf_n) = \vbf_{n+1}$.

   Knowing $\Dp \vbf_{n+1}$, we can calculate $\Dp \ubf_{n+1}$ through
   \begin{eqnarray*}
           \ubf_{n+1} &=& \vbf_{n+1} + \sum_{k=1}^n \vbf_{n-k+1} \pro \ubf_k. 
   \end{eqnarray*}
   We obtain
   \begin{eqnarray*}
           && \Dp \ubf_{n+1} = \Dp \vbf_{n+1} + \sum_{k=1}^n 
           \Dp\vbf_{n-k+1} \pro \Dp\ubf_k \\
           && = \|\otimes \ubf_{n+1} + \ubf_{n+1}\otimes \| 
           + \sum_{m=1}^n \Big(R^{n-1}_m + \ubf_m + \vbf_m
           +\sum_{k=1}^{m-1} \vbf_k \pro R^{n-k+1}_{m-k} + 
           R^{n-k-1}_{m-k}\ubf_k\Big) \otimes \ubf_{n+1-m} \\  
           && + \sum_{l=1}^{n-1}\sum_{i=1}^{n-l}
           \Big(-R^{n-1}_l - \ubf_l +R^{l+i}_l + R^{n-i-1}_l 
            +\sum_{k=1}^{l-1} -R^{n-l+k-1}_k \pro \ubf_{l-k} 
           +R^{n-l-i+k-1}_k \pro R^{i+l-k}_{l-k}\Big)\otimes 
           \vbf_{n+1-l-i} \pro \ubf_i.
   \end{eqnarray*}

   In the above equation, the coefficient of $\ubf_{n+1-m}$
   is denoted by $X$. It can be simplified by using Eq.(\ref{recurxx}):
   \begin{eqnarray*}
           X &=& R^{n+1}_m + \ubf_m +R^{n-1}_m +\sum_{k=1}^{m-1}
           R^{n-k-1}_{m-k} \pro \ubf_k - R^n_m.
   \end{eqnarray*}
   Equation (\ref{mpnpl}) of the next lemma (\ref{quadraticR}), for $l=0$, 
   together with $R^m_m=\ubf_m$ gives us $X=R^{n+1}_m$.
   The coefficient of $\vbf_{n+1-l-i} \ubf_i$ is
   treated similarly. Using Eq.(\ref{mpnpl}), it is shown
   that the sum of the positive terms is $R^n_l$ and
   the sum of the negative terms is $-R^n_l$. Therefore,
   the coefficient of $\vbf_{n+1-l-i} \ubf_i$
   is zero and the proposition is proved.
   \end{proof of} 

   To show lemma (\ref{quadraticR}), we shall use several lemmas. 
   The proof of these lemmas is very similar to the
   proofs given for the non commutative Hopf algebra
   of formal diffeomorphisms in section \ref{junediff}.
   Since we think that the results of section \ref{junediff}
   can be interesting independently of QED or renormalization
   theory, the proofs are given with all details in section \ref{junediff}
   and not here. We apologize to the reader for this sacrifice of the
   chronological order.

   The first lemma is a relation for $\Dp \vbf_n$
   which is deduced from the definition of the coproduct
   over photon trees.

   \begin{lemma}
   For $n\geq 1$, the photon coproduct of $\vbf_{n+1}$
   satisfies the following relation:
   \begin{eqnarray}
           \Dp \vbf_{n+1} &=& 
           \vbf_{n+1}\otimes 1 + (\Id\otimes V) \Big( \Dp \ubf_n
           -\sum_{k=0}^{n-1} \big( \vbf_{k+1}\otimes \| \big) 
           \pro \Dp \ubf_{n-k-1}\Big)
   \label{DeltaV}
   \end{eqnarray} 
   \end{lemma}

   \begin{proof}
   The property described by Eq.(\ref{property})
   together with the fact that $t^l\vee t^r=V(t^r)t^l$, yield
   a useful identity relating $\ubf_n$ and $\vbf_n$:
   \begin{eqnarray}
           \ubf_n = \sum_{k=0}^{n-1}\sum_{|t^l|=k} 
           \sum_{|t^r|=n-k-1} V(t^r) \pro t^l
           =\sum_{k=0}^{n-1}\vbf_{n-k} \pro \ubf_k.
   \label{formuledinversion}
   \end{eqnarray}

   From the recursive definition of $\Dp V(t)$ (cf. Eq.(\ref{photoncoproduct})) 
   and Eq.(\ref{property}) we find, for $t=t^l\vee t^r$, 
   \begin{eqnarray*}
           \Dp \vbf_{n+1} &=& 
           \sum_{|t|=n} \Dp V(t)=\sum_{|t|=n} V(t)\otimes 1 \\ 
           && +\sum_{|t|=n}(\Id\otimes V)\Dp t-\sum_{k=0}^{n-1}
           (\Id\otimes V) \Big( \sum_{|t^r|=k} V(t^r)\otimes \| \Big) 
           \pro \sum_{|t^l|=n-k-1} \Dp t^l \\
           &=& \vbf_{n+1}\otimes 1 +(\Id\otimes V)\Dp \ubf_n
           -\sum_{k=0}^{n-1} (\Id\otimes V) (\vbf_{k+1}\otimes \|) 
           \pro \Dp \ubf_{n-k-1}. 
   \end{eqnarray*} 
   This gives our lemma.
   \end{proof}

   Then, we need a recursive relation for $R^n_m$:

   \begin{lemma}
   The following recursive relations hold for any $n\geq 1$: 
   \begin{eqnarray}
           R^n_m(\vbf_*) &=& R^{n-1}_m(\vbf_*) + \vbf_m 
           + \sum_{l=1}^{m-1} \vbf_k R^{n-l}_{m-l}(\vbf_*), 
           \qquad m \geq 2 \label{recurxx} \\ 
           R^n_1(\vbf_*) &=& R^{n-1}_1(\vbf_*) + \vbf_1.  
   \end{eqnarray}
   \end{lemma}

   \begin{proof}
   From the obvious identity
   \begin{eqnarray*}
           P^{(k)}_m(\vbf_*) = 
           \sum_{l=1}^{m-k+1} \vbf_l \pro P^{(k-1)}_{m-l}(\vbf_*),
   \end{eqnarray*}
   we proceed exactly as for the proof of lemma \ref{recursiveQ}. 
   We start from the definition
   \begin{eqnarray*}
           R^n_m(\vbf_*) = 
           \sum_{k=1}^m \binom{n-m+k}{k} P^{(k)}_m(\vbf_*)
   \end{eqnarray*}
   and use Pascal's relation for binomial coefficients to obtain
   the lemma.
   \end{proof}

   From the recursive relation, we can find the generating function
   of $R^n_m$. 

   \begin{lemma} 
   The generating function 
   \begin{eqnarray}
           R(x,y) &=& \sum_{m=1}^\infty\sum_{n=0}^\infty 
           x^n y^{2m} R^{n+m}_m(\vbf_*) +\frac{1}{1-x}
   \label{defG22}
   \end{eqnarray}
   is the Green function of the hamiltonian 
   \begin{eqnarray}
   \label{Z3bf}
           \Zpbf(y) &=& 1-\sum_{n=1}^\infty y^{2n} \vbf_{n} 
           \in \Hp[[y^2]]
   \end{eqnarray}
   at energy $x$. Thus, the following resolvent equation is satisfied 
   \begin{eqnarray}
           R(x,y) = R(z,y) + (x-z) R(x,y) \pro R(z,y). 
   \label{resolventr}
   \end{eqnarray}
   \end{lemma}

   \begin{proof}
   The proof follows closely the proof of the corresponding lemma 
   (\ref{quadraticQ}) of section \ref{junediff}. 
   Using the recursive relation (\ref{recurxx}), we obtain a
   relation for $R(x,y)$ which leads to the equation
   \begin{eqnarray}
           R(x,y) &=& \big( \Zpbf(y)-x \big)^{-1}, 
   \label{Rgenfunc}
   \end{eqnarray}
   from which the lemma follows.
   \end{proof}

   It may be noticed that the explicit expression (\ref{defG22})
   can be interesting for a classical problem of quantum mechanics.
   Assume that a Hamiltonian $H_0$ is completely degenerate (i.e.
   all the eigenvectors of $H_0$  have the same eigenvalue,
   that we take equal to 1). 
   What happens to the system when
   the Hamiltonian is perturbed as $H(y)=H_0+V(y)$,
   where $V(y)$ can be expanded $V(y)=\sum_n \vbf_n y^{2n}$ ?
   The answer is given by the Green function
   $R(x,y)={(H(y)-x)}^{-1}$ which can be calculated from 
   Eq.(\ref{defG22}).

   The final lemma that we need is quite similar to lemma \ref{quadraticQ}.

   \begin{lemma}
   \label{quadraticR}
   The following quadratic relations hold for all $l,n \geq 0$:
   \begin{eqnarray}
           R^{m+n+l+1}_m &=& R^{m+n}_m + R^{m+l}_m 
           + \sum_{k=1}^{m-1} R^{m-k+n}_{m-k} \pro R^{k+l}_{k}, 
           \qquad m \geq 2 \label{mpnpl} \\ 
           R^{n+l}_1 &=& R^n_1 + R^l_1, \qquad m=1. 
   \end{eqnarray}
   \end{lemma}

   \begin{proof}
   We now use the resolvent equation in the form
   \begin{eqnarray*}
   \frac{R(x,y)}{1-z/x} &=& \frac{R(z,y)}{1-z/x} +x R(x,y) \pro R(z,y),
   \end{eqnarray*}
   and follow the proof of lemma \ref{quadraticQ}. 
   \end{proof}

   \begin{remark}
   As a consequence of the coassociativity of $\Dp$ we find 
   \begin{eqnarray*}
           \Dp R^n_m &=& R^n_m \otimes \| + \| \otimes R^n_m 
           +\sum_{k=1}^{m-1} R^n_k\otimes R^{n-k}_{m-k}.
   \end{eqnarray*}
   \end{remark}


   \subsection{Hopf structure for electrons at order $n$}

   Let $\xbf_n$ and $\ybf_n$ be the elements of $\He$ defined by 
   \begin{eqnarray*}
           \xbf_n &=& \sum_{|t|=n} t , \\ 
           \ybf_n &=& \sum_{|t|=n} \SP(t) , 
   \end{eqnarray*} 
   for any $n \geq 0$. 
   From the definition of the product antipode $\SP$ over trees, 
   it is straightforward to show that
   \begin{eqnarray}
           \ybf_n &=& \SP(\xbf_n) 
           = -\xbf_n - \sum_{m=1}^{n-1} \SP(\xbf_m) \xbf_{n-m}
           = -\xbf_n - \sum_{m=1}^{n-1} \xbf_m \SP(\xbf_{n-m}) 
           \label{antipode}\\
           &=& \sum_{k=1}^n (-1)^k P^{(k)}_n(\xbf_*). \nonumber
   \end{eqnarray}
   Notice that, on $\xbf_n$ and $\ybf_n$, we have $\SP^2=\Id$,
   because the pruning operator on the elements $\xbf_n$ and
   $\ybf_n$ is cocommutative \cite{BF}. Hence we also have
   \begin{eqnarray*}
           \xbf_n &=& \SP(\ybf_n) 
           = \sum_{k=1}^n (-1)^k P^{(k)}_n(\xbf_*). 
   \end{eqnarray*}

   The following proposition describes explicitly the
   coproduct over $\xbf_n$. 

   \begin{proposition}
   \label{Deltaxbf}
   The electron coproduct over trees induces the following coproduct
   on $\xbf_n\in \He$:
   \begin{eqnarray}
           \De \xbf_{n} &=& (\|,\xbf_{n}) \otimes \| + 1 \otimes \xbf_{n}
           + \sum_{m=1}^{n-1} T^{n}_m(\vbf_*,\xbf_*) \otimes \xbf_{n-m}, 
           \nonumber\\
           &=& (\|,\xbf_{n}) \otimes \| + 1 \otimes \xbf_{n}
           + \sum_{m=1}^{n-1} \Tbar^{n}_m(\vbf_*,\ybf_*) \otimes \xbf_{n-m}, 
           \label{deltaxbf}
   \end{eqnarray}
   where the polynomials on the left-hand side have mixed terms 
   in both components $\Hp$ and $\He$ of the algebra $\Hq$, 
   defined as 
   \begin{eqnarray}
           T^n_m(\vbf_*,\xbf_*) &=& (R^{n-1}_m(\vbf_*),\|) + 
           (\|,\xbf_m) +
           \sum_{k=1}^{m-1} (R^{n-k-1}_{m-k}(\vbf_*),\xbf_k), 
           \qquad 1 \leq m <n  \label{T(v,x)} \\ 
           T^m_m(\vbf_*,\xbf_*) &=& (\|,\xbf_m),  \nonumber   
   \end{eqnarray}
   and 
   \begin{eqnarray}
           \Tbar^{n}_m(\vbf_*,\ybf_*) &=& T^n_m(\vbf_*,\SP(\ybf_*)) 
           \nonumber \\ 
           &=& (R^{n-1}_m(\vbf_*),\|) 
           + \sum_{l=1}^m (-1)^l (\|,P_m^{(l)}(\ybf_*)) 
           + \sum_{k=1}^{m-1} \sum_{l=1}^k (-1)^l 
           (R^{n-k-1}_{m-k}(\vbf_*),P_k^{(l)}(\ybf_*)), 
           \label{Tbar(v,y)}.
   \end{eqnarray}
   \end{proposition}

   Before starting the proof, we give the simplest examples
   \begin{eqnarray*}
   \De \xbf_1 &=& 1\otimes \xbf_1 + (\|,\xbf_1)\otimes \| \\
   \De \xbf_2 &=& 1\otimes \xbf_2 + (\|,\xbf_2)\otimes \|
     + [(\vbf_1,\|)+(\|,\xbf_1)]\otimes \xbf_1\\
   \De \xbf_3 &=& 1\otimes \xbf_3 + (\|,\xbf_3)\otimes \|
     + [(2\vbf_1,\|)+(\|,\xbf_1)]\otimes \xbf_2 
     + [(\vbf_2,\|)+(\|,\xbf_2)+(\vbf_1,\xbf_1)]\otimes \xbf_1.
   \end{eqnarray*}

   \begin{proof}
   The proof will only be sketched, since it follows very closely
   the proof of the corresponding proposition (\ref{Deltaubf}) 
   for the photon case. 

   Equation (\ref{deltaxbf}) can be checked by hand up to
   order 3, by summing the known values of $\De t$ for
   $t\le 3$. We assume that it was proved up to order $n$,
   then we start from the recursive definition of the electron
   coproduct as
   \begin{eqnarray*}
           \De (t\vee s) & = & (\|,t\vee s)\otimes \|
           + (\Id\otimes \vee ) \Dp t \De s,  
   \end{eqnarray*}
   to obtain the coproduct of $\xbf_n$ as
   \begin{eqnarray*}
           \De \xbf_{n+1} & = & (\|,\xbf_{n+1})\otimes \|
           + \sum_{k=0}^n (\Id\otimes \vee ) \Dp \ubf_k \De \xbf_{n-k}.
   \end{eqnarray*}
   Then we expand $\Dp \ubf_k$, as given by Eq.~(\ref{Deltaubf}), 
   and $\De \xbf_{n-k}$ using Eq.~(\ref{T(v,x)}) to obtain
   \begin{eqnarray*}
           \De \xbf_{n+1} &=& (\|,\xbf_{n+1})\otimes \| 
           + 1 \otimes \xbf_{n+1} \\ 
           &&\hspace*{-10mm} + \sum_{m=1}^n \Big[ (R^n_m, \|) + 
           (\|,\xbf_m) + \sum_{k=1}^{m-1} R(^{n-k}_{m-k},\xbf_k) \Big]
           \otimes \vee (\ubf_{n-m}, \|) \\ 
           &&\hspace*{-10mm} + \sum_{m=1}^{n-1} \Big[ (\|,T^n_m) 
           + (\ubf_m,\|) + \sum_{k=1}^{m-1} (\|,\ubf_k T^{n-k}_{m-k}) \Big]
           \otimes \vee (\|, \xbf_{n-m}) \\ 
           &&\hspace*{-10mm} + \sum_{m=1}^{n-1} \sum_{k=m+1}^{n-1} 
           \Big[ (\|,T^k_m) + (R^{n-k+m}_m,\|) + \sum_{l=1}^{m-1}  
           (R^{n-k+l}_l,T^{k-l}_{m-l}) \Big]
           \otimes \vee (\ubf_{n-k},\xbf_{k-m}).
   \end{eqnarray*}

   From Eq.(\ref{mpnpl}) and the definition of $T^k_m$ 
   given by Eq.(\ref{T(v,x)}), it can be shown that
   \begin{eqnarray}
           (\|,T^k_m) + (R^{n-k+m}_m,\|) +
           \sum_{l=1}^{m-1}  (R^{n-k+l}_l,T^{k-l}_{m-l})
           &=& (\|,T^{n+1}_m) .\label{tnpl1m}
   \end{eqnarray}
   Moreover, the factor of $\otimes\vee (\ubf_{n-m},\|)$
   is Eq.(\ref{tnpl1m}) for $k=m$, and the factor of
   $\otimes \vee (\|, \xbf_{n-m})$ is Eq.(\ref{tnpl1m}) for $k=n$.
   Therefore,
   \begin{eqnarray*}
           \De \xbf_{n+1} & = & (\|,\xbf_{n+1})\otimes \| + 
           1 \otimes \xbf_{n+1} + (\|,T^{n+1}_n) \otimes \xbf_1 \\ 
           &&\hspace*{-8mm} + \sum_{m=1}^{n-1} (\|,T^{n+1}_m) \otimes 
           \Big( \sum_{k=m}^{n} \vee (\ubf_{n-k}, \xbf_{k-m}) \Big) \\
           &=& (\|,\xbf_{n+1}) \otimes \| + 1 \otimes \xbf_{n+1}
           + \sum_{m=1}^n (\|,T^{n+1}_m) \otimes \xbf_{n+1-m}.
   \end{eqnarray*}
   The last line was obtained because
   \begin{eqnarray*}
           \sum_{k=m}^{n} \vee (\ubf_{n-k}, \xbf_{k-m}) &=&
           \sum_{k=0}^{n-m} \vee (\ubf_{k}, \xbf_{n-m-k}) \\
           &=& \sum_{k=0}^{n-m} \sum_{|s|=k} \sum_{|t|=n-m-k} 
           \vee (s, t) \\ 
           &=& \sum_{k=0}^{n-m} \sum_{|s|=k} \sum_{|t|=n-m-k}
           s\vee t =\sum_{|t|=n-m+1} t = \xbf_{n-m+1}.
   \end{eqnarray*}
   \end{proof}

   From the expression for $T^n_m(\vbf_*,\xbf_*)$ given in 
   Eq.~(\ref{T(v,x)}), and that for the photon generating function
   $R(x,y)$ given in Eq.~(\ref{Rgenfunc}), it is straightforward to show 
   the following lemma. 

   \begin{lemma}
   The generating function for the terms $T^n_m(\vbf_*,\xbf_*)$ is 
   \begin{eqnarray*}
           T(x,y) &=& \sum_{m \geq 1}\sum_{n \geq 0} 
           x^n y^{2m} T^{m+n}_m(\vbf_*,\xbf_*) + \frac{1}{1-x} 
   \end{eqnarray*}
   and satisfies the equation 
   \begin{eqnarray*}
           T(x,y)  &=& (\Zpbf(y) {\big( \Zpbf(y)-x \big)}^{-1},\Zebf(y)),
   \end{eqnarray*}
   where $\Zpbf(y)$ was defined in (\ref{Z3bf}) and 
   \begin{eqnarray*}
           \Zebf(y)=1+\sum_{m \geq 1} y^{2n} \xbf_n \in \He[[y^2]].
   \end{eqnarray*}
   \end{lemma}

   We can now determine the action of the electron coproduct
   $\De$ on the elements $\ybf_n=\SP(\xbf_n)$. 

   \begin{proposition}
   \label{Deltaybf}
   The product antipode interwines with the electron coproduct
   according to the formula
   \begin{eqnarray}
           \De\SP = \big((\Id , \SP) \otimes\SP)\De.
   \end{eqnarray}
   In other words, the electron coproduct of $\ybf_n$
   is given by
   \begin{eqnarray}
           \De \ybf_{n} &=& (\|,\ybf_{n}) \otimes \| + 1 \otimes \ybf_{n}
           + \sum_{m=1}^{n-1} T^{n}_m(\vbf_*,\ybf_*) \otimes \ybf_{n-m}. 
           \label{deltaybf} 
   \end{eqnarray}
   \end{proposition}

   \begin{proof}
   The proof of Eq.(\ref{deltaybf}) is recursive and
   follows the now usual scheme. The electron coproduct
   is applied to Eq.(\ref{antipode}), and gives 
   \begin{eqnarray*}
   \De\ybf_n = -\De\xbf_n - \sum_{m=1}^{n-1} \De\ybf_m \De\xbf_{n-m}.
   \end{eqnarray*}
   We expand all coproducts on the right-hand side. 
   The expansion yields a term $1\otimes \ybf_n+ (\|,\ybf_n)\otimes\|$,
   plus a term in $\otimes \xbf_{n-m}$, a term in
   $\otimes \ybf_{n-m}$ and a term in
   $\otimes \ybf_{k-m}\xbf_{n-k}$. 
   From Eq.(\ref{antipode}) and  Eq.(\ref{mpnpl}), it turns
   out that the coefficient of
   $\xbf_{n-m}$ is $-(R^{n-1}_m,\|)$, the
   coefficient of $\ybf_{k-m}\xbf_{n-k}$ is
   also $-(R^{n-1}_m,\|)$, and the coefficient of
   $\ybf_{n-m}$ is 
   $(\|,\ybf_{m})+\sum_k (R^{n-k-1}_{m-k},\ybf_k)$.
   Using again Eq.(\ref{antipode}) we obtain
   \begin{eqnarray*}
           \De \ybf_{n} &=& (\|,\ybf_{n}) \otimes \| + 1 \otimes \ybf_{n} 
           + \sum_{m=1}^{n-1} \big[ (R^{n-1}_m,\|)+(\|,\ybf_{m})
           +\sum_{k=1}^{m-1} (R^{n-k-1}_{m-k},\ybf_k) \big] \otimes \ybf_{n-m}, 
   \end{eqnarray*}
   which is Eq.(\ref{deltaybf}) at order $n$.
   \end{proof}

   The elements $\xbf_n$ are used to calculate the electron propagator and
   the elements $\ybf_n$ are used to calculate the electron self-energy.

    
   \subsection{Renormalization at order $n$} 
   \label{ren-order}

   In ref.\cite{BF}, the renormalized amplitudes were defined 
   recursively on each tree. However, they are {\em not} a finite quantity. 
   As shown in Ref.\cite{BF}, 
   finite renormalized amplitudes are obtained by summing
   all trees of a given order. 
   Applying the results of section \ref{ren-tree}  
   and the formulas of propositions (\ref{Deltaubf}), 
   (\ref{Deltavbf}) and (\ref{Deltaxbf}), (\ref{Deltaybf}) 
   for the coproducts, we can describe the renormalization 
   of the propagators at each order of interaction. 

   The coefficients $Z_{3,n}$ and $Z_{2,n}$ 
   of the perturbative expansion defined by Eqs.~(\ref{Z2}) and 
   (\ref{Z3}), are now expanded over the elements $\vbf_*$ and $\ybf_*$ 
   as  
   \begin{eqnarray}
           Z_{3,n} &=& \sum_{|t|=n} \zeta_3(V(t)) = \zeta_3(\vbf_n),  
           \label{zeta3n} \\ 
           Z_{2,n} &=& \sum_{|t|=n} \zeta_2(\SP(t)) = \zeta_2(\ybf_n),  
           \label{zeta2n} 
   \end{eqnarray}
   where the multiplicative maps $\zeta_3: \Hp \longrightarrow \C$ and 
   $\zeta_2: \He \longrightarrow \C$ were defined in section \ref{ren-tree}. 

   \begin{theorem}
   The photon propagator at each order $n$ of interaction 
   is renormalized as 
   \begin{eqnarray*}
           \bar{D}_n(q) &=& D_n(q) 
           + \sum_{m=1}^{n} R_m^n(Z_{3,*}) D_{n-m}(q) , 
   \end{eqnarray*} 
   where the noncommutative polynomial $R_m^n$ was defined in 
   (\ref{R(v)}). Similarly, the electron propagator at each 
   order $n$ of interaction is renormalized as 
   \begin{eqnarray*}
           \bar{S}_n(q) &=& S_n(q) 
           + \sum_{m=1}^{n} \Tbar_m^n(Z_{3,*},Z_{2,*}) S_{n-m}(q) , 
   \end{eqnarray*}
   where the noncommutative polynomial $T_m^n$ was defined 
   in (\ref{Tbar(v,y)}). 
   \end{theorem}

   \begin{proof}
   Start with the definitions 
   \begin{eqnarray*}
           D_n(q) = \phip(\ubf_n) , &\qquad&  
           \bar{D}_n(q) = \bar\phip(\ubf_n) , \\ 
           S_n(q) = \phie(\xbf_n) , &\qquad&  
           \bar{S}_n(q) = \bar\phie(\xbf_n) , 
   \end{eqnarray*}
   with renormalized amplitudes 
   \begin{eqnarray*}
           \bar\phip &=& \zeta_3 \star \phip , \\
           \bar\phie &=& (\zeta_3 , \zeta_2) \star \phie . 
   \end{eqnarray*}
   Then, for the photon propagator we have: 
   \begin{eqnarray*}
           \bar{D}_n(q) &=& \bar\phip(\ubf_n) \\ 
           &=& (\zeta_3 \star \phip)(\ubf_n) \\ 
           &=& \zeta_3(\ubf_n)\phip(\|) + \zeta_3(\|)\phip(\ubf_n) 
           + \sum_{m=1}^{n-1} R_m^n(\zeta_3(\vbf_*))\phip(\ubf_{n-m}) \\ 
           &=& \zeta_3(R_n^n(\vbf_*))\phip(\|) + \zeta_3(\|)\phip(\ubf_n) 
           + \sum_{m=1}^{n-1} R_m^n(\zeta_3(\vbf_*))\phip(\ubf_{n-m}) \\ 
           &=& D_n(q) + \sum_{m=1}^{n} R_m^n(Z_{3,*})D_{n-m}(q) . 
   \end{eqnarray*} 
   A similar computation yields the renormalization 
   of the electron propagator. 
   \end{proof}

   \begin{theorem}
   The vacuum polarization at each order $n$ of interaction 
   is renormalized as 
   \begin{eqnarray*}
           \bar\Pi_n(q) &=& \Pi_n(q) + Z_{3,n} 
           + \sum_{m=1}^{n-2} R_m^{n-2}(Z_{3,*}) \Pi_{n-m}(q) , 
   \end{eqnarray*} 
   where the noncommutative polynomial $R_m^n$ was defined in 
   (\ref{R(v)}). Similarly, the self-energy at each 
   order $n$ of interaction is renormalized as 
   \begin{eqnarray*}
           \bar\Sigma_n(q) &=& \Sigma_n(q) - Z_{2,n} 
           + \sum_{m=1}^{n-1} 
           T_m^n\big( Z_{3,*},Z_{2,*} \big) 
           \Sigma_{n-m}(q) , 
   \end{eqnarray*}
   where the noncommutative polynomial $T_m^n$ was defined 
   in (\ref{T(v,x)}). 
   \end{theorem}

   \begin{proof}
   Start with the definitions 
   \begin{eqnarray*}
           \Pi_n(q) = \phip(\vbf_n) , &\qquad&  
           \bar\Pi_n(q) = \bar\phip(\vbf_n) , \\ 
           \Sigma_n(q) = - \phie(\ybf_n) , &\qquad&  
           \bar\Sigma_n(q) = - \bar\phie(\ybf_n) , 
   \end{eqnarray*}
   with renormalized amplitudes as above. 
   Then, for the vacuum polarization we have: 
   \begin{eqnarray*}
           \bar\Pi_n(q) &=& \bar\phip(\vbf_n) \\ 
           &=& (\zeta_3 \star \phip)(\vbf_n)  \\ 
           &=& \zeta_3(\vbf_n)\phip(\|) +\zeta_3(\|)\phip(\vbf_n) 
           +\sum_{m=1}^{n-2} R_m^{n-2}(\zeta_3(\vbf_*))\phip(\vbf_{n-m}) \\ 
           &=& Z_{3,n} + \Pi_n(q)  
           +\sum_{m=1}^{n-2} R_m^{n-2}(Z_{3,*}) \Pi_{n-m}(q).   
   \end{eqnarray*}
   A similar computation gives the formula for the 
   renormalization of the self-energy. 
   \end{proof}

   %
   %
    
   \section{Noncommutative Hopf algebra of formal diffeomorphism}
   \label{junediff} 

   It is well known that the composition of series with 
   non-commutative coefficients is not an associative operation. 
   However, in quantum field theory, the compositions of 
   matrix-valued series are necessary, 
   and the associativity of the operations is usually imposed by  
   replacing the usual composition with a ``normal ordering'',  
   which re-establishes the position of the misplaced variables. 
   Such an operation is not binary, and needs to be defined 
   ``a priori'' on each set of $n$ variables. 

   In this section, we define a dual operation to 
   the normal orderning, which surprisingly 
   is a binary co-operation, that is, a coprodut. 

   In fact, it turns out that this coproduct is part of the structure 
   of a non-commutative nor co-commutative Hopf algebra 
   which plays the role of ``dual'' to the space of formal diffeomorphisms
   $\fbf(x)=x+\sum_n x^{n+1} \abf_n$ where $\abf_n$ belongs to a unital
   associative (non-commutative) algebra and $x$ commutes with all $\abf_n$'s. 
   That is, its abelianization is the commutative Hopf algebra of functions 
   on formal diffeomorphisms of $\C$. 

   As the group of formal diffeomorphisms acts on the set of 
   formal power series, we finally find a ``dual'' non-commutative 
   version given by a coaction of the Hopf algebra on the free 
   associative algebra on infinitely many variables. 
   This coaction will be used to compose non-commutative formal series 
   and non-commutative formal diffeomorphisms in normal ordering. 


   \subsection{The Hopf algebra}
   \label{Hdiff} 

   \begin{definition}
   Let $\Hd=\C\< \abf_1,\abf_2,\abf_3 ... \>$ be the algebra 
   of noncommutative polynomials on infinitely many variables 
   $\abf_n$, for $n \geq 1$, with unit $1$. 
   For any $m \geq 1$ and any $k \leq m$, let 
   \begin{eqnarray}
   \label{polynomial}
       P_m^{(k)}(\abf_*) &=& \sum_{ {j_1 > 0,\dots, j_k > 0}\atop
       {j_1+\cdots+j_k=m}} \abf_{j_1} \cdots \abf_{j_k} 
   \end{eqnarray}
   be the symmetric (noncommutative) polynomial
   of degree $m$ in $k$ variables $\abf_*$. 
   For instance, 
   \begin{eqnarray*}
           && P_1^{(1)}(\abf_*) = \abf_1, \\
           && P_2^{(1)}(\abf_*) = \abf_2, 
           \qquad P_2^{(2)}(\abf_*) = \abf_1^2, \\ 
           && P_3^{(1)}(\abf_*) = \abf_3, 
           \qquad P_3^{(2)}(\abf_*) = \abf_2 \abf_1 + \abf_1 \abf_2, 
           \qquad P_3^{(3)}(\abf_*) = \abf_1^3 . 
   \end{eqnarray*}
   \end{definition}

   \begin{proposition}
   \label{Hd=Hopf}
   The algebra $\Hd$ is a Hopf algebra with coproduct defined 
   on the generators as 
   \begin{eqnarray}
   \label{deltaabf}
       \Dd \abf_n &=& 1 \otimes \abf_n + \abf_n \otimes 1
       +\sum_{m=1}^{n-1} Q^{n-m}_m(\abf_*) \otimes \abf_{n-m},  
   \end{eqnarray}
   where we set 
   \begin{eqnarray}
   \label{polQ}
      Q^{l}_m(\abf_*) = 
      \sum_{k=1}^{m} \binom{l+1}{k} P^{(k)}_m(\abf_*), 
   \qquad l \geq 0, \quad m \geq 1, 
   \end{eqnarray}
   and where we assume that $\binom{l+1}{k}$ is zero for $k>l+1$. 
   The counit is given by $\epsilon(\abf_n)=0$ and $\epsilon(1)=1$, and 
   the antipode is given by the standard recursive formula 
   \begin{eqnarray*}
           \Sd \abf_n &=& -\abf_n - \sum_{m=1}^{n-1} 
           \Sd \big(Q^{n-m}_m(\abf_*)\big) \abf_{n-m} \\ 
           &=& -\abf_n - \sum_{m=1}^{n-1} 
           Q^{n-m}_m(\abf_*) \Sd \abf_{n-m}.  
   \end{eqnarray*}
   \end{proposition}

   Here are some examples of coproducts: 
   \begin{eqnarray*}
           \Dd \abf_1 &=& \abf_1 \otimes 1 + 1 \otimes \abf_1, \\ 
           \Dd \abf_2 &=& \abf_2 \otimes 1 + 1 \otimes \abf_2 
           + 2 \abf_1 \otimes \abf_1 , \\ 
           \Dd \abf_3 &=& \abf_3 \otimes 1 + 1 \otimes \abf_3 
           + 3 \abf_1 \otimes \abf_2 + 2 \abf_2 \otimes \abf_1 
           + \abf_1^2 \otimes \abf_1 , 
   \end{eqnarray*}
   and some examples of antipodes: 
   \begin{eqnarray*}
           \Sd \abf_1 &=& -\abf_1 , \\ 
           \Sd \abf_2 &=& -\abf_2 + 2 \abf_1^2 , \\ 
           \Sd \abf_3 &=& -\abf_3 + 3 \abf_1 \abf_2 
           + 2 \abf_2 \abf_1 - 5\abf_1^3. 
   \end{eqnarray*}
   The coproduct is clearly not cocommutative, 
   and the square of the antipode is equal to the identity 
   only for $\abf_1$ and $\abf_2$. 

   The proof of the proposition is given after a few remarks. 

   \begin{corollary}
   Let $\Ho=\Hd/(\abf_{2n+1},n\geq 0)$ be the quotient algebra of $\Hd$ 
   obtained by setting all the odd coefficients to zero. 
   Then $\Ho$ is a Hopf algebra with coproduct 
   \begin{eqnarray}
   \label{deltaabfodd}
           \Do \abf_{2n} &=& 1 \otimes \abf_{2n} + \abf_{2n} \otimes 1
           +\sum_{m=1}^{n-1} \bar{Q}^{n-m}_m(\abf_{2*}) \otimes
           \abf_{2(n-m)},  
   \end{eqnarray}
   where we set 
   \begin{eqnarray}
   \label{polQbar}
           \bar{Q}^{l}_m(\abf_{2*}) = Q^{2l}_m(\abf_{2*}) = 
           \sum_{k=1}^{m} \binom{2l+1}{k} \bar P^{(k)}_m(\abf_{2*}), 
           \qquad l \geq 0, \quad m \geq 1,  
   \end{eqnarray}
   and
   \begin{eqnarray*}
           \bar P^{(k)}_m(\abf_{2*}) &=& \sum_{ {j_1 > 0,\dots, j_k >
   0}\atop
       {j_1+\cdots+j_k=m}} \abf_{2j_1} \cdots \abf_{2j_k}.
   \end{eqnarray*}
   \end{corollary}

   \begin{remark}
   \label{Hdif-ab}
   The coproduct $\Dd$ on $\Hd$ induces a Hopf structure 
   on the commutative algebra 
   $\Hd_{ab} = \Hd / ([\Hd,\Hd]) \cong \C[a_1,a_2,a_3,...]$. 
   Since on commutative variables the polynomial $P_m^{(k)}(a_*)$ becomes 
   \begin{eqnarray*}
           P_m^{(k)}(a_*) &=& 
           \sum_{{p_1+2p_2+\dots +mp_m=m}\atop{p_1+p_2+\dots +p_m=k}}
           \frac{k!}{p_1!\cdots p_m!} 
           \ a_1^{p_1}\cdots a_m^{p_m}\footnote{
   Remark that these combinatorial factors are well known
   from Planck's quantum theory of blackbody radiation.
   }, 
   \end{eqnarray*}
   where the sum runs over $p_1,\dots,p_m\geq 0$, 
   the coproduct on commutative variables becomes 
   \begin{eqnarray*}
           \Dd a_n &=& 1 \otimes a_n + a_n \otimes 1 \nonumber \\ 
           && +\sum_{m=1}^{n-1} \sum_{k=1}^{m}
           \frac{(n-m+1)!}{(n-m+1-k)!} 
           \sum_{{p_1+\dots +mp_m=m}\atop{p_1+p_2+\dots +p_m=k}}
           \frac{1}{p_1!\cdots p_m!} 
           \ a_1^{p_1}\cdots a_m^{p_m} \otimes a_{n-m}. 
   \end{eqnarray*}
   \end{remark}

   \begin{remark}
   \label{Gdif}
   The commutative Hopf algebra $\Hd_{ab}$ is the dual
   Hopf algebra of the group 
   \begin{eqnarray*}
           \Gd &=& \big\{ f: \C \longrightarrow \C, \ 
           f(x)= x + \sum_{n \geq 1} x^{n+1} a_n(f), a_n \in \C \big\} 
   \end{eqnarray*}
   of formal diffeomorphisms $f$ of $\C$ such that $f(0)=0$ and
   $f'(0)=1$. 
   Hence $\Hd$ can be considered as a non commutative analogue of (the
   commutative part of) the Connes-Moscovici Hopf algebra used in
   \cite{ConnesM}, \cite{Connes}, \cite{CKI}. 

   To see this, we show that the generators $a_n$ of $\Hd_{ab}$ behave as 
   the functions on $\Gd$ defined as 
   \begin{eqnarray*} 
            \< a_n,f\> = a_n(f) &=& \frac{1}{(n+1)!} \frac{d^{n+1}
   f(0)}{dx^{n+1}} . 
   \end{eqnarray*}
   That is, we show that the composition $f \circ g$ of two series 
   $f,g \in \Gd$ gives rise to the coproduct $\Dd$ by the duality 
   $\< \Dd a_n, g\otimes f\>=\< a_n, f\circ g\>$, where the pairing 
   on the left hand-side is defined as 
   $\<a_n \otimes a_m, g \otimes f \> = a_n(g) a_m(f)$. 

   Let $f$ and $g$ be two series in $\Gd$, then 
   \begin{eqnarray*}
           f\circ g(x) &=& g(x) + \sum_{n=1}^\infty g(x)^{n+1} a_n(f)\\
           &=&x+\sum_{n=1}^\infty x^{n+1} a_n(g)+
           \sum_{n=1}^\infty x^{n+1}
           (1+\sum_{j=1}^\infty x^j a_j(g))^{n+1} a_n(f)\\
           &=& x+\sum_{n=1}^\infty x^{n+1} a_n(g)+
           \sum_{n=1}^\infty x^{n+1}\sum_{k=0}^{n+1} \binom{n+1}{k}
           (\sum_{j=1}^\infty x^j a_j(g))^{k} a_n(f)\\
           &=& x+\sum_{n=1}^\infty x^{n+1} (a_n(g)+a_n(f))+
           \sum_{n=1}^\infty \sum_{k=1}^{n+1} \sum_{m=k}^{\infty} x^{n+m+1} 
           \binom{n+1}{k}P^{(k)}_m(a_*(g)) a_n(f).
   \end{eqnarray*}
   If we identify the terms of order $x^{n+1}$ we obtain precisely 
   \begin{eqnarray*}
           a_n(f\circ g) &=& a_n(f) + a_n(g)+
           \sum_{m=1}^{n-1} \sum_{k=1}^{m} \binom{n-m+1}{k} 
           P^{(k)}_m(a_*(g))a_{n-m}(f).
   \end{eqnarray*}
   \end{remark}

   \begin{remark}
   Similarly to (\ref{Hdif-ab}) and (\ref{Gdif}), the abelianized Hopf algebra 
   $\Ho_{ab} = \Hd / ([\Hd,\Hd]) \cong \C[a_2,a_4,a_6,...]$,  
   with the coproduct induced by $\Do$, is dual to the group 
   \begin{eqnarray*}
           \Go &=& \big\{ f: \C \longrightarrow \C, \ 
           f(x)= x + \sum_{n \geq 1} x^{2n+1} a_{2n}(f), a_{2n} \in \C \big\} 
   \end{eqnarray*}
   of ``odd'' formal diffeomorphisms considered in \cite{CKII}. 
   \end{remark}

   \begin{proof of}{Hd=Hopf}
   The only non trivial statement is the coassociativity of the 
   coproduct $\Dd$. Applying the definition (\ref{deltaabf}), we obtain 
   \begin{eqnarray*}
           (\Dd \otimes 1) \Dd \abf_n - (1 \otimes \Dd) \Dd \abf_n &=& \\ 
           && \hspace*{-4cm}=
           \sum_{m=1}^{n-1} \Big(\Dd Q^{n-m}_m -1\otimes Q^{n-m}_m 
           -Q^{n-m}_m \otimes 1
           -\sum_{k=1}^{m-1} Q^{n-k}_k \otimes Q^{n-m}_{m-k} \Big) 
           \otimes \abf_{n-m}.  
   \end{eqnarray*}
   The right-hand side is zero because of lemma (\ref{DeltaQ}),
   hence the coproduct is coassociative. 
   \end{proof of} 

   We prove lemma (\ref{DeltaQ}) in several steps, starting 
   with a recursion formula for $Q^n_m$. 
   First, let us give some examples of the polynomials $Q^n_m$.
   We have $Q^0_m=\abf_m$ for any $m \geq 1$, and 
   \begin{eqnarray*}
      Q^l_1 &=& (l+1) \abf_1,\\
      Q^l_2 &=& (l+1) \abf_2 + \frac{l(l+1)}{2} \abf_1^2,\\
      Q^l_3 &=& (l+1) \abf_3 + \frac{l(l+1)}{2} 
      (\abf_1\abf_2+\abf_2\abf_1) + 
      \frac{(l+1)l(l-1)}{6} \abf_1^3 
   \end{eqnarray*}
   for any $l \geq 1$. 

   \begin{lemma}
   \label{recursiveQ}
   The following recursive identities are satisfied: 
   \begin{eqnarray}
           Q^n_m &=& Q^{n-1}_m + \abf_m 
           + \sum_{l=1}^{m-1} \abf_l Q^{n-1}_{m-l}, 
           \qquad n \geq 1, m \geq 2 \label{lemma1} \\ 
           Q^n_1 &=& Q^{n-1}_1 + \abf_1, \qquad n \geq 1, m=1 
           \nonumber .  
   \end{eqnarray}
   \end{lemma}
    
   \begin{proof} 
   From the definition of $P^{(k)}_m$ it is clear that,
   for $k>1$,
   \begin{eqnarray*}
           P^{(k)}_m &=& \sum_{l=1}^{m+1-k} \abf_l P^{(k-1)}_{m-l}.
   \end{eqnarray*}
   Therefore,
   \begin{eqnarray*}
   Q^n_m &=& \sum_{k=1}^{m} \binom{n+1}{k} P^{(k)}_m
          =  \sum_{k=1}^{m} \binom{n}{k} P^{(k)}_m
            +\sum_{k=1}^{m} \binom{n}{k-1} P^{(k)}_m\\
         &=& Q^{n-1}_m + \abf_m + \sum_{k=1}^{m-1} \binom{n}{k}
   P^{(k+1)}_m\\
         &=& Q^{n-1}_m + \abf_m + \sum_{k=1}^{m-1} \sum_{l=1}^{m-k}
                                   \binom{n}{k} \abf_l P^{(k)}_{m-l}\\
         &=& Q^{n-1}_m + \abf_m + \sum_{l=1}^{m-1} \abf_l Q^{n-1}_{m-l}.
   \end{eqnarray*}
   The identity $Q^n_1 = Q^{n-1}_1 + \abf_1$ comes from the definition
   of $Q^n_1=(n+1)\abf_1$.
   \end{proof}
    
   \begin{lemma} 
   \label{genQ(x,y)}
   The generating function 
   \begin{eqnarray}
           Q(x,y) &=& \sum_{m=1}^\infty\sum_{n=0}^\infty 
           x^n y^m Q^n_m(\abf_*) +\frac{1}{1-x}
   \label{defG}
   \end{eqnarray}
   is the Green function of the Hamiltonian $y/\fbf(y)$
   at $x$, where 
   \begin{eqnarray*}
           \fbf(y) = y + \sum_{n=1}^\infty y^{n+1} \abf_n \in \Hd[[y]].
   \end{eqnarray*}
   That is, the generating function is 
   \begin{eqnarray}
   \label{GenGreen}
           Q(x,y) &=& (y - x \fbf(y))^{-1} \fbf(y) 
           = (y \fbf(y)^{-1} - x)^{-1} ,  
   \end{eqnarray}
   and satisfies the following resolvent equation
   \begin{eqnarray}
   \label{resolvent}
           Q(x,y) = Q(z,y) + (x-z) Q(x,y) Q(z,y). 
   \end{eqnarray}
   \end{lemma}
    
   \begin{proof} 
   Using lemma \ref{recursiveQ}, we can write $Q(x,y)$ as
   \begin{eqnarray*}
           Q(x,y) &=& \frac{1}{1-x} 
           + \sum_{m=1}^\infty\sum_{n=1}^\infty x^n y^m Q^{n-1}_m 
           + \sum_{m=1}^\infty\sum_{n=0}^\infty y^m x^n \abf_m \\
           && + \sum_{m=2}^\infty\sum_{n=1}^\infty \sum_{l=1}^{m-1} 
           x^n y^m \abf_l Q^{n-1}_{m-l}.
   \end{eqnarray*}
   Setting $\fbf(y) = y + \sum_{n=1}^\infty y^{n+1} \abf_n$, 
   we can rewrite the above equation as
   \begin{eqnarray*}
           Q(x,y) &=& \frac{1}{1-x} 
           +x\big(Q(x,y)-\frac{1}{1-x}\big)
           + \big(\frac{\fbf(y)}{y}-1\big)\frac{1}{1-x} \\
           && +x\big(\frac{\fbf(y)}{y}-1\big)\big(Q(x,y)-\frac{1}{1-x}\big),  
   \end{eqnarray*}
   and obtain the following identity 
   \begin{eqnarray}
   \label{GGreen}
           [y - x \fbf(y)] Q(x,y) &=& \fbf(y), 
   \end{eqnarray}
   from which follows Eq.~(\ref{GenGreen}). 

   Now, multiply the left hand-side of Eq.~(\ref{resolvent}) 
   by $[y - x \fbf(y)]$ from the left, and using Eq.~(\ref{GGreen}), 
   we obtain: 
   \begin{eqnarray*}
           && [y - x \fbf(y)] [Q(z,y) + (x-z) Q(x,y) Q(z,y)] = \\  
           && \hspace{2cm} = [y - x \fbf(y)] Q(z,y) + (x-z) [y - x \fbf(y)] Q(x,y)
   Q(z,y) \\ 
           && \hspace{2cm} = [y - z \fbf(y)] Q(z,y) + (z-x) \fbf(y) Q(z,y) 
           + (x-z) \fbf(y) Q(z,y) \\ 
           && \hspace{2cm} = \fbf(y) = [y - x \fbf(y)] Q(x,y).
   \end{eqnarray*} 
   This proves the resolvent equation (\ref{resolvent}). 
   \end{proof}
    
   \begin{lemma}
   \label{quadraticQ}
   The following quadratic equations are satisfied for all $l,n\geq 0$:  
   \begin{eqnarray}
           Q^{n+l+1}_m &=& Q^n_m + Q^l_m 
           + \sum_{k=1}^{m-1} Q^l_k Q^n_{m-k}, 
           \qquad m \geq 2, \label{lemma3} \\ 
           Q^{n+l+1}_1 &=& Q^n_1 + Q^l_1. \nonumber 
   \end{eqnarray}
   \end{lemma}
    
   \begin{proof} 
   To prove this equations, we rewrite the resolvent equation 
   (\ref{resolvent}) as
   \begin{eqnarray*}
           \frac{Q(z,y)}{1-x/z} &=& \frac{Q(x,y)}{1-x/z} + z Q(x,y) Q(z,y).
   \end{eqnarray*}
   On the left-hand side we have 
   \begin{eqnarray*}
           \frac{Q(z,y)}{1-x/z} &=& \frac{z}{(z-x)(1-z)}
           +\sum_{m=1}^\infty\sum_{n=0}^\infty \sum_{l=0}^{\infty}
           y^m x^l z^{n-l} Q^n_m , 
   \end{eqnarray*}
   while on the right-hand side the first term is 
   \begin{eqnarray*}
           \frac{Q(x,y)}{1-x/z} &=& \frac{z}{(z-x)(1-x)}
           +\sum_{m=1}^\infty\sum_{n=0}^\infty \sum_{l=0}^{\infty}
           y^m x^{n+l} z^{-l} Q^n_m , 
   \end{eqnarray*}
   and from definition (\ref{defG}) we obtain the second term 
   \begin{eqnarray*}
           zQ(x,y)Q(z,y) &=& 
           \sum_{m=2}^\infty \sum_{n=0}^\infty \sum_{l=0}^{\infty} 
           \Big(\sum_{k=1}^{m-1} y^m x^l z^{n+1} Q^l_k Q^n_{m-k}\Big) \\ 
           && +\sum_{m=1}^\infty\sum_{n=0}^\infty \sum_{l=0}^{\infty}
           y^m x^l z^{n+1} \Big(Q^l_m+Q^n_m\Big) +\frac{z}{(1-x)(1-z)} . 
   \end{eqnarray*}
    The fractions cancel out and the coefficient of 
   $y^m x^l z^{n+1}$ ($m>0$, $l\ge0$, $n\ge0$) gives Eq.(\ref{lemma3}).
   \end{proof}
    
   \begin{lemma}
   \label{DeltaQ}
   The coproduct of $Q^n_m(\abf_*)$ is given by 
   \begin{eqnarray}
           \Dd Q^{n}_{m}(\abf_*) &=& 1 \otimes Q^{n}_{m} 
           + Q^{n}_{m} \otimes 1 
           + \sum_{k=1}^{m-1} Q^{n+m-k}_{k} \otimes Q^{n}_{m-k}, 
           \qquad m \geq 2, n \geq 0 \label{deltaQ} \\ 
           \Dd Q^{n}_{1}(\abf_*) &=& 1 \otimes Q^{n}_{1} 
           + Q^{n}_{1} \otimes 1 , \qquad m=1, n \geq 0 
           \nonumber. 
   \end{eqnarray}
   \end{lemma}
    
   \begin{proof}
   The proof is given by a recursion argument.
   From the definition of the coproduct given in Eq.~(\ref{deltaabf}),
   the identity (\ref{deltaQ}) is true for $n=0$.
   Now assume that relation (\ref{deltaQ}) holds
   up to order $n$.
   Then, using Eq.(\ref{lemma1}), we have 
   \begin{eqnarray*}
           \Dd Q^{n+1}_{m} = \Dd Q^{n}_{m} +\Dd \abf_m
            + \sum_{l=1}^{m-1} \Dd \abf_l \Dd Q^{n}_{m-l}.
   \end{eqnarray*}
   Expanding these coproducts we obtain
   \begin{eqnarray*}
           \Dd Q^{n+1}_{m} &=& 1 \otimes Q^{n+1}_{m} 
           + Q^{n+1}_{m} \otimes 1
           +\sum_{k=1}^{m-1} \Big(Q^{m-k}_{k}+Q^{n}_{k}+
           \sum_{p=1}^{k-1} Q^{m-k}_{p}Q^{n}_{k-p}\Big)
           \otimes \abf_{m-k} \\ 
           && + \sum_{k=1}^{m-1} \Big(Q^{n+m-k}_{k} + \abf_k +
           \sum_{l=1}^{k-1} \abf_l Q^{n+m-k}_{k-l} \Big)
           \otimes Q^n_{m-k} \\ 
           && + \sum_{k=1}^{m-1}\sum_{l=1}^{m-k-1}
           \Big(Q^{n+m-k-l}_{k}+Q^{l}_{k}+
           \sum_{p=1}^{k-1} Q^{l}_{p}Q^{n+m-k-l}_{k-p}\Big)
           \otimes \abf_{l}Q^{n}_{m-k-l}.
   \end{eqnarray*}
   From Eq.(\ref{lemma3}), we see that the each sum in parenthesis is
   equal to $Q^{n+1+m-k}_k$. Therefore, using Eq.(\ref{lemma1}),
   we obtain Eq.(\ref{deltaQ}) for $n+1$. 
   \end{proof}


   \subsection{Coaction on the dual of series}
   \label{coaction} 

   \begin{proposition}
   \label{Ad=coaction}
   Let $\Ad=\C\< \bbf_1,\bbf_2,\bbf_3 ... \>$ denote the algebra 
   of noncommutative polynomials on infinitely many variables 
   $\bbf_n$, for $n \geq 1$, with unit $1$. 
   Then, the multiplicative map 
   $\ddd:\Ad \longrightarrow \Hd \otimes \Ad$ defined 
   on the generators by 
   \begin{eqnarray}
   \label{deltabbf}
       \ddd \bbf_n &=& 1 \otimes \bbf_n 
       +\sum_{m=1}^{n-1} Q^{n-m-1}_m(\abf_*) \otimes \bbf_{n-m},  
   \end{eqnarray}
   where the polynomials $Q^{l}_m(\abf_*)$ where defined in 
   (\ref{polQ}), gives a left coaction of $\Hd$ on $\Ad$. 
   \end{proposition}

   Here are some examples of coactions: 
   \begin{eqnarray*}
           \ddd \bbf_1 &=& 1 \otimes \bbf_1, \\ 
           \ddd \bbf_2 &=& 1 \otimes \bbf_2 + \abf_1 \otimes \bbf_1 , \\ 
           \ddd \bbf_3 &=& 1 \otimes \bbf_3 + 2 \abf_1 \otimes \bbf_2 
           + \abf_2 \otimes \bbf_1 .  
   \end{eqnarray*}

   \begin{proof}
   We want to show that 
   \begin{eqnarray*} 
           (\Id \otimes \ddd) \ddd &=& (\Dd \otimes \Id) \ddd. 
   \end{eqnarray*}
   On the left-hand side, we have
   \begin{eqnarray*} 
           (\Id \otimes \ddd) \ddd \bbf_n 
           &=& 1 \otimes \ddd \bbf_n 
           + \sum_{m=1}^{n-1} Q^{n-m-1}_m(\abf_*) \otimes \ddd \bbf_{n-m} \\ 
           &=& 1 \otimes 1 \otimes \bbf_n 
           +\sum_{m=1}^{n-1} 1 \otimes Q^{n-m-1}_m(\abf_*) \otimes \bbf_{n-m}
           +\sum_{m=1}^{n-1} Q^{n-m-1}_m(\abf_*) \otimes 1 \otimes \bbf_{n-m} \\
           && +\sum_{p=1}^{n-1} \sum_{q=1}^{n-p-1} Q^{n-p-1}_p(\abf_*) 
           \otimes Q^{n-p-q-1}_q(\abf_*) \otimes \bbf_{n-p} \\ 
           &=& 1 \otimes 1 \otimes \bbf_n 
           +\sum_{m=1}^{n-1} \big[ 1 \otimes Q^{n-m-1}_m(\abf_*) 
           + Q^{n-m-1}_m(\abf_*) \otimes 1 \\ 
           && + \sum_{p=1}^{m-1} 
           Q^{n-p-1}_p(\abf_*) \otimes Q^{n-m-1}_{m-p}(\abf_*) \big] 
           \otimes \bbf_{n-m}. 
   \end{eqnarray*}
   Because of lemma (\ref{DeltaQ}), this term coincide with the term 
   on the right-hand side, namely 
   \begin{eqnarray*} 
           (\Dd \otimes \Id) \ddd \bbf_n 
           &=& 1 \otimes 1 \otimes \bbf_n 
           + \sum_{m=1}^{n-1} \Dd Q^{n-m-1}_m(\abf_*) \otimes \bbf_{n-m}.  
   \end{eqnarray*}
   \end{proof}

   \begin{remark}
   The abelianization $\Ad_{ab}\cong \C[b_1, b_2,\ldots]$
   is the set of functions on the formal series
   \begin{eqnarray*}
   f(x) &=& 1+\sum_n x^n b_n(f),
   \end{eqnarray*}
   where the $b_n$ are functions on the series
   defined as
   \begin{eqnarray*}
   b_n(f) &=& \frac{1}{n!} \frac{\dd^n f(0)}{\dd x^n}.
   \end{eqnarray*}

   Then, the coaction  of $\Hd_{ab}$ on $\Ad_{ab}$
   induced by the 
   coaction $\ddd$ of $\Hd$ on $\Ad$,
   is dual to the natural right action
   of $\Gd$ on the set of series starting with $1$.

   In fact, if we start from two formal series 
   $f(x)=1+\sum_n x^n b_n(f)$
   and $g(x)=x+\sum_n x^{n+1} a_n(g)$
   with commutative entries, the 
   calculation of the composition 
   is very similar to that of \ref{Gdif} 
   and yields 
   \begin{eqnarray*}
           f(g(x)) &=& 1+ 
           \sum_{n=1}^\infty {\big(g(x)\big)}^{n} b_n(f) \\ 
           &=& 1+\sum_{n=1}^\infty x^n \Big( a_n(f) 
           + \sum_{m=1}^{n-1} Q^{n-m-1}_m(a_*(g)) b_{n-m}(f)\Big)\\
           &=& 1+\sum_{n=1}^\infty x^n \langle g \otimes f, 1\otimes b_n+
           \sum_{m=1}^{n-1} Q^{n-m-1}_m(a_*)\otimes b_{n-m}\rangle.
   \end{eqnarray*}
   \end{remark}


   \subsection{Normal ordering}
   \label{normal ordering} 

   Given two matrix-valued series 
   $\fbf(x) = x+\sum_{n=1}^\infty x^{n+1} a_n(\fbf)$ and 
   $\gbf(x) = x+\sum_{n=1}^\infty x^{n+1} a_n(\gbf)$, 
   we can not calculate $\fbf \circ \gbf(x)$ by replacing 
   $x$ by $\gbf(x)$ in the series defining $\fbf(x)$,
   before having fixed the position of the variable $x$
   in the series $\fbf(x)$.
   In fact, the problem is 
   to extend the domain of definition of a function 
   from a commutative domain to a non-commutative one. 
   For instance, consider the function 
   $\fbf: \C \longrightarrow M_4(\C)$ defined 
   as the multipication by a non-diagonal matrix $A\in M_4(\C)$, 
   $\fbf(x) = x A = A x$. 
   When we try to extend it to the larger domain $M_4(\C)$, 
   which contains $\C \cong \C \cdot \mathbf{1}$, 
   we observe a contradiction in the evaluation 
   on any matrix $B$: 
   $\tilde\fbf(B) = BA \neq AB = \tilde\fbf(B)$. 
   In other words, there are two possible ways to 
   extend the function $\fbf$ and the normal 
   ordering is just one of the two, chosen by fixing 
   the position of the extended (non-commutative) variable. 
   For instance, we can set $:\fbf(X): = X A$, 
   for any matrix $X$. 

   Such a problem arizes in QED when we want to consider 
   the bare propagators, which depend on the bare electric charge $e_0$, 
   as functions of the renormalized electric charge $e$. 
   In section \ref{junechar} we show that the perturbative expression 
   of $e_0=e_0(e)$ depends on the noncommutative variables 
   of the Hopf algebra $\Hp$. In section \ref{junedyso} 
   we will use the normal ordering to give a meaning to the 
   composition $D(e_0) = D(e_0(e))$ in the noncommutative context.

   Therefore, given any two matrix-valued formal series 
   $\fbf(x)$ and $\gbf(x)$ starting with $x$,
   we define the normally ordered composition as
   \begin{eqnarray}
           :\fbf \circ \gbf(x) : &=& 
           \gbf(x) \otimes 1 
           + \sum_{n=1}^\infty \gbf(x)^{n+1} \otimes a_n(\fbf)\\
   &=& x + \sum_{n=1}^\infty x^{n+1} \< \Dd \abf_n,\gbf \otimes
   \fbf\>.
   \label{NOextension}
   \end{eqnarray}

   Similarly, we can define the action of a series
   $\gbf(x)$ starting with $x$ on a series $\fbf(x)$ 
   starting with 1 as
   \begin{eqnarray}
           :\fbf \circ \gbf(x) :  &=& 1+ 
           \sum_{n=1}^\infty {\big(\gbf(x)\big)}^{n} \otimes b_n(\fbf) \nonumber\\ 
   &=& 1 + \sum_{n=1}^\infty x^{n} \< \ddd \bbf_n,\gbf \otimes
   \fbf\>\nonumber\\
           &=& 1+\sum_{n=1}^\infty x^n \langle \gbf \otimes \fbf, 
      1\otimes \bbf_n+
           \sum_{m=1}^{n-1} Q^{n-m-1}_m(\abf_*)\otimes \bbf_{n-m}\rangle.\label{NOaction}
   \end{eqnarray}

   \begin{remark} 
   When the formal series converge to an analytical
   function, a non-perturbative expression for the
   normally ordered composition can be given as
   \begin{eqnarray*}
           :\fbf \circ \gbf(x) : &=& 
           \frac{1}{2\pi i} \int_\gamma 
           \frac{\dd z}{z-\gbf(x)}\otimes \fbf(z), 
   \end{eqnarray*}
   where $\gamma$ is a path in the complex plane surrounding the origin.
   This can be shown by adapting Kato's definition
   of the scalar functions on operators to matrix-valued
   functions $\fbf(z)$ (see \cite{Kato}).
   \end{remark}


   %
   %
    
   \section{Renormalization of the electric charge}
   \label{junechar}

   \subsection{Renormalization of $e_0$} 

   As we said in section \ref{juneren}, 
   Dyson's formulas (\ref{Z2},\ref{Z3}) for the renormalization of QED 
   (plus the Ward identities) imply that the relation between 
   the bare electric charge $e_0$ and the renormalized electric 
   charge $e$ is given only by means of the photon renormalization 
   factor $Z_3(e)$, according to Eq.~(\ref{Z3e0}), 
   \begin{eqnarray*}
   \label{rencharge}
           e_0 &=& Z_3(e)^{-1/2} e. 
   \end{eqnarray*}
   We show that such an equation makes sense in a noncommutative
   context. 

   Recall that $Z_3(e)$ is a formal power series on $e$ with 
   entries which depend on the elements of the noncommutative 
   Hopf algebra $\Hp$. 
   More precisely, from Eqs.(\ref{Z3}) and (\ref{zeta3n}) we know that it is 
   expanded over the photon generating trees $V(t)=\|\vee t$, 
   so that
   \begin{eqnarray}
           Z_3(e) &=& 1 - \sum_{n=1}^\infty e^{2n} \zeta_3(\vbf_n) 
           \nonumber \\
           &=& \zeta_3(1 - \sum_{n=1}^\infty e^{2n} \vbf_n) , 
           \label{Z3(e)}
   \end{eqnarray}
   where $\vbf_n = \sum_{|t|=n-1} V(t) \in \Hp$ was defined in 
   (\ref{defvbf}) and the $\C$-valued function $\zeta_3$ 
   is extended linearly from $\Hp$ to $\Hp[[e]]$. 

   \begin{proposition}
   \label{chargeNCR}
   For any $n \geq 1$, define an element $\wbf_n$ in $\Hp$ as 
   \begin{eqnarray}
           \wbf_n &=& \sum_{k=1}^n \frac{(2k-1)!!}{k! 2^k}
           P_n^{(k)}(\vbf_*), \label{defwbf} 
   \end{eqnarray} 
   and two elements $\Zpbf(e)$ and $\ebf(e)$ 
   in $\Hp[[e^2]]$ as  
   \begin{eqnarray} 
           \Zpbf(e) &=& 1 - \sum_{n=1}^\infty e^{2n} \vbf_n , 
           \label{defZpbf} \\ 
           \ebf(e) &=& e + \sum_{n=1}^\infty e^{2n+1} \wbf_n .  
           \label{defebf} 
   \end{eqnarray}
   Then the $Z_3$ factor and the bare electric charge become 
   \begin{eqnarray}
           Z_3(e) &=& \zeta_3(\Zpbf(e)), \\ 
           e_0(e) &=& \zeta_3(\ebf(e)),  
   \end{eqnarray} 
   and the charge renormalization (\ref{Z3e0}) 
   can be deduced from a noncommutative analogue in $\Hp[[e]]$, 
   \begin{eqnarray}
           \ebf(e) &=& e\Zpbf(e)^{-1/2}. \label{Zobf=Zopf}
   \end{eqnarray}
   \end{proposition}

   \begin{proof}
   Starting from the definitions (\ref{defwbf}), 
   (\ref{defZpbf}) and (\ref{defebf}), we show that 
   equation (\ref{Zobf=Zopf}) holds.  
   \begin{eqnarray*}
           \Zpbf(e)^{-1/2} 
           &=& \big(1-\sum_{p=1}^\infty e^{2p} 
           \vbf_p \big)^{-1/2} \\ 
           &=& 1 + \sum_{k=1}^\infty \frac{(2k-1)!!}{k! 2^k} 
           \big( \sum_{p=1}^\infty e^{2p} \vbf_p \big)^k \\ 
           &=& 1 + \sum_{k=1}^\infty \frac{(2k-1)!!}{k! 2^k} 
           \sum_{p_1,...,p_k \geq 1} e^{2(p_1+...+p_k)} 
           \vbf_{p_1} \cdots \vbf_{p_k} \\ 
           &=& 1 + \sum_{k=1}^\infty \frac{(2k-1)!!}{k! 2^k} 
           \sum_{n \geq k} e^{2n} \sum_{p_1+\cdots+p_k=n} 
           \vbf_{p_1} \cdots \vbf_{p_k} \\ 
           &=& 1 + \sum_{n=1}^\infty e^{2n} 
           \sum_{k=1}^n \frac{(2k-1)!!}{k! 2^k} 
           P_n^{(k)}(\vbf_*) \\ 
           &=& 1 + \sum_{n=1}^{\infty} e^{2n} \wbf_n = \frac{\ebf(e)}{e}.  
   \end{eqnarray*}
   \end{proof}

   Remark that inverting Eq.~(\ref{defwbf}) we obtain 
   \begin{eqnarray*}
           \vbf_n &=& \sum_{k=1}^n (-1)^{k+1} (k+1) 
           P_n^{(k)}(\wbf_*). 
   \end{eqnarray*} 

   \begin{theorem}
   \label{iso}
   The Connes-Kreimer map $\Psi : \Ho \longrightarrow \Hp$ given by 
   \begin{eqnarray*}
           \Psi(\abf_{2n}) &=& \wbf_n 
   \end{eqnarray*} 
   is an injective morphism of Hopf algebras. Therefore its
   image identifies the set of renormalization transformation 
   of the electric charge as the noncommutative Hopf algebra 
   of odd formal diffeomorphisms. 
   \end{theorem}

   First we find a useful relation between the coefficients 
   $\ubf_n$ and $\wbf_n$. 

   \begin{lemma}
   For any $n \geq 1$, the following relations hold: 
   \begin{eqnarray}
           \wbf_n &=& \frac{1}{2} \ubf_n 
           - \frac{1}{2} \sum_{p=1}^{n-1} \wbf_p \wbf_{n-p}, 
           \label{w-u} \\ 
           \ubf_n &=& 2 \wbf_n + \sum_{p=1}^{n-1} \wbf_p \wbf_{n-p} 
           \label{u-w}.  
   \end{eqnarray} 
   \end{lemma}

   \begin{proof}
   Relation (\ref{w-u}) is the inverse of relation (\ref{u-w}). 
   Let us prove the second one. 

   If we introduce the definitions of $\ubf_n$
   and $\vbf_n$ (Eqs.(\ref{defubf},\ref{defvbf})) into 
   Eq.(\ref{inversephoton}), we obtain
   \begin{eqnarray*}
   \big(1+\sum_{n=1}^\infty e^{2n} \ubf_n \big)^{-1} 
   &=& 1-\sum_{n=1}^\infty e^{2n}\vbf_n.
   \end{eqnarray*}
   Thus
   \begin{eqnarray*}
   \frac{1}{\Zpbf(e)} = 1+\sum_{n=1}^\infty e^{2n} \ubf_n.
   \end{eqnarray*}
   We use the fact
   \begin{eqnarray*}
   \frac{\ebf^2}{e^2} = \frac{1}{\Zpbf(e)} 
   \end{eqnarray*}
   to have the equality 
   \begin{eqnarray*}
           1 +  \sum_{n \geq 1} \ubf_n e^{2n}&=& 
           \big(1 +  \sum_{n \geq 1} \wbf_n e^{2n} \big)^2 \\ 
           &=& 1 + \sum_{n \geq 1} \left[ 2 \wbf_n 
           + \sum_{p=1}^{n-1} \wbf_{p}\wbf_{n-p} \right]
           e^{2n},  
   \end{eqnarray*} 
   from which follows relation (\ref{u-w}). 
   \end{proof}

   \begin{proof of}{iso}
   We have to prove that 
   \begin{eqnarray*}
           (\Psi \otimes \Psi) \Dd(\abf_{2n}) &=& \Dp(\Psi(\abf_{2n})), 
   \end{eqnarray*} 
   that is, 
   \begin{eqnarray*}
           \Dp(\wbf_n) &=& \wbf_n \otimes \| + \| \otimes \wbf_n 
           + \sum_{m=1}^{n-1} \Qbar_m^n(\wbf_*) \otimes \wbf_{n-m}, 
   \end{eqnarray*} 
   where the polynomials $\Qbar_m^n(\wbf_*) = Q_m^{2n-2m}(\wbf_*)$ 
   were defined in (\ref{polQ}). 
   We prove it by induction, using the quadratic 
   expression on the $\wbf$'s and $\ubf$'s of Eq.~(\ref{w-u}). 

   Suppose that the formula is true for all $\wbf_k$ with $k<n$. 
   Write $\wbf_n$ as in (\ref{w-u}), look at  
   \begin{eqnarray*}
           \Dp(\wbf_n) &=& \frac{1}{2} \Dp(\ubf_n) - \frac{1}{2} 
           \sum_{p=1}^{n-1} \Dp(\wbf_p) \Dp(\wbf_{n-p}) 
   \end{eqnarray*} 
   and compute each summand separately. 

   Take the term $\Dp(\ubf_n)$ as computed in (\ref{Deltaubf}), it yields  
   \begin{eqnarray*}
           \frac{1}{2} \Dp \ubf_{n} &=& \frac{1}{2} \ubf_{n} \otimes \| 
           + \| \otimes \frac{1}{2} \ubf_{n} 
           + \sum_{p=1}^{n-1} R^{n}_p(\vbf_*) \otimes \frac{1}{2} \ubf_{n-p} \\ 
           &=& \wbf_{n} \otimes \| 
           + \frac{1}{2} \sum_{p=1}^{n-1} \left[ \wbf_p \wbf_{n-p} \otimes \| 
           + \| \otimes \wbf_p \wbf_{n-p} \right] 
           + \| \otimes \wbf_{n} \\ 
           && + \sum_{p=1}^{n-1} R^{n}_p(\vbf_*) \otimes 
           \Big[ \wbf_{n-p} + \frac{1}{2} \sum_{k=1}^{n-p-1} 
           \wbf_{k} \wbf_{n-p-k} \Big]. 
   \end{eqnarray*}

   The second term, by inductive hypothesis, is: 
   \begin{eqnarray*}
           \frac{1}{2} \sum_{p=1}^{n-1} \Dp(\wbf_p) \Dp(\wbf_{n-p}) &=&  \\
           && \hspace*{-4cm}= \frac{1}{2} \sum_{p=1}^{n-1} 
           \left[ \wbf_p \otimes \| + \| \otimes \wbf_p 
           + \frac{1}{2} \sum_{m=1}^{p-1} \Qbar_m^p \otimes \wbf_{p-m} \right] 
           \left[ \wbf_{n-p} \otimes \| + \| \otimes \wbf_{n-p} 
           + \frac{1}{2} \sum_{l=1}^{n-p-1} 
           \Qbar_l^{n-p} \otimes \wbf_{n-p-l} \right].
   \end{eqnarray*}
   Expanding the product gives 
   \begin{eqnarray*} 
           \frac{1}{2} \sum_{p=1}^{n-1} \Dp(\wbf_p) \Dp(\wbf_{n-p}) &=&  \\ 
           && \hspace*{-4cm}= \frac{1}{2} \sum_{p=1}^{n-1} 
           \left[ \wbf_p\wbf_{n-p} \otimes \| 
           + \| \otimes \wbf_p \wbf_{n-p} \right] 
           + \sum_{p=1}^{n-1} \wbf_{n-p}\otimes \wbf_p \\ 
           && \hspace*{-3cm}+ \frac{1}{2} \sum_{p=1}^{n-1} 
           \left[ \sum_{m=1}^{p-1} \Qbar_m^p \wbf_{n-p} \otimes \wbf_{p-m} 
           + \sum_{l=1}^{n-p-1} \wbf_p \Qbar_l^{n-p} \otimes 
           \wbf_{n-p-l} \right] \\ 
           && \hspace*{-3cm}+ \frac{1}{2} \sum_{p=1}^{n-1}
           \sum_{m=1}^{p-1} \sum_{l=1}^{n-p-1} 
           \left[ \Qbar_m^p \otimes \wbf_{p-m} \wbf_{n-p} 
           + \Qbar_l^{n-p} \otimes \wbf_p \wbf_{n-p-l} 
           + \Qbar_m^p \Qbar_l^{n-p} \otimes \wbf_{p-m} \wbf_{n-p-l} \right] . 
   \end{eqnarray*} 
   Hence we obtain 
   \begin{eqnarray*}
           \Dp(\wbf_n) &=& \wbf_{n} \otimes \| + \| \otimes \wbf_{n} \\ 
           && + \sum_{m=1}^{n-1} \left[ 
           R^{n}_{m} - \wbf_{m} 
           - \frac{1}{2} \sum_{l=1}^{m-1} \left( 
           \Qbar^{n-l}_{m-l} \wbf_l + \wbf_l \Qbar^{n-l}_{m-l} \right) \right] 
           \otimes \wbf_{n-m} \\ 
           && +  \frac{1}{2} \sum_{m=1}^{n-1} \sum_{p=1}^{n-m-1} \left[ 
           R^{n}_{m} - \Qbar_{m}^{m+p} - \Qbar_{m}^{n-p} 
           - \sum_{l=1}^{m-1} \Qbar_{l}^{l+p} \Qbar_{m-l}^{n-p-l} \right] 
           \otimes \wbf_p \wbf_ {n-m-p}    
   \end{eqnarray*} 
   and therefore 
   \begin{eqnarray*}
           \Dp(\wbf_n) &=& \wbf_{n} \otimes \| + \| \otimes \wbf_{n} 
           + \sum_{m=1}^{n-1} \Qbar_{m}^{n-m} \otimes \wbf_{n-m}
   \end{eqnarray*}  
   because of the two relations between the $R$'s and the $\Qbar$'s 
   proved in the following lemmas (\ref{uffa1}) and (\ref{uffa2}). 
   \end{proof of}

   \begin{lemma}
   The generating function 
   \begin{eqnarray}
           \Qbar(x,y) &=& \sum_{m=1}^\infty\sum_{n=0}^\infty 
           x^n y^{2m} \Qbar^{n+m}_m(\wbf_*) +\frac{1}{1-x}
   \end{eqnarray}
   satisfies the equation 
   \begin{eqnarray}
           \Qbar(x,y) = {\big(1-x\Zobf(y)^2\big)}^{-1}\Zobf(y), 
       \label{defQbar}
   \end{eqnarray}
   where $\Zobf(y) = 1+\sum_{n=1}^\infty y^{2n} \wbf_n \in \Hp[[y^2]]$. 
   \end{lemma}

   \begin{proof}
   For the proof, we use the combinatorial identity
   \begin{eqnarray*}
           \binom{2n-2m+1}{k} = 
           \binom{2n-2m-1}{k}+2\binom{2n-2m-1}{k-1}+\binom{2n-2m-1}{k-2}
   \end{eqnarray*}
   to derive the recurrence relation
   \begin{eqnarray*}
           \Qbar^{n}_m &=& \Qbar^{n-1}_m + 2 \wbf_m + P^{(2)}_m(\wbf_*)
           +2 \sum_{k=1}^{m-1} \wbf_k \Qbar^{n-k-1}_{m-k}
           + \sum_{k=2}^{m-2} P^{(2)}_k(\wbf_*) \Qbar^{n-k-1}_{m-k}.
   \end{eqnarray*}
   The expression for $\Qbar(x,y)$ follows by the usual method:
   in the last equation, we write $n=m+\alpha$, we multiply
   by $x^\alpha y^{2m}$ and we sum over $\alpha$ and $m$ from
   $1$ to $\infty$. This gives us
   \begin{eqnarray*}
   \Qbar(x,y)-\frac{1}{1-x}-\Zobf(y)+1 &=&
   \sum_{m=1}^\infty \sum_{\alpha=1}^\infty x^\alpha y^{2m} 
   \Qbar^{m+\alpha}_m \\
   &=& x\big(\Qbar(x,y)-\frac{1}{1-x}\big)
      +\frac{2x(\Zobf(y)-1)}{1-x}
      +\frac{x(\Zobf(y)-1)^2}{1-x}
    \\&&
      +2x(\Zobf(y)-1)\big(\Qbar(x,y)-\frac{1}{1-x}\big)
      +x(\Zobf(y)-1)^2\big(\Qbar(x,y)-\frac{1}{1-x}\big)\\
   &=& 1-\frac{1}{1-x} + x \big(\Zobf(y)\big)^2\Qbar(x,y).
   \end{eqnarray*}
   Solving for $\Qbar(x,y)$ yields Eq.(\ref{defQbar}).
   \end{proof}

   \begin{lemma}
   \label{uffa1} 
   The following identity holds for any $n,m \geq 1$: 
   \begin{eqnarray*}
           R^{n}_m(\vbf_*) &=& \Qbar^{n}_{m}(\wbf_*) + \wbf_{m} 
           + \frac{1}{2} \sum_{l=1}^{m-1} \left( 
           \Qbar_{m-l}^{n-l}(\wbf_*) \wbf_l + \wbf_l \Qbar_{m-l}^{n-l}(\wbf_*)
       \right). 
   \end{eqnarray*}
   \end{lemma}

   \begin{proof}
   Using Eq.(\ref{defQbar}) and $\Zobf(y)=(\Zpbf(y))^{-1/2}$ we
   can write
   \begin{eqnarray*}
   \Qbar(x,y)=\Big(1-\frac{x}{\Zpbf(y)}\Big) (\Zpbf(y))^{-1/2}
             =(\Zpbf(y)-x)^{-1}(\Zpbf(y))^{1/2}
             = R(x,y)(\Zpbf(y))^{1/2}.
   \end{eqnarray*}
   Therefore, $R(x,y)=\Qbar(x,y) \Zobf(y)$. Since
   $\Zobf(y)$ commutes with $\Qbar(x,y)$ (because of Eq.(\ref{defQbar})),
   we obtain
   \begin{eqnarray*}
   R(x,y)=\frac{1}{2} \big( \Qbar(x,y)\Zobf(y)+\Zobf(y)\Qbar(x,y)\big).
   \end{eqnarray*}
   In the expansion of this equality over $x$ and $y$,
   the term $x^n y^{2m}$, for any $n \geq 0$ and $m\geq 1$, 
   gives the equality 
   \begin{eqnarray*}
           R^{n+m}_m &=& \Qbar^{n+m}_{m} + \wbf_{m} 
           + \frac{1}{2} \sum_{l=1}^{m-1} \left( 
           \Qbar_{m-l}^{n+m-l} \wbf_l + \wbf_l \Qbar_{m-l}^{n+m-l} 
       \right)
   \end{eqnarray*}
   from which the lemma follows.
   \end{proof}

   \begin{lemma}
   \label{uffa2} 
   The following identity holds for any $n \geq 0$ and $m \geq 1$: 
   \begin{eqnarray*}
           R^n_{m}(\vbf_*) &=& \Qbar^{m+p}_{m}(\wbf_*) 
           + \Qbar^{n-p}_{m}(\wbf_*) + \sum_{l=1}^{m-1} 
           \Qbar^{p+l}_{l}(\wbf_*) \Qbar^{n-p-l}_{m-l}(\wbf_*).
   \end{eqnarray*}
   \end{lemma}

   \begin{proof}
   The definition of the generating function enables us to
   write
   \begin{eqnarray*}
           \Qbar(x,y)\Qbar(z,y) -\frac{1}{(1-x)(1-z)}&=&
           \sum_{m=1}^\infty \sum_{p=0}^\infty \sum_{q=0}^\infty  
           x^p y^{2m} z^q \Big(\Qbar^{m+p}_m +\Qbar^{m+q}_m 
           + \sum_{l=1}^{m-1} \Qbar^{l+p}_l\Qbar^{m-l+q}_{m-l}\Big) .
   \end{eqnarray*}
   On the other hand, from the relation between
   $\Qbar(x,y)$ and $R(x,y)$ we find
   \begin{eqnarray*}
           \Qbar(x,y)\Qbar(z,y)=\Zpbf(y)R(x,y)R(z,y).
   \end{eqnarray*}
   Now we use Eq.(\ref{resolventr}), the resolvent identity for
   $R(x,y)R(z,y)$, to write
   \begin{eqnarray*}
           \Zpbf(y)R(x,y)R(z,y) &=& \Zpbf(y) \frac{R(x,y)-R(z,y)}{x-z}
           =\frac{x R(x,y)-z R(z,y)}{x-z},
   \end{eqnarray*}
   where we have used the relations
   $\Zpbf(y) R(x,y) = 1+x R(x,y)$.
   Finally, we obtain
   \begin{eqnarray*}
           \Qbar(x,y)\Qbar(z,y) -\frac{1}{(1-x)(1-z)}&=&
           \frac{x R(x,y)-z R(z,y)}{x-z}-\frac{1}{(1-x)(1-z)} \\
           &=& \sum_{m=1}^\infty \sum_{k=0}^\infty y^{2m} 
           \frac{x^{k+1}-z^{k+1}}{x-z} R^{m+k}_m \\
           &=& \sum_{m=1}^\infty \sum_{p=0}^\infty \sum_{q=0}^\infty
           y^{2m} x^p z^q R^{m+p+q}_m ,
   \end{eqnarray*}
   since
   \begin{eqnarray*}
           \frac{x^{k+1}-z^{k+1}}{x-z} &=& \sum_{p+q=k} x^p z^q.
   \end{eqnarray*} 
   If we identify the coefficient of $x^p y^{2m} z^q$ on
   both sides, we obtain the identity 
   \begin{eqnarray*}
           R^{m+p+q}_{m}(\vbf_*) &=& \Qbar^{m+p}_{m}(\wbf_*) 
           + \Qbar^{m+q}_{m}(\wbf_*) + \sum_{l=1}^{m-1} 
           \Qbar^{l+p}_{l}(\wbf_*) \Qbar^{m-l+q}_{m-l}(\wbf_*).
   \end{eqnarray*}
   from which follows the lemma. 
   \end{proof}

    
   \subsection{Renormalization of $\alpha_0$ \label{alpha0}} 

   The Connes-Kreimer morphism $\Psi$ between the Hopf algebras 
   $\Ho$ and $\Hp$ was chosen according to the prescription
   $\ebf=e \Zpbf(e)^{-1/2}$, which corresponds to the choice
   of Ref.\cite{CKII}. However, a simpler morphism
   can be given between the Hopf algebras $\Hd$ and $\Hp$. 
   This is the subject of the present section.

   In QED, the propagators and renormalization factors
   depend on the squares $e_0^2$ and $e^2$ of the coupling 
   constants. Therefore, it is tempting to expand them 
   over the fine structure constants $\alpha_0=e_0^2/4\pi$ 
   and $\alpha=e^2/4\pi$.

   Let us set up the notations with respect to the new variable
   $\alpha$: 
   \begin{eqnarray}
           \Zpbf(\alpha) &=& 1 - \sum_{n=1}^\infty 
           (4\pi)^n \alpha^n\vbf_n , \nonumber \\
           \frac{1}{\Zpbf(\alpha)} &=& 1 + \sum_{n=1}^\infty 
           (4\pi)^n \alpha^n\ubf_n, \label{zpbfa} \\
           \mbox{\boldmath$\alpha$}_0 =\frac{\alpha}{\Zpbf(\alpha)}
           &=& \alpha + \sum_{n=1}^\infty (4\pi)^n \alpha^{n+1}\ubf_n.
   \nonumber
   \end{eqnarray}

   \begin{theorem}
   \label{renalpha}
   Let us define the Connes-Kreimer map 
   $\hat\Psi:\Hd \longrightarrow \Hp$ 
   as the algebra morphism given on the generators by 
   $\hat\Psi(\abf_n)=(4\pi)^n \ubf_n$ and $\hat\Psi(1)=\|$. 
   Then $\hat\Psi$ is an injective Hopf algebra morphism,
   and the Hopf algebra $\Hd$ can be seen as the set 
   of renormalization transformations of the 
   fine structure constant. 
   \end{theorem}

   \begin{lemma}
   \label{R=Q}
   For any $n \geq 0$ and any $m>0$ we have an identity 
   \begin{eqnarray*}
           R^{n}_m(\vbf_*) &=& Q^{n-m}_m(\ubf_*).
   \end{eqnarray*}
   \end{lemma}

   \begin{proof}
   To obtain this result, we start from the generating function
   $R(x,y)$, defined by Eq.~(\ref{Rgenfunc}), that we rewrite as  
   \begin{eqnarray*}
           R(x,y) &=& \big(\Zpbf(y)-x\big)^{-1} 
              = \Zpbf(y)^{-1} \pro \big[1-x\Zpbf(y)^{-1}\big]^{-1}. 
   \end{eqnarray*}
   From lemma (\ref{Inverseelectron}), by taking the sum 
   over all trees in Eq.~(\ref{inverseelectron}), 
   we see that the inverse of the series 
   $\Zpbf(y)= 1+\sum_{n=1}^\infty y^{2n} \vbf_n$ is 
   \begin{eqnarray*}
           \frac{1}{\Zpbf(y)} &=& 
           1-\sum_{n=1}^\infty y^{2n} \ubf_n. 
   \end{eqnarray*} 
   Now, in lemma (\ref{genQ(x,y)}) we show that if we set 
   $\fbf(y)= y +\sum_{n=1}^\infty y^{2n+1} \ubf_n = y \Zpbf(y)^{-1}$, 
   then $Q(x,y)$ satisfies the equation 
   \begin{eqnarray*}
           Q(x,y) &=& \big(\frac{y}{\fbf(y)}-x\big)^{-1} 
           = \fbf(y)\big[y-x \fbf(y)\big]^{-1} \\ 
           &=& \Zpbf(y)^{-1} \big[1-x \Zpbf(y)^{-1} \big]^{-1} 
           = R(x,y). 
   \end{eqnarray*}
   If we expand the generating functions $Q(x,y)$ and
   $R(x,y)$ over $x$ and $y$ and consider the
   coefficient of $x^n y^{2m}$ we obtain the expected result. 
   \end{proof}

   \begin{proof of}{renalpha}
   The map $\hat\Psi$ is obviously compatible with 
   the coproduct, because of lemma (\ref{R=Q}) and 
   proposition (\ref{Deltaubf}). In fact, 
   \begin{eqnarray*}
           (\hat\Psi \otimes \hat\Psi) \circ \Dd \abf_n 
           &=& \hat\Psi(\abf_n) \otimes 1 + 1 \otimes \hat\Psi(\abf_n)
           + \sum_{m=1}^{n-1} Q^{n-m}_m(\hat\Psi(\abf_*)) 
           \otimes \hat\Psi(\abf_{n-m}) \\ 
           &=& (4\pi)^n \ubf_n \otimes 1 + 1 \otimes (4\pi)^n \ubf_n 
           + \sum_{m=1}^{n-1} Q^{n-m}_m((4\pi)^* \ubf_*) 
           \otimes (4\pi)^{n-m} \ubf_{n-m} \\ 
           &=& \Dp ((4\pi)^n \ubf_n) = \Dp (\hat\Psi(\abf_n)). 
   \end{eqnarray*}
   \end{proof of}

   %
   %

   \section{Noncommutative Dyson's formulas}
   \label{junedyso}

   In this section, we show that Dyson's equations 
   (\ref{DysonS},\ref{DysonD}) can be seen as a
   consequence of the Connes-Kreimer map 
   of section \ref{junediff}, 
   between the Hopf algebra of formal diffeomorphisms
   and the Hopf algebra of QED renormalization.  

   But we shall prove more: Dyson's formulas are still valid when 
   $Z_2(e)$ and $Z_3(e)$ are matrices, that is, for their 
   noncommutative versions $\Zebf(e)$ and $\Zpbf(e)$. 
   In fact, scalar Dyson's formulas can be deduced from 
   analogue identities in the algebra $\Hq\otimes\Hq$.
   This is the essence of noncommutative renormalization.

   Dyson's formulas involve the bare and renormalized propagators, 
   which are functions respectively of the bare and of the 
   renormalized electric charge. In order to confront them, 
   we are forced to describe the explicit dependency of 
   the bare propagators on the renormalized charge, that is, 
   we are forced to perform a composition between the bare 
   propagators, as functions of $e_0$, and $e_0$ as a function 
   of $e$. 

   In section \ref{junechar} we showed that the bare electric charge 
   is the evaluation of a multiplicative scalar map $\zeta_3$ 
   on a non-commutative formal series, 
   \begin{eqnarray*}
           \ebf(e) &=& e - \sum_{n=1}^\infty e^{2n+1} \wbf_n 
           \in \Hp[[e]].  
   \end{eqnarray*} 
   Wishing to isolate some non-commutative analogue  
   of the bare propagators, we are forced to regard  
   the series in $\Hq[[e_0]]$ as series in $\Hq[[\ebf]]$, 
   and then compose them with $\ebf(e)$ as a series in 
   $\Hq[[e]]$. 
   As remarked in (\ref{NOextension}), when we want to extend 
   the domain of definition of a series, from the set 
   of complex numbers to a non-commutative set, 
   the normal ordering is required to fix the position of the
   variables.
   Therefore, in this section we make use of the notation
   of section \ref{normal ordering}. 


   \subsection{Dyson's formula for photons}

   \begin{proposition} 
   \label{DysonPhoton}
   The operators  
   \begin{eqnarray}
           :\Dbf(\ebf): &=& \|\otimes\| 
           + \sum_{n=1}^\infty \ebf^{2n} \otimes\ubf_n 
           \in \Hp[[\ebf^2]], \\
           \bar \Dbf(e) &=& \| \otimes \| 
           + \sum_{n=1}^\infty e^{2n} \Dp \ubf_n 
           \in \Hp\otimes\Hp[[e^2]]  
   \end{eqnarray}
   satisfy the noncommutative Dyson formula
   \begin{eqnarray}
       (\Zpbf(e) \otimes \|) \bar \Dbf(e) &=& :\Dbf(\ebf):
   \label{Dysonp}
   \end{eqnarray}
   for $\ebf = e \Zpbf(e)^{-1/2}$.

   Moreover, the bare and renormalized photon propagators can be 
   recovered as
   \begin{eqnarray*}
           D(e_0) &=& 
           \langle \zeta_3 \otimes \phip, :\Dbf(\ebf): \rangle , \\
           \bar D(e) &=& 
           \langle \zeta_3 \otimes \phip, \bar \Dbf(e) \rangle ,   
   \end{eqnarray*}
   and Dyson's formula (\ref{DysonD}) follows from Eq.~(\ref{Dysonp}). 
   \end{proposition}

   \begin{proof}
   Equation (\ref{Dysonp}) is quite easy to prove. 
   Since 
   \begin{eqnarray*}
           \Dp \ubf_n &=& \ubf_n \otimes \| + \| \otimes \ubf_n 
           + \sum_{m=1}^{n-1} R_m^n(\vbf_*) \otimes \ubf_{n-m} ,   
   \end{eqnarray*}
   if we expand the left hand-side of (\ref{Dysonp}) we obtain
   \begin{eqnarray*}
     (\Zpbf(e) \otimes \|) \bar \Dbf(e) &=& \|\otimes\|
     +\sum_{n=1}^\infty e^{2n} \Big[ 
       (\ubf_n-\vbf_n-\sum_{k=1}^{n-1}\vbf_k\ubf_{n-k})\otimes \|
       + \|\otimes \ubf_n \\&&
       + \sum_{m=1}^{n-1} ( R^n_m-\vbf_m-\sum_{k=1}^{m-1}\vbf_k R^{n-k}_{m-k})
           \otimes \ubf_{n-m} \Big].
   \end{eqnarray*}
   The factor of $\otimes \|$ is zero because of Eq.(\ref{formuledinversion}), 
   and the factor of $\otimes \ubf_{n-m}$ is equal to $R^{n-1}_m$ because
   of Eq.(\ref{recurxx}). Hence,
   \begin{eqnarray*}
     (\Zpbf(e) \otimes \|) \bar \Dbf(e) &=& \|\otimes\|
     +\sum_{n=1}^\infty e^{2n} \Big( \|\otimes \ubf_n
       + \sum_{m=1}^{n-1}  R^{n-1}_m \otimes \ubf_{n-m} \Big)\\
     &=& \|\otimes\|
     +\sum_{n=1}^\infty e^{2n} \Big( \|\otimes \ubf_n
       + \sum_{m=1}^{n-1}  Q^{n-m-1}_m(\ubf_*) \otimes \ubf_{n-m} \Big).
   \end{eqnarray*} 
   On the right hand-side of (\ref{Dysonp}), apply the normally 
   ordered composition. Because the series $:\Dbf(\ebf):$ starts with 
   $1$, we apply the action (\ref{NOaction}). 
   Since the Connes-Kreimer map 
   transforms the coproduct on series into the photon coproduct, 
   and because $:\Dbf(\ebf):$ depends on the square of $\ebf$, 
   where 
   \begin{eqnarray*}
           \ebf^2 &=& \frac{e^2}{\Zpbf(e)} = 
           e^2 + \sum_{n=1}^\infty (e^2)^{n+1} \ubf_n,
   \end{eqnarray*}
   the series $:\Dbf(\ebf):$ is
   \begin{eqnarray*}
            :\Dbf(\ebf): &=& \|\otimes \| +\sum_{n=1}^\infty 
           e^{2n}\big( 1 \otimes \ubf_n 
           + \sum_{m=1}^{n-1}  Q^{n-m-1}_m(\ubf_*) \otimes \ubf_{n-m} \big)
   \end{eqnarray*}
   because the Connes-Kreimer map is a morphism of Hopf algebras. 
   Therefore, modulo the free action $\mu$, 
   we obtain 
   \begin{eqnarray*}
     (\Zpbf(e) \otimes \|) \bar \Dbf(e) &=& :\Dbf(\ebf):\, .
   \end{eqnarray*}

   To recover the classical Dyson's formula, 
   recall that the bare and renormalized photon propagators 
   were expanded in section \ref{juneren} as 
   \begin{eqnarray*}
           D(q,e_0) &=& 1 + \sum_{n=1}^\infty e_0^{2n} D_n(q), \\
           \bar D(q,e_0) &=& 1 + \sum_{n=1}^\infty e_0^{2n} \bar D_n(q), 
   \end{eqnarray*}
   and that the results of section \ref{juneorde} tell us how to find 
   the coefficients: 
   \begin{eqnarray*}
           D_n(q) &=& \phip(\ubf_n) = \<\phip , \ubf_n \>, \\
           \bar D_n(q) &=& \bar\phip(\ubf_n) 
           = \< \zeta_3 \otimes \phip , \Dp \ubf_n \> . 
   \end{eqnarray*}
   \end{proof}

   Since Dyson's formula is noncommutative, we can use
   operators and not only scalars as $Z_3$.
   An example is $Z_3=1+\Pi(q)$. Dyson's formula becomes now
   \begin{eqnarray*}
     (1+\Pi(q)) \bar D(q;e) = \,
            : D \big(q;\frac{e}{\sqrt{1+\Pi(q)}}\big) : \, .
   \end{eqnarray*}

   Similar algebraic relations can be established
   for the vacuum polarization. 

   \begin{proposition}
   The operators  
   \begin{eqnarray}
           :\Pibf(\ebf): &=& \sum_{n=1}^\infty \ebf^{2n} \otimes \vbf_n 
           \in \Hp[[\ebf^2]], \\
           \bar\Pibf(e) &=& \sum_{n=1}^\infty e^{2n} \Dp \vbf_n 
           \in \Hp\otimes\Hp[[e^2]]   
   \end{eqnarray}
   satisfy the noncommutative Dyson formula
   \begin{eqnarray}
           \| \otimes \| - \bar\Pibf(e) &=& 
           (\Zpbf(e) \otimes \|) (\| \otimes \| - :\Pibf(\ebf):), 
   \end{eqnarray}
   for $\ebf = e \Zpbf(e)^{-1/2}$.

   The bare and renormalized vacuum polarizations can be 
   recovered as
   \begin{eqnarray*}
           \Pi(e_0) &=& 
           \langle \zeta_3 \otimes \phip, :\Pibf(\ebf): \rangle , \\
           \bar\Pi(e) &=& 
           \langle \zeta_3 \otimes \phip, \bar\Pibf(e) \rangle .    
   \end{eqnarray*}
   \end{proposition}

   \begin{proof}
   The proof is similar to that of (\ref{DysonPhoton}). 
   \begin{eqnarray}
     :\Pibf(e_0): &=& \sum_{n=1}^\infty e_0^{2n} \otimes\vbf_n  \\
     &=& \sum_{n=1}^\infty e^{2n} \big( \|\otimes\vbf_n 
         +\sum_{m=1}^{n-1} Q^{n-m-1}_m(\ubf_*)\otimes \vbf_{n-m}\big)\\
     &=& \sum_{n=1}^\infty e^{2n} \big( \|\otimes\vbf_n 
         +\sum_{m=1}^{n-1} R^{n-1}_m\otimes \vbf_{n-m}\big)\\
     \bar \Pibf(e) &=& \sum_{n=1}^\infty e^{2n} \big[ 
      \vbf_n \otimes \| + \| \otimes \vbf_n 
      + \sum_{m=1}^{n-2} R_m^{n-2}(\vbf_*) \otimes \vbf_{n-m} \big].  
   \end{eqnarray}

   Then, using Eq.(\ref{recurxx}), it is straightforward to show that
   \begin{eqnarray*}
   \|\otimes\| - \bar\Pibf(e) = (\Zpbf(e)\otimes\|)(\|\otimes\| -
   \Pibf(e_0)).
   \end{eqnarray*} 
   \end{proof}

   As a noncommutative example of this second type of Dyson's formula,
   we take $Z_3=1-\Pi(p;e)$:
   \begin{eqnarray*}
   1 - \bar\Pi(p,q;e) = \big(1 - \Pi(p;e)\big)\Big(1 -
   :\Pi(q;\frac{e}{\sqrt{1 - \Pi(p;e)}}):\Big).
   \end{eqnarray*} 


   \subsection{Dyson's formula for electrons}

   The case of the electron is very similar. We give only
   the definitions and the results. 

   \begin{proposition} 
   \label{DysonElectron}
   The operators  
   \begin{eqnarray}
           :\Sbf(\ebf): &=& \|\otimes\| 
           + \sum_{n=1}^\infty \ebf^{2n} \otimes \xbf_n 
           \in \He[[\ebf^2]], \\
           \bar\Sbf(e) &=& \| \otimes \| 
           + \sum_{n=1}^\infty e^{2n} \De \xbf_n 
           \in \Hq\otimes\He[[e^2]] \\ 
           \Zebf(e) &=& \| + \sum_{n=1}^\infty e^{2n} \ybf_n 
           \in \He[[e^2]] 
   \end{eqnarray}
   satisfy the noncommutative Dyson formula in $\Hq\otimes\He[[e^2]]$ 
   \begin{eqnarray}
       (\Zebf(e) \otimes \|) \bar\Sbf(e) &=& :\Sbf(\ebf):
   \label{Dysone}
   \end{eqnarray}
   for $\ebf = e \Zpbf(e)^{-1/2}$.

   Moreover, the bare and renormalized electron propagators can be 
   recovered as
   \begin{eqnarray*}
           S(e_0) &=& 
           \langle (\zeta_3,\zeta_2) \otimes \phie, :\Sbf(\ebf): \rangle , \\
           \bar S(e) &=& 
           \langle (\zeta_3,\zeta_2) \otimes \phie, \bar \Sbf(e) \rangle ,   
   \end{eqnarray*}
   and Dyson's formula (\ref{DysonS}) follows from Eq.~(\ref{Dysone}). 
   \end{proposition}

   \begin{proposition} 
   The operators  
   \begin{eqnarray}
           :\Sigmabf(\ebf): &=& - \sum_{n=1}^\infty \ebf^{2n} \otimes \ybf_n 
           \in \He[[\ebf^2]], \\
           \bar\Sigmabf(e) &=& - \sum_{n=1}^\infty e^{2n} \De \ybf_n 
           \in \Hq\otimes\He[[e^2]]
   \end{eqnarray}
   satisfy the noncommutative Dyson formula in $\Hq\otimes\He[[e^2]]$ 
   \begin{eqnarray}
           \|\otimes\|-\bar \Sigmabf(e) &=& 
           (\Zebf(e)\otimes\|) \big(\|\otimes\|-:\Sigmabf(\ebf):\big) .
   \label{DysonSigma}
   \end{eqnarray}

   The bare and renormalized electron self-energies can be 
   recovered as
   \begin{eqnarray*}
           \Sigma(e_0) &=& 
           \langle (\zeta_3,\zeta_2) \otimes \phie, :\Sigmabf(\ebf): \rangle , \\
           \bar\Sigma(e) &=& 
           \langle (\zeta_3,\zeta_2) \otimes \phie, \bar \Sigmabf(e) \rangle .   
   \end{eqnarray*}
   \end{proposition}

   %
   %

   \section{Conclusion and perspective}

   In this paper we provided explicit formulas, valid at all orders,
   for the renormalization of massless QED. These formulas
   are equivalent to Zimmermann's forest formula
   for the propagators, self-energy or vacuum polarization
   but they use the coefficients of $e^{2n}$ in the perturbative
   expansion instead of the Feynman diagrams. 
   The coefficients of $e^{2n}$ 
   make full use of Ward's identity $Z_1=Z_2$ and 
   decrease the number of integrals to calculate.
   This number grows as a polynomial in $n$ instead of 
   an exponential of $n$ as is the case for Feynman
   diagrams.

   Besides this practical aspect, this paper presented
   a noncommutative extension of renormalization which
   enables us to consider the three lepton generations
   as a single quantity and to renormalize them
   in one stroke. This provides a proper framework 
   to investigate lepton transmutation \cite{Bordes}.
   The Hopf algebra structure of noncommutative renormalization
   preserves the main properties of renormalization, such
   as its coassociativity, which is the root of the renormalization
   group.
   The Connes-Kreimer map still exists, but the
   objects corresponding to the group of diffeomorphisms
   remains to be explored in the noncommutative
   setting.

   In forthcoming publications, we shall present two
   extensions of the present work. Firstly, we shall
   consider the case of massive QED. The formulas
   for renormalized massive propagators are simple
   extensions of those for massless propagators,
   involving the derivatives of the bare propagators
   with respect to the mass $m$. However, the extension of
   the Hopf algebra of renormalization is more involved.
   Secondly, we shall investigate more fully the
   bigrading of QED, where photon and electron loops
   are distinguished. In particular, we shall give
   explicit formulas for the renormalization at
   a fixed number of photon and electron loops.
   The loop bigrading is interesting
   because the series over electron loops can be
   summed to all orders, providing a systematic way
   to obtain semi-perturbative (or semi-nonperturbative)
   equations for the propagators, where the electron
   loops are summed to all orders and the photon loops
   are introduced one by one.

   \section*{Acknowledgements}
   We are grateful to the Institut Girard Desargues of the University of
   Lyon~I for kind hospitality and support.
   We thank Dirk Kreimer and Jean-Louis Loday
   for helpful discussions.
   This is IPGP contribution \#0000.

   %
   %

   \end{fmffile}
   \end{document}